\documentclass[12pt,letterpaper,twopage]{article}

\PassOptionsToPackage{final}{graphicx}
\usepackage{amsmath}
\usepackage{times}
\usepackage{graphicx}
\usepackage{hyperref}            
\usepackage[natbibapa]{apacite}  
\renewcommand{\abstract}[1]{
\thispagestyle{empty}
\markboth{\runauthor}{\runtitle}
\ \vspace{-0mm}\\
\begin{center} {\bf Abstract} \end{center}
#1
}

\PassOptionsToPackage{noabbrev,capitalise}{cleveref}
\PassOptionsToPackage{font=normalsize}{subcaption}

\makeatletter
\newcommand{\providepackage}[2][]{
  \@ifpackageloaded{#2}{}{\usepackage[#1]{#2}}}
\makeatother

\providepackage{amsmath}
\providepackage{amssymb}
\providepackage{bm}
\providepackage[T1]{fontenc}
\providepackage[utf8]{inputenc}
\providepackage{graphicx}
\providepackage{mathtools} 
\providepackage{braket}  
\providepackage{etoolbox}
\providepackage{xcolor}
\providepackage{algorithm}
\providepackage{algpseudocode}
\providepackage{booktabs}
\providepackage{placeins}  
\providepackage{float}
\providepackage{changepage}  

\providepackage{siunitx}
\providepackage{cite}
\providepackage{twoopt}
\providecommandtwoopt{\citep}[3][][]{\cite{#3}}
\providecommandtwoopt{\citet}[3][][]{\cite{#3}}

\providecommand{\thickhline}{\hline}   
\providecommand{\toprule}{\thickhline}
\providecommand{\bottomrule}{\thickhline}
\providecommand{\midrule}{\hline}
\providecommand{\cmidrule}[1]{\cline{#1}}

\providepackage{nameref}
\providepackage{hyperref}
\hypersetup{draft=false,colorlinks,breaklinks=true,
            linkcolor=black,citecolor=black,filecolor=black}
\providepackage{cleveref}
\makeatletter
\if@cref@capitalise
  \if@cref@abbrev
    \crefname{subfigure}{Fig.}{Figs.}
  \else
    \crefname{subfigure}{Figure}{Figures}
  \fi
\else
  \if@cref@abbrev
    \crefname{subfigure}{fig.}{figs.}
  \else
    \crefname{subfigure}{figure}{figures}
  \fi
\fi
\makeatother

\Crefname{subfigure}{Figure}{Figures}
\newcommand{\creffloat}[1]{\namecref{#1}~\ref{#1}}
\newcommand{\creffloats}[2]{\namecrefs{#1}~\ref{#1} and~\ref{#2}}
\newcommand{\creffloatrange}[2]{\namecrefs{#1}~\ref{#1}--\ref{#2}}

\newlength{\fullwidth}
\newlength{\colwidth}
\newlength{\maxwidth}
\newlength{\maxheight}
\patchcmd{\smallmatrix}{\thickspace}{\kern.6em}{}{}
\patchcmd{\smallmatrix}{\baselineskip6}{\baselineskip9}{}{}

\newlength{\oldtabcolsep}
\providecommand{\thetitle}{}
\providecommand{\runtitle}{}
\providecommand{\runauthor}{}
\renewcommand{\title}[1]{\renewcommand{\thetitle}{#1}}
\providecommand{\titlerunning}[1]{\renewcommand{\runtitle}{#1}}
\providecommand{\authorrunning}[1]{\renewcommand{\runauthor}{#1}}

\makeatletter
\renewcommand{\fps@figure}{!htp}
\renewcommand{\fps@table}{!htbp}
\makeatother

\newcommand{\captitle}[1]{\textbf{#1}}
\providepackage{subcaption}
\newcommand{\subrefformat}[1]{\textbf{#1}}
\newcommand{\subreflabel}[1]{\subrefformat{\subref{#1}.}}
\newcommand{\subreflabels}[2]{\subrefformat{\subref{#1},\subref{#2}.}}
\DeclareCaptionSubType[Alph]{figure}  
\DeclareCaptionSubType[Alph]{table}

\let\oldcaption\caption
\newenvironment{inlinetable}[1][]  
               {\renewcommand{\caption}[1]{
                  \let\caption\oldcaption
                  \captionof{table}{##1} \vspace{0.5ex}}
                \vspace{1ex} \noindent\begin{minipage}{\textwidth}}
               {\end{minipage} \vfill\null}
\newenvironment{inlinefigure}[1][]
               {\renewcommand{\caption}[1]{
                  \let\caption\oldcaption
                  \vspace{0.5ex}\captionof{figure}{##1}}
                \vspace{1ex} \noindent\begin{minipage}{\textwidth}}
               {\end{minipage} \vfill\null}

\makeatletter
\newcommand{\eqname}{\cref@equation@name}
\newcommand{\Eqname}{\Cref@equation@name}
\newcommand{\eqnames}{\cref@equation@name@plural}
\newcommand{\Eqnames}{\Cref@equation@name@plural}
\newcommand{\figname}{\cref@figure@name}
\newcommand{\Figname}{\Cref@figure@name}
\newcommand{\fignames}{\cref@figure@name@plural}
\newcommand{\Fignames}{\Cref@figure@name@plural}
\makeatother

\DeclareUnicodeCharacter{00A0}{~}
\DeclareUnicodeCharacter{2010}{-}
\DeclareUnicodeCharacter{2013}{--}
\DeclareUnicodeCharacter{2014}{---}
\DeclareUnicodeCharacter{2009}{\,}
\DeclareUnicodeCharacter{2026}{\textellipsis}

\DeclareUnicodeCharacter{0391}{A}
\DeclareUnicodeCharacter{0392}{B}
\DeclareUnicodeCharacter{0393}{\Gamma}
\DeclareUnicodeCharacter{0394}{\Delta}
\DeclareUnicodeCharacter{0395}{E}
\DeclareUnicodeCharacter{0396}{Z}
\DeclareUnicodeCharacter{0397}{H}
\DeclareUnicodeCharacter{0398}{\Theta}
\DeclareUnicodeCharacter{0399}{I}
\DeclareUnicodeCharacter{039A}{K}
\DeclareUnicodeCharacter{039B}{\Lambda}
\DeclareUnicodeCharacter{039C}{M}
\DeclareUnicodeCharacter{039D}{N}
\DeclareUnicodeCharacter{039E}{\Xi}
\DeclareUnicodeCharacter{039F}{O}
\DeclareUnicodeCharacter{03A0}{\Pi}
\DeclareUnicodeCharacter{03A1}{P}
\DeclareUnicodeCharacter{03A3}{\Sigma}
\DeclareUnicodeCharacter{03A4}{T}
\DeclareUnicodeCharacter{03A5}{Y}
\DeclareUnicodeCharacter{03A6}{\Phi}
\DeclareUnicodeCharacter{03A7}{X}
\DeclareUnicodeCharacter{03A8}{\Psi}
\DeclareUnicodeCharacter{03A9}{\Omega}

\DeclareUnicodeCharacter{03B1}{\alpha}
\DeclareUnicodeCharacter{03B2}{\beta}
\DeclareUnicodeCharacter{03B3}{\gamma}
\DeclareUnicodeCharacter{03B4}{\delta}
\DeclareUnicodeCharacter{03B5}{\epsilon}
\DeclareUnicodeCharacter{03B6}{\zeta}
\DeclareUnicodeCharacter{03B7}{\eta}
\DeclareUnicodeCharacter{03B8}{\theta}
\DeclareUnicodeCharacter{03B9}{\iota}
\DeclareUnicodeCharacter{03BA}{\kappa}
\DeclareUnicodeCharacter{03BB}{\lambda}
\DeclareUnicodeCharacter{03BC}{\mu}
\DeclareUnicodeCharacter{03BD}{\nu}
\DeclareUnicodeCharacter{03BE}{\xi}
\DeclareUnicodeCharacter{03BF}{\omicron}
\DeclareUnicodeCharacter{03C0}{\pi}
\DeclareUnicodeCharacter{03C1}{\rho}
\DeclareUnicodeCharacter{03C3}{\sigma}
\DeclareUnicodeCharacter{03C4}{\tau}
\DeclareUnicodeCharacter{03C5}{\upsilon}
\DeclareUnicodeCharacter{03C6}{\phi}
\DeclareUnicodeCharacter{03C7}{\chi}
\DeclareUnicodeCharacter{03C8}{\psi}
\DeclareUnicodeCharacter{03C9}{\omega}

\DeclareUnicodeCharacter{0229}{\c{e}}

\providepackage[strings]{underscore}

\newcommand\Tstrut{\rule{0pt}{2.6ex}}         

\makeatletter
\newcommand{\subalign}[1]{%
  \vcenter{%
    \Let@ \restore@math@cr \default@tag
    \baselineskip\fontdimen10 \scriptfont\tw@
    \advance\baselineskip\fontdimen12 \scriptfont\tw@
    \lineskip\thr@@\fontdimen8 \scriptfont\thr@@
    \lineskiplimit\lineskip
    \ialign{\hfil$\m@th\scriptstyle##$&$\m@th\scriptstyle{}##$\hfil\crcr
      #1\crcr
    }%
  }%
}
\makeatother

\DeclareMathOperator{\Expect}{E}

\DeclareMathOperator{\Binom}{Binom}
\DeclareMathOperator{\Bernoulli}{Bernoulli}
\DeclareMathOperator{\Uniform}{Uniform}

\DeclareMathOperator*{\RMSE}{RMSE}

\providepackage{mathrsfs}

\title{\noindent Inference of a Mesoscopic Population Model from Population Spike Trains}
\titlerunning{Inference of a mesoscopic population model}
\authorrunning{A. René, A. Longtin, J. Macke}

\begin{document}


\hspace{13.9cm}1

\ \vspace{20mm}\\

{\LARGE \thetitle}

\ \\
{\bf \large Alexandre~René$^{\displaystyle 1, \displaystyle 2, \displaystyle 3}$,
 André~Longtin$^{\displaystyle 1, \displaystyle 4}$,
 Jakob~H.~Macke$^{\displaystyle 2, \displaystyle 5}$}\\
{$^{\displaystyle 1}$Department of Physics, University of Ottawa, Ottawa, Canada}\\
{$^{\displaystyle 2}$Max Planck Research Group Neural Systems Analysis, Center of Advanced European Studies and Research (caesar), Bonn, Germany}\\
{$^{\displaystyle 3}$Institute of Neuroscience and Medicine (INM-6) and Institute for Advanced Simulation (IAS-6) and JARA-Institute Brain Structure-Function Relationships (INM-10), Jülich Research Centre, Jülich, Germany}\\
{$^{\displaystyle 4}$Brain and Mind Research Institute, University of Ottawa, Ottawa, Canada}\\
{$^{\displaystyle 5}$Computational Neuroengineering, Department of Electrical and Computer Engineering, Technical University of Munich, Germany}\\
%



\abstract{
To understand how rich dynamics emerge in neural populations, we require models exhibiting a wide range of activity patterns while remaining interpretable in terms of connectivity and single-neuron dynamics. However, it has been challenging to fit such mechanistic spiking networks at the single neuron scale to empirical population data. To close this gap, we propose to fit such data at a meso scale, using a mechanistic but low-dimensional and hence statistically tractable model.
The mesoscopic representation is obtained by approximating a population of neurons as multiple homogeneous `pools' of neurons, and modelling the dynamics of the aggregate population activity within each pool. We derive the likelihood of both single-neuron and connectivity parameters given this activity, which can then be used to either optimize parameters by gradient ascent on the log-likelihood, or to perform Bayesian inference using Markov Chain Monte Carlo (MCMC) sampling. We illustrate this approach using a model of generalized integrate-and-fire neurons for which mesoscopic dynamics have been previously derived, and show that both single-neuron and connectivity parameters can be recovered from simulated data. In particular, our inference method extracts posterior correlations between model parameters, which define parameter subsets able to reproduce the data. We compute the Bayesian posterior for combinations of parameters using MCMC sampling and investigate how the approximations inherent to a mesoscopic population model impact the accuracy of the inferred single-neuron parameters.
}


\section{Introduction}\label{introduction}

Neuron populations produce a wide array of complex collective dynamics. Explaining how these emerge requires a mathematical model that not only embodies the network interactions, but that is also parameterized in terms of interpretable neuron properties. Just as crucially, in order to draw data-supported conclusions, we also need to be able to infer those parameters from empirical observations. These requirements tend to involve a trade-off between model expressiveness and tractability. Low-dimensional state-space models \citep{pillowSpatiotemporalCorrelationsVisual2008,mackeEmpiricalModelsSpiking2011,zhaoInterpretableNonlinearDynamic2016,pandarinathInferringSingletrialNeural2018} are simple enough to allow for inference, but achieve that simplicity by focussing on phenomenology: any mechanistic link to the individual neurons is ignored.
Conversely, microscopic mechanistic models with thousands of simulated neurons do provide that link between parameters and output \citep{potjansCelltypeSpecificCortical2014,hawrylyczInferringCorticalFunction2016}; however, this complexity makes the analysis difficult and limited to networks with highly simplified architectures \citep{doironMechanicsStatedependentNeural2016,martiCorrelationsSynapsesPairs2018}. Since methods to fit these models to experimental data are limited to single neurons \citep{mensiParameterExtractionClassification2012}, it is also unclear how to set their parameters such that they capture the dynamics of large heterogeneous neural populations.

\begin{figure}
  \includegraphics{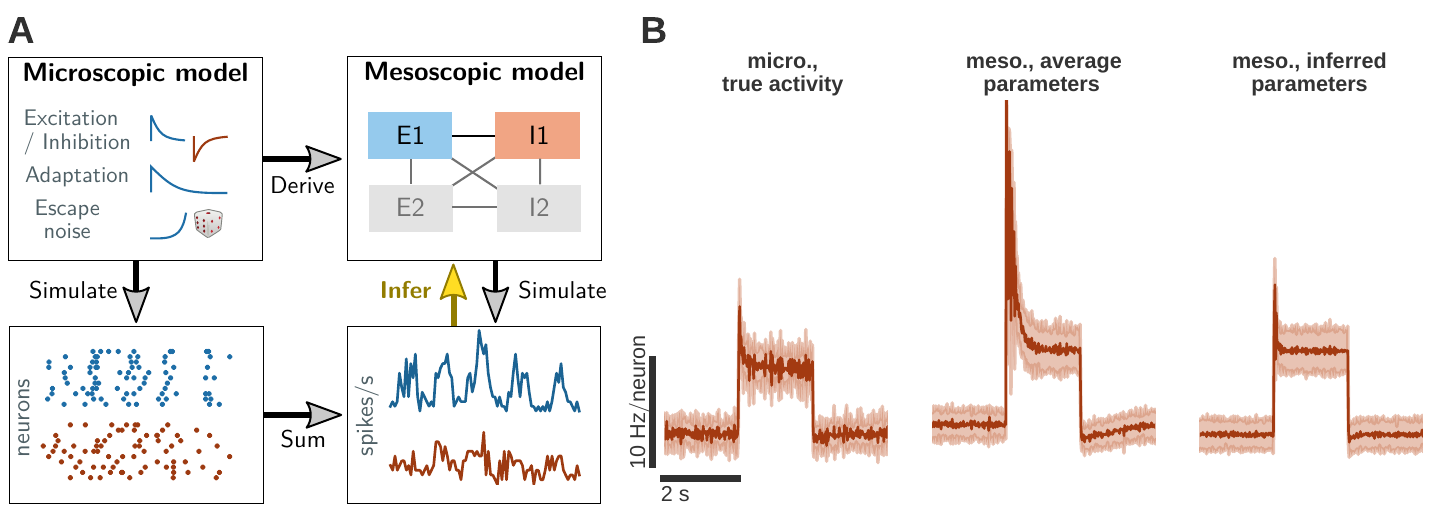}
 {\phantomsubcaption\label{fig:visual-abstract}}
 {\phantomsubcaption\label{fig:intro-inferred-vs-mean}}
 \caption{
  \subreflabel{fig:visual-abstract} \captitle{General procedure to infer parameters of a mesoscopic population model from microscopic data.} A microscopic model of GIF neurons is used to generate spike trains, which are averaged to obtain traces of population activity; these traces constitute our data. A
  mesoscopic model of either two or four populations is then fit to these traces. Simulating the mesoscopic model with the inferred parameters allows us to evaluate how well it reproduces the true dynamics.
  \subreflabel{fig:intro-inferred-vs-mean}
  \captitle{For heterogeneous systems, average parameters might not predict mean activity.} Mean activity (line) and its standard deviation (shaded area) for a heterogeneous microscopic model (left) and mesoscopic models attempting to approximate it (middle, right). A mesoscopic model constructed by averaging parameters across the microscopic population overestimates the population's variability (middle). Inferred parameters in this case deviate from these averages and provide a better representation of the true activity (right). Models are as in \creffloat{fig:hetero-params}; traces are for the inhibitory population. Means and standard deviations are computed from \num{50} realizations and averaged over disjoint bins of \SI{10}{ms}.
  \label{fig:intro}
 }
\end{figure}

To reduce the problem to a manageable size and scale, one can consider models that provide a mesoscopic dynamical description founded on microscopic single-neuron dynamics \citep{wallaceEmergentOscillationsNetworks2011,dumontStochasticfieldDescriptionFinitesize2017,nykampPopulationDensityApproach2000}. Specifically, we will focus on the model described in \citet{schwalgerTheoryCorticalColumns2017}, where neurons are grouped into putative excitatory (E) and inhibitory (I) populations in a cortical column. The key approximation is to replace each population with another of equal size, but composed of identical neurons, resulting in an effective mesoscopic model of homogeneous populations. In contrast with previous work on population rate dynamics \citep{gerstnerPopulationDynamicsSpiking2000,wilsonExcitatoryInhibitoryInteractions1972,nykampPopulationDensityApproach2000}, \citet{schwalgerTheoryCorticalColumns2017} correct their mean-field approximations for the finite size of populations. They are thus able to provide stochastic equations for the firing rate of each population with explicit dependence on the population sizes, neuron parameters, and connectivities between populations (\cref{fig:visual-abstract}, top). We use these equations to fit the model to traces of population activity.

Directly inferring mesoscopic model parameters has a number of advantages compared to extrapolating from those obtained by fitting a microscopic model. For one, it allows the use of data that do not have single-neuron resolution. In addition, since neuron parameters in a mesoscopic model represent a whole population, there may not be a clear way to relate micro- and mesoscopic parameters if the former are heteregeneous. By inferring population parameters from population recordings, we target the values that best compensate for the mismatch between the data and the idealized mesoscopic model (\cref{fig:intro-inferred-vs-mean}).


The method we present assumes that the model to be inferred can be expressed as a set of stochastic equations and that we have access to time series for both the observed (and possibly aggregated) neural activities and external input. It is thus not limited to mesoscale models, and could also be applied to e.g.\ Hodgkin-Huxley type neurons in isolation or networks. Nevertheless, in this paper, the underlying microscopic model does make the inferred parameters more readily interpretable, and provides a good idea of what values an inference algorithm should find for the parameters.

Methods have recently been developed for inferring models where stochastic equations are treated as a black box simulator
\citep{greenbergAutomaticPosteriorTransformation2019,papamakariosSequentialNeuralLikelihood2018,lueckmannFlexibleStatisticalInference2017,papamakariosFastEpsilonFree2016}. In such a case, one does not have access to the internal variables of the model and thus cannot compute the likelihood of its parameters; instead, these methods make use of repeated simulations to find suitable parameters. While this makes them applicable to a wide range of models, the repeated simulations can make them computationally expensive, and best suited to optimizing a set of statistical features rather than full time traces.
Moreover, for the models of interest here, the likelihood can be derived from the stochastic evolution equations.

We show in this work that the likelihood can indeed be used to infer model parameters using  non-convex optimization. The resulting optimization problem shares many similarities with training recurrent neural networks (RNNs) popular in machine learning \citep{waibelPhonemeRecognitionUsing1989,iangoodfellowDeepLearning2016}, and allows us to leverage optimization tools from that field. However, RNNs in machine learning are typically based on generic, non-mechanistic models, which implies that interpretation of the resulting network can be challenging (but see e.g.\ work on RNN visualization by Barak et~al. \citep{sussilloOpeningBlackBox2012,barakRecurrentNeuralNetworks2017,havivUnderstandingControllingMemory2019}). Thus, our approach can be regarded as complementary to RNN approaches, as we directly fit a mechanistically interpretable model.

This paper is organized as follows. In \cref{sec:model-summary,sec:recovering-parameters} we establish that maximum likelihood inference for our chosen mesoscopic model is sound, and in \cref{sec:data-requirements} provide empirical estimates for the amount of data this procedure requires. Using the example of heterogeneous populations, \cref{sec:inference-hetero-model} then shows how inference can find effective parameters which compensate for the mismatch between data and model. In \cref{sec:posteriors} we identify co-dependence between multiple model parameters by recovering the full Bayesian posterior. Finally, \cref{sec:4pop-fit} demonstrates that the approach scales well by considering a more challenging four population model with thirty-six free parameters.
\Cref{sec:discussion} discusses our results, with an emphasis on circumscribing the class of models amenable to our approach. Method details are provided in \cref{sec:methods}, along with technical insights gained as we adapted likelihood inference to a detailed dynamical model. Additional details, including a full specification of parameter values used throughout the paper, are given in \crefrange{sec:priors-parameters}{sec:stationary-state-derivation}.


\section{Results}
\label{sec:results}

\subsection{Model summary}
\label{sec:model-summary}

We studied the pair of microscopic and mesoscopic models presented in \citet{schwalgerTheoryCorticalColumns2017}, which is designed to represent excitatory (E) and inhibitory (I) populations of a putative cortical column of four neural layers \citep{potjansCelltypeSpecificCortical2014}. For this study we only considered layers 2/3 and 4, and made minor parameter adjustments to maintain realistic firing rates (c.f.\ \cref{sec:priors-parameters}). We also reduced all population sizes by a factor of 50 to ease the simulation of the microscopic model. This increases the variance of population activities, and so does not artificially simplify the task of inferring mesoscopic parameters.

\begin{table}
  \centering
  \caption{Key variable definitions.
           \label{tbl:variable-definitions}}
  \begin{tabular}{cl}
    \toprule
    Variable & Definition \tabularnewline\midrule
    $N_α$ & No. of neurons in population $α$. \tabularnewline
    $M$   & No. of populations, $α = 1,\dotsc,M$. \tabularnewline
    $L$   & No. of time steps used to compute the likelihood. \tabularnewline
    $Δt$  & Time step. \tabularnewline
    $\mathcal{I}_α$ & Set of indices of neurons belonging to population $α$. \tabularnewline
    $s_i(t)$ & 1 if neuron $i$ spiked within time window $[t, t+Δt)$, 0 otherwise. \tabularnewline
    $A_α(t)$ & Activity in population $α$ averaged over time window $[t, t+Δt)$. \tabularnewline
    $a_α(t)$ & Expectation of $A(t)$ conditioned on $\{A(t')\}_{t' < t}$. \tabularnewline
    \bottomrule
  \end{tabular}
\end{table}

The microscopic model is composed of either two or four populations of generalized integrate-and-fire (GIF) neurons. Neurons are randomly connected, with connectivity probabilities depending on the populations. The combination of excitatory and inhibitory input, along with internal adaptation dynamics, produces for each neuron $i$ a time-dependent firing rate $λ_i(t | \mathcal{H}_t)$; this rate is conditioned on the spike history up to $t$, denoted $\mathcal{H}_t$ (for equations see \cref{sec:details-micro-model}). Whether or not that neuron spikes within a time window $[t, t+Δt)$ is then determined by sampling a Bernoulli random variable \citep{schwalgerTheoryCorticalColumns2017}:
\begin{equation}
  \label{eq:s-lambda-relation}
  s_i(t | \mathcal{H}_t) \sim \Bernoulli(λ_i(t | \mathcal{H}_t) Δt) \,,
\end{equation}
where $Δt$ is chosen such that $λ_i(t | \mathcal{H}_t) Δt \ll 1$ is always true; we later refer to this stochastic process as \emph{escape noise}.
If all parameters are shared across all neurons within each population, we call this a \emph{homogeneous} microscopic model. Conversely, we call a model \emph{heterogeneous} if at least one parameter is unique to each neuron. We denote $\mathcal{I}_α$ the set of indices for neurons belonging to a population $α$.


The \emph{expected activity} $a_α$ of a population $α$ is the normalized expected number of spikes,
\begin{equation}
  \label{eq:exp-activity-definition}
  a_α(t | \mathcal{H}_t) = \frac{1}{N_α} \sum_{i \in \mathcal{I}_α} λ_i(t | \mathcal{H}_t) \,,
\end{equation}
which is a deterministic variable once we know the history up to $t$. In contrast, the \emph{activity} $A_α$ of that population is a random variable corresponding to the number of spikes actually observed,
\begin{align}
  \label{eq:activity-definition}
  A_α(t | \mathcal{H}_t) &\coloneqq \frac{1}{N_α} \sum_{i \in \mathcal{I}_α} s_i(t | \mathcal{H}_t) \,.
\end{align}

In practice data is discretized into discrete time steps $\{t_k\}_{k=1}^{L}$, which we assume to have uniform lengths $Δt$ and to be short enough for spike events of different neurons to be independent within one time step (this condition is always fulfilled when the time step is less than the synaptic transmission delay). Under these assumptions, \cref{eq:activity-definition} can be approximated by a binomial distribution \citep{schwalgerTheoryCorticalColumns2017},
\begin{equation}
  \label{eq:A-a-relation}
  A_α^{(k)} \coloneqq A_α(t_k | \mathcal{H}_{t_k}) \sim \frac{1}{N_α Δt} \Binom(N_α a_α(t_k | \mathcal{H}_{t_k}) Δt; N_α) \,.
\end{equation}
If we repeat a simulation $R$ times with the same input, we obtain an ensemble of histories $\{\mathcal{H}_{t_k}^r\}_{r=1}^R$ (due to the escape noise). Averaging over these histories yields the \emph{trial-averaged activity},
\begin{equation}
  \label{eq:trial-averaged-A}
  \bar{A}_{α}^{(k)} \coloneqq \frac{1}{R} \sum_{r=1}^R A_α(t_k | \mathcal{H}_{t_k}^r) \,,
\end{equation}
the theoretical counterpart to the peristimulus time histogram (PSTH).

For the \textbf{microscopic model}, the history is the set of all spikes,
\begin{equation}
  \label{eq:hist-micro}
  \mathcal{H}_{t_k} = \{s_i(t_l)\}_{\subalign{&i =1\dotsc N\\&t_l < t_k}} \,.
\end{equation}
To generate activities, we first generate spikes with \cref{eq:s-lambda-relation} and use \cref{eq:activity-definition} to obtain activities (c.f.\ \cref{fig:visual-abstract}).

For the \textbf{mesoscopic model}, hereafter referred to as ``mesoGIF{}'', the history only contains population activities:
\begin{equation}
  \label{eq:hist-meso}
  \mathcal{H}_{t_k} = \{A_α^{(l)}\}_{\subalign{&α =1\dotsc M\\&t_l < t_k}} \,.
\end{equation}
The expected activity is then an expectation over all spike sequences consistent with that history, for which a closed form expression was derived in \citet{schwalgerTheoryCorticalColumns2017} (the relevant equations are given in \cref{sec:meso-equations}). Activities are generated by using this expression to compute $a_α(t)$ and then sampling \cref{eq:A-a-relation}.
Unless mentioned otherwise, for the results reported in the sections below we used the microscopic model for data generation and the mesoscopic model for inference.

In addition to homogeneity of populations and independence of spikes within a time step, the mesoscopic model depends on one more key approximation: that neuron populations can be treated as \emph{quasi-renewal} \citep{naudCodingDecodingAdapting2012,schwalgerTheoryCorticalColumns2017}. If neurons are viewed as having both refractory and adaptation dynamics, this is roughly equivalent to requiring that the latter be either slow or weak with respect to the former. (A typical example where this approximation does not hold is bursting neurons \citep{naudCodingDecodingAdapting2012}.) Under these approximations, the unbounded history $\mathcal{H}_{t_k}$ can be replaced by a finite state vector ${S}^{(k)}$, which is updated along with the expected activity $a^{(t)}$ (c.f.\ \cref{sec:details-meso-model}). Since the update equations only depend on ${S}^{(k-1)}$, they are then Markovian in ${S}$. This in turn allows the probability of observations $P\left(A^{(L)}, A^{(L-1)},\dotsc,A^{(1)}\right)$ to be factorized as
$P\left(A^{(L)}|S^{(L)}\right) \cdot
P\left(A^{(L-1)}|S^{(L-1)}\right) \dotsb
P\left(A^{(1)}|S^{(1)}\right)$,
 which is key to making the inference problem tractable.

\subsection{Recovering population model parameters}
\label{sec:recovering-parameters}

\begin{figure}
  \includegraphics{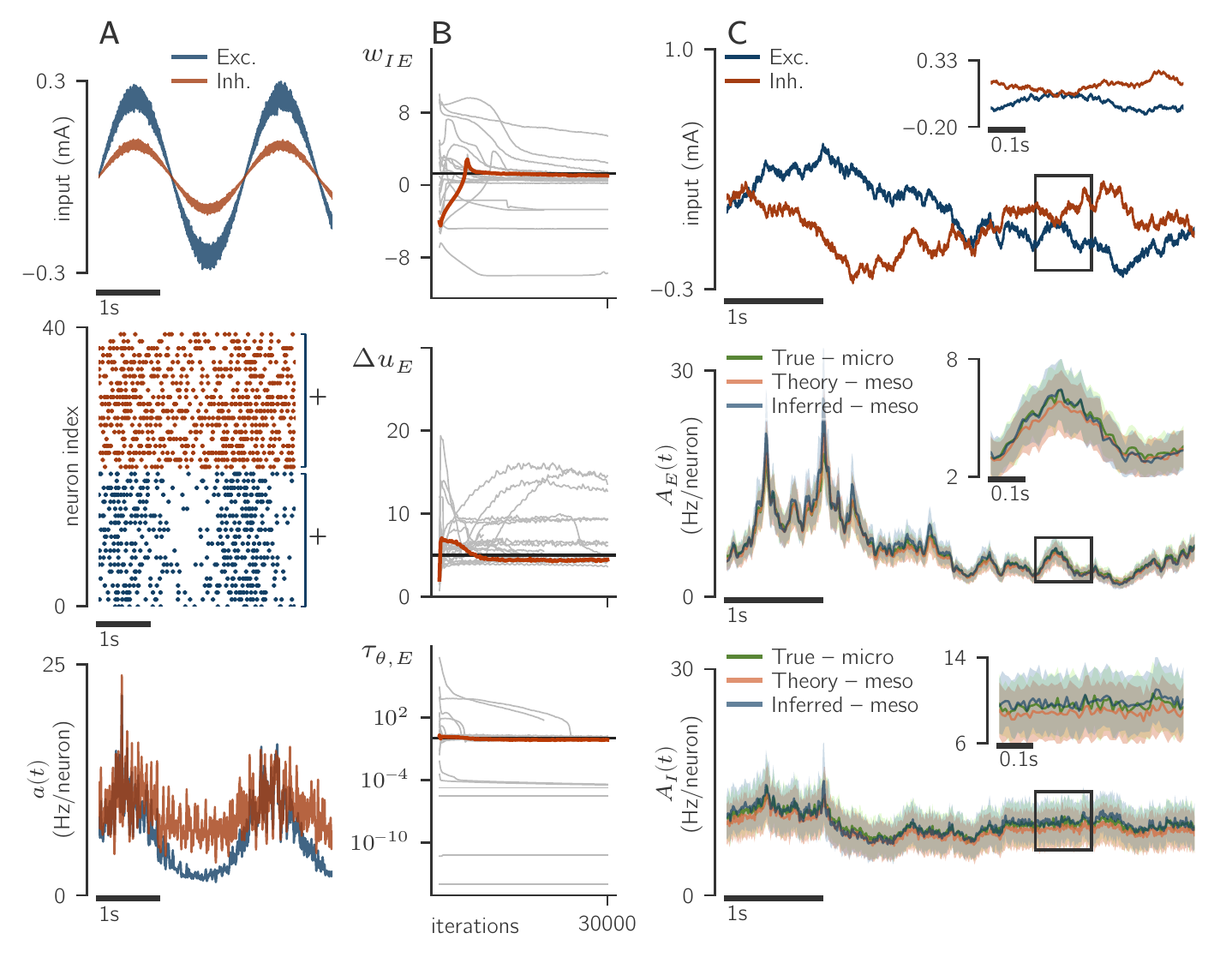}
  {\phantomsubcaption\label{fig:meso-micro-data}}
  {\phantomsubcaption\label{fig:meso-micro-fits}}
  {\phantomsubcaption\label{fig:meso-micro-sims}}
  \caption{
  \captitle{Inferred model generalizes to different inputs.}
  \subreflabel{fig:meso-micro-data} \captitle{Data generation.} Microscopic E and I populations receive a noisy sinusoidal input (\cref{eq:sine-modulated-wn-input}, \creffloat{tbl:sin-wn-Iparams}), which is shared across populations (top). Generated spikes (middle) are summed across each population, such that the inference algorithm sees only the total activity in each.
  Despite being deterministic given the history $\mathcal{H}$, the population-averaged expected activity (\cref{eq:exp-activity-definition}) still shows substantial fluctuations due to stochasticity of the history itself (bottom).
  \subreflabel{fig:meso-micro-fits} \captitle{Inference recovers parameter values close to those used to generate the data.} We performed a total of \num{25}~fits, retaining the one which found the local optimum with the highest likelihood (shown in red). Black lines indicate the prediction of the mesoscopic theory of \citet{schwalgerTheoryCorticalColumns2017}, based on ground truth values of the microscopic model. Fits for all \num{14}~parameters are shown in \creffloat{fig:2pop-fits}.
  \subreflabel{fig:meso-micro-sims} \captitle{Inferred mesoscopic model reproduces input-driven variations in population activity.}
  For testing we used low-pass-filtered frozen white noise input (\creffloat{eq:OU-input}, \creffloat{tbl:OU-params}) (top) to simulate the inferred mesoscopic model; middle and bottom plots respectively show the activity of the E and I populations. Each model was simulated \SI{100}{times}; we show the mean and standard deviation over these realizations as lines and shading of corresponding colors. (Values were averaged over disjoint bins of \SI{10}{ms}.) Performance measures are $\bar{ρ}=\num{0.950},\num{0.946},\num{0.918}$ and $\RMSE=\num{3.42+-0.07},\num{3.55+-0.09},\num{3.4+-0.08}$ for the true, theory and inferred models respectively (c.f.\ \cref{sec:performance-measures}).
  \label{fig:meso-micro}
  }
\end{figure}

We first consider a two-population model composed of E and I neurons.
We use the homogeneous microscopic model to generate activity traces (\creffloat{fig:meso-micro-data}), with a frozen noise input which is shared within populations; this input is sine-modulated to provide longer term fluctuations (c.f. \cref{eq:sine-modulated-wn-input}). A maximum a posteriori (MAP) estimate ${\hat{η}}_{\scriptscriptstyle \mathrm{MAP}}$ of \num{14} model parameters is then obtained by performing stochastic gradient descent on the posterior (c.f.\ \cref{sec:methods}). Because the likelihood is non-convex, we perform multiple fits, initializing each one by sampling from the prior (\cref{fig:meso-micro-fits}). We then keep the one which achieves the highest likelihood, which in practice is often sufficient to find a near-global optimum \citep{meyerModelsNeuronalStimulusResponse2017}.

An important note is that one can only fit parameters which are properly constrained by our data. For example, in the mesoGIF{} model, the firing probability is determined by the ratio (c.f.\ \cref{eq:firing-rate-general})
\begin{equation}
  \label{eq:lambda-ratio-example}
  \frac{u(t) - \vartheta(t)}{{Δ_{\mathrm{u}}}} \,,
\end{equation}
where $u$ is the membrane potential, $\vartheta$ the firing threshold and ${Δ_{\mathrm{u}}}$ a parameter describing the level of noise. All of these quantities are computed in units of millivolts, and the terms in the numerator depend on the resting potential $u_{\mathrm{rest}}$ and threshold $u_{\mathrm{th}}$. However, since \cref{eq:lambda-ratio-example} is dimensionless, the choice of millivolts is arbitrary: after changing ${Δ_{\mathrm{u}}}$, one can rescale $u_{\mathrm{rest}}$ and $u_{\mathrm{th}}$ (along with the synaptic weights $w$ and reset potential $u_{\mathrm{r}}$) to recover exactly the same dynamics. The set of parameters $w$, ${Δ_{\mathrm{u}}}$, $u_{\mathrm{rest}}$, $u_{\mathrm{th}}$ and $u_{\mathrm{r}}$ is thus degenerate, and they cannot all be inferred simultaneously; for this paper, we set the voltage scale to millivolts by fixing $u_{\mathrm{rest}}$ and $u_{\mathrm{th}}$ to the values proposed by \citet{schwalgerTheoryCorticalColumns2017}. Other parameters are similarly ill-constrained, and in total we inferred \num{14}~model parameters; these are listed in \creffloat{tbl:parameters}.

\begin{figure}
  \includegraphics{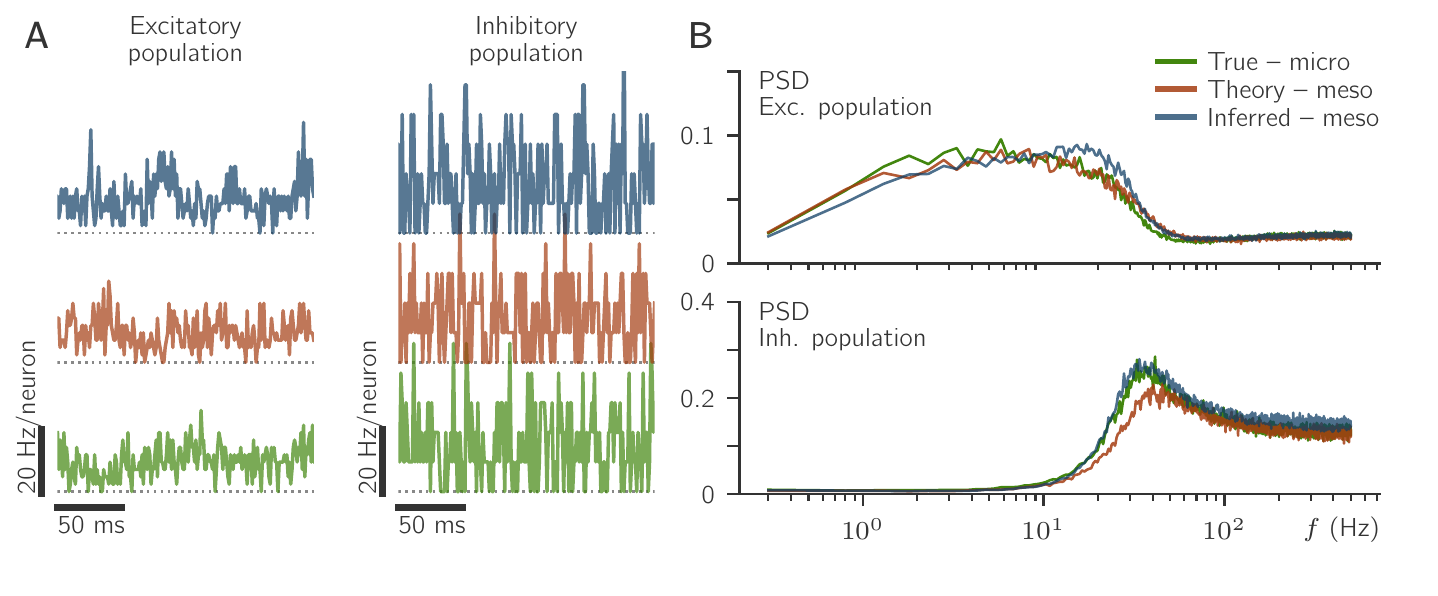}
  {\phantomsubcaption\label{fig:cst-sim}}
  {\phantomsubcaption\label{fig:psd-panel}}
  \caption{
  \captitle{Inferred model reproduces expected power spectral density.}
  \subreflabel{fig:cst-sim} Segment of simulations of the same three models shown in \creffloat{fig:meso-micro-sims} under constant \SI{0.5}{mA} input to both E and I populations. Dotted line indicates ordinate zero.
  \subreflabel{fig:psd-panel} Power spectral density for the excitatory (top) and inhibitory (bottom) populations. For each model, spectra were computed for \num{50} distinct realizations of \num{9}{s} each and averaged. To reduce the error due to finite number of realizations, the frequency axis was then coarsened to steps of \SI{0.5}{Hz} by averaging non-overlapping bins.
  \label{fig:psd}}
\end{figure}

We tested the inferred model on frozen low-pass-filtered white-noise of the same form as in \citet{augustinLowdimensionalSpikeRate2017} (\creffloat{fig:meso-micro-sims}, top), ensuring that a range of relevant time scales are tested. Despite the frozen input, variability between realizations does remain: for the GIF model this is due to sampling the escape noise (\cref{eq:s-lambda-relation}), while for the mesoGIF{} model it is due to sampling the binomial in \cref{eq:A-a-relation}. We thus we compare models based on the statistics of their response rather than single realizations: each model is simulated \num{100} times with different internal noise sequences (for each neuron in the case of the GIF model, and for each population in the case of the mesoGIF{} model) to produce an ensemble of realizations, from which we estimate the time-dependent mean and standard deviation of $A(t)$. Mean and standard deviation are then averaged over disjoint \num{10}{ms} windows to reduce variability due to the finite number of realizations. The results are reported as respectively lines and shading in \creffloat{fig:meso-micro-sims}, and show agreement between true and inferred models; we also find good agreement in the power spectrum of the response to constant input (\creffloat{fig:psd}).
Parameterizations for the training and test inputs are given in \cref{sec:stimulation}, and the full set of fits is shown in \creffloat{fig:2pop-fits}.

\subsection{Quantifying data requirements}
\label{sec:data-requirements}

\begin{figure}
 \includegraphics{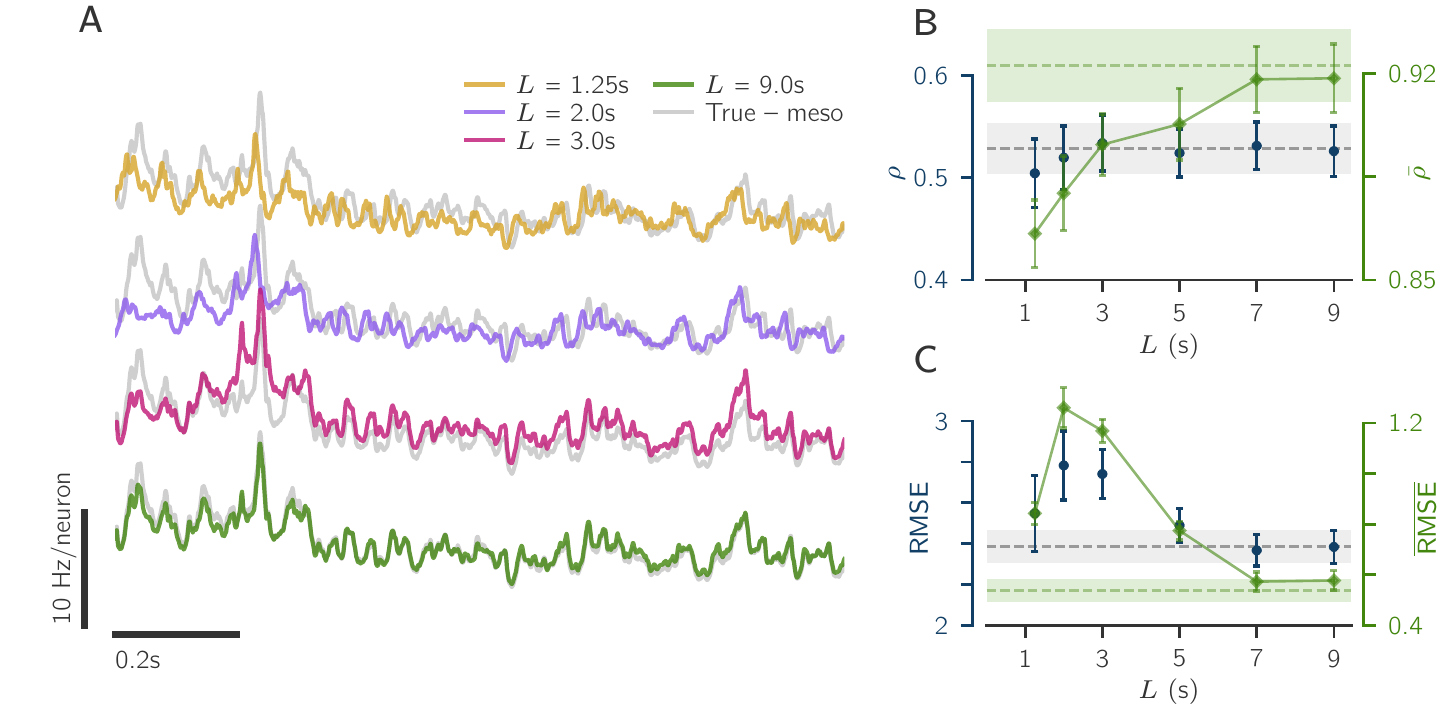}
 {\phantomsubcaption\label{fig:param-error-trace}}
 {\phantomsubcaption\label{fig:param-error-corr}}
 {\phantomsubcaption\label{fig:param-error-rms}}
 \caption{
  \captitle{Inferred model no longer improves after \SI[detect-weight]{20000}{spikes}.}
  Model parameters were inferred from data generated using shared frozen noisy sinusoidal input and tested on low-pass-filtered frozen white noise. \subreflabel{fig:param-error-trace} \captitle{Sample portion of the simulated traces used to compute discrepancy measures.} Traces of the expected activity $a(t)$ of the excitatory population in a two population E-I model, using parameters inferred from increasing amounts $L$ of data; all simulations are done on test input using the same random seed to sample the binomial in \cref{eq:A-a-relation}. Note that the model did not see this input during training.
  \subreflabels{fig:param-error-corr}{fig:param-error-rms} \captitle{Inference performance of the inferred model.} Inferrence performance, measured as either Pearson correlation $ρ$ (\subref{fig:param-error-corr}) or RMSE (\subref{fig:param-error-rms}) between \num{20}~simulations of the inferred and true mesoscopic models. Dashed lines indicate maximum achievable performance, estimated by computing the measures on a different set of \num{20} realizations of the ground truth model; shading indicates standard deviation of that value. Blue points: per-trial statistics (\cref{eq:corr-def,eq:rms-def}); green points: trial-averaged traces (\cref{eq:corrbar-def,eq:rmsbar-def}). Trial-averaged errors were estimated by bootstrapping. Results suggests that performance is well summarized by $\bar{ρ}$ and $\RMSE$.
  \label{fig:param-error}
 }
\end{figure}

While simulated data can be relatively cheap and easy to obtain, this is rarely the case of experimental data. An important question therefore is the amount required to infer the parameters of a model. To this end, we quantify in \creffloat{fig:param-error} the accuracy of the inferred dynamics as a function of the amount of data.

In order to be certain our ground truth parameters were exact, for this \lcnamecref{sec:data-requirements} we used the mesoGIF{} for both data generation and inference. This allows us to quantify the error on the inferred parameters, rather than just on the inferred dynamics. In a more realistic setting, data and model are not perfectly matched, and this will likely affect data requirements.
Testing and training were done with different external inputs to avoid overfitting; as in \cref{sec:recovering-parameters}, we used a sinusoidal frozen white noise for training and a low-pass-filtered frozen white noise for testing. During training, E and I neurons had respective average firing rates of \num{5.9} and \SI{8.4}{Hz}, which translates to approximately \SI{3500}{spikes} per second for the whole population.

We measured the accuracy of inferred dynamics by simulating the model with both the ground truth and inferred parameters, generating \num{20} different realizations for each model. These were used to calculate both the per-trial and trial-averaged Pearson correlation ($ρ$, $\bar{ρ}$) and root-mean-square error ($\RMSE$, $\overline{\RMSE}$) between models. An additional \num{20} simulations of the ground truth model were used to estimate the best achievable performance for each measure. For per-trial measures, the reported standard deviation provides an estimate of the variability between realizations; for trial-averaged measures, the standard deviation is obtained by bootstrapping, and is purely an uncertainty on the statistic (it vanishes in the limit of large number of realizations). The calculations for these measures are fully described in \cref{sec:performance-measures}. In subsequent \lcnamecrefs{sec:data-requirements}, we report only the values of $\bar{ρ}, \RMSE$ to avoid redundant information.

Consistent with the observations of \citet{augustinLowdimensionalSpikeRate2017}, we found that $ρ$ (in contrast to $\bar{ρ}$) does not allow to differentiate between models close to ground truth. The $\RMSE$ and $\overline{\RMSE}$ on the other hand showed similar sensitivity, but may be unreliable far from ground-truth (as evidenced by the data point at $L$=\num{1.25}{s} in \creffloat{fig:param-error-rms}). Since the per-trial $\RMSE$ additionally quantifies the variability between realizations (through its standard deviation), we preferred it over its trial-averaged analog.

As we would expect, the inferred model better reproduces the dynamics of the true model when the amount of data is increased (\creffloat{fig:param-error}); when fitting all \num{14}~parameters of the mesoGIF{} model, the inferred model no longer improves when more than \SIrange{5}{7}{s} of data are provided (\creffloatrange{fig:param-error-corr}{fig:param-error-rms}) – corresponding to a total of about \SIrange{17500}{24500}{spikes}.
In \cref{sec:app:fit-statistics}, we repeat the test described here with smaller parameter sets (achieved by clamping certain parameters to their known ground truth values). We find that this has only a modest effect on the achieved performance, but does significantly improve the consistency of fits (compare \creffloats{fig:data-reqs-scatter}{fig:data-reqs-scatter-eta5}). Inferring larger parameter sets is thus expected to require more fits (and consequent computation time) before a few of them find the MAP.
Certain parameters are also more difficult to infer: for the case shown in \cref{fig:param-error}, relative errors on the inferred parameters range from \num{5}\% to \num{22}\% (c.f.\ \cref{sec:app:fit-statistics}, \creffloat{tbl:all-fits-deltas}). Parameters describing the inhibitory population ($τ_{m,I}$, $w_{IE}$, $w_{II}$) show the highest relative error, as well as the escape rates ($c_E$, $c_I$) and the adaptation time constant ($τ_{θ,E}$).

\subsection{Modelling high-dimensional heterogeneous populations with an effective low-dimensional homogeneous model}
\label{sec:inference-hetero-model}

\begin{figure}
  \includegraphics{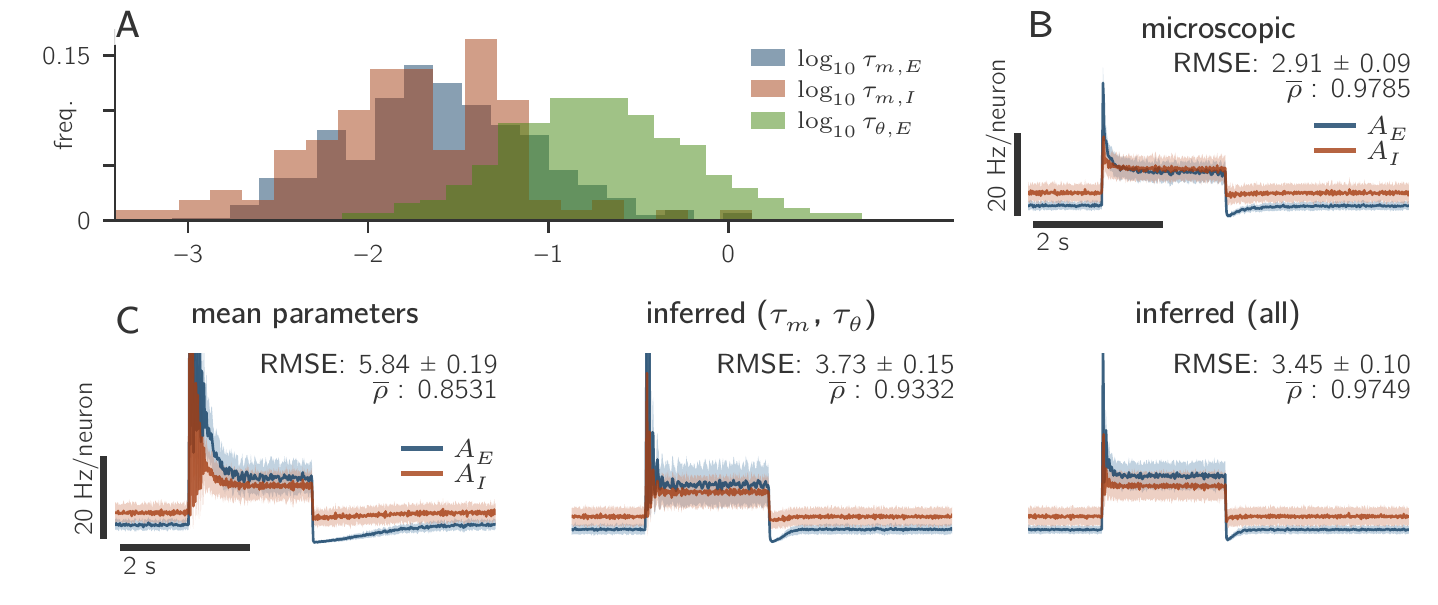}
  {\phantomsubcaption\label{fig:hetero-params-dists}}
  {\phantomsubcaption\label{fig:hetero-params-true-step}}
  {\phantomsubcaption\label{fig:hetero-params-inferred-steps}}
  \caption{
  \captitle{Inferred effective parameters can compensate for mismatch between microscopic and mesoscopic models.}
  \subreflabel{fig:hetero-params-dists}~A heterogeneous microscopic model of two populations was constructed by sampling three time constants from log-normal distributions (c.f.\ \creffloat{tbl:hetero-model-dist-params}). All other parameters are as in \cref{sec:recovering-parameters} and \creffloat{fig:meso-micro}, and the same sine-modulated white noise as in \cref{fig:meso-micro-data} was used to train the model.
  \subreflabel{fig:hetero-params-true-step}~\captitle{Heterogeneous microscopic model driven by a single step current.} Shown are the mean (line) and standard deviation (shading) of the model's response, computed from \num{60} realizations and averaged over disjoint windows of $\SI{10}{ms}$. Realizations differ due to sampling the escape noise.
  \subreflabel{fig:hetero-params-inferred-steps}~Simulations of the mesoscopic model with the same step input as in (\subref{fig:hetero-params-true-step}), using mean parameters (left), inferred $τ_m$ and $τ_θ$ (middle -- all other parameters homogeneous and set to ground truth), and the inferred full (14) parameter set (right). Line and shading have the same meaning as in (\subref{fig:hetero-params-true-step}) and are based on \num{50} realizations for each model; these differ by the sampling of the binomial in \cref{eq:A-a-relation}. We see that inferred models more closely reproduce the trace in (\subref{fig:hetero-params-true-step}), which is confirmed by the decreased $\RMSE$ and increased $\bar{ρ}$.
  \label{fig:hetero-params}
  }
\end{figure}

A frequently understated challenge of meso- and macroscale population models is that of choosing their parameters such that the dynamics of the modeled neuron populations are consistent with the high-dimensional dynamics of networks of individual neurons. A typical approach, when measurements of microscopic single neuron parameters are available, is to assign each parameter its mean across the population \citep[\S\,12][]{gerstnerNeuronalDynamicsSingle2014}. However, as alluded to in \cref{introduction}, mean parameters do not always make good predictors for nonlinear systems; this is evidenced by \creffloat{fig:hetero-params}, which expands upon \creffloat{fig:intro-inferred-vs-mean}.

An alternative approach would be to fit the population model to observed population activities, such as to ensure maximum consistency with data -- for example, by finding the maximum a posteriori (MAP) parameters. In this way we obtain effective parameters which compensate for the mismatch between data and population model.

To show that this can work, we made the microscopic model heterogeneous in three parameters: $τ_{m,E}$, $τ_{m,I}$ and $τ_{θ,E}$. These parameters were set individually for each neuron by sampling from a log-normal distribution (\creffloat{fig:hetero-params-dists}, \creffloat{tbl:hetero-model-dist-params}). As in previous sections, output from the microscopic model under sine-modulated frozen white noise input was then used to train the mesoscopic one. For testing we used a single step input (\creffloat{fig:hetero-params-true-step}); this allowed us to test the performance of the inferred model both in the transient and steady-state regimes. The per-trial $\RMSE$ and trial-averaged correlation $\bar{ρ}$ were computed on ensembles of realizations, as described in \cref{sec:performance-measures}.

We considered three sets of parameters for the mesoscopic model. For the first, we set $τ_{m,E}$, $τ_{m,I}$ and $τ_{θ,E}$ to their sample averages. This produced rather poor results (\creffloat{fig:hetero-params-inferred-steps}, left); in particular, the transient response to the step is much more dramatic and long-lived than that of the ground truth model. As the neural model is highly nonlinear in its parameters, linearly averaging parameters is not guaranteed to produce optimal results.

The test results are improved when the heterogeneous parameters are inferred (\creffloat{fig:hetero-params-inferred-steps}, middle). However, fitting only the heterogeneous parameters gives the mesoscopic model only three degrees of freedom to compensate for approximating a heterogeneous model by a homogeneous one, and it still produces traces with too high variance. Indeed, giving the model full freedom over the parameters provides another step improvement (\creffloat{fig:hetero-params-inferred-steps}, right), with output from the mesoscopic model differing from the target output only by a higher transient peak and slightly different mean activities (obtained parameter values are listed in \creffloat{tbl:hetero-fit-result}). Thus while fitting more parameters may incur additional computational cost (\cref{sec:app:fit-statistics}), it also provides more opportunities to accommodate model mismatch.

The results of this section show the necessity of inferring population parameters rather than simply averaging single neuron values. It also demonstrates the ability of population models to reproduce realistic activities when we provide them with good effective parameters; in order to compensate for modelling assumptions, those parameters will in general differ from those of a more detailed microscopic model.

\pagebreak[2]
\subsection{Full posterior estimation over parameters}
\label{sec:posteriors}

\begin{figure}
  \includegraphics{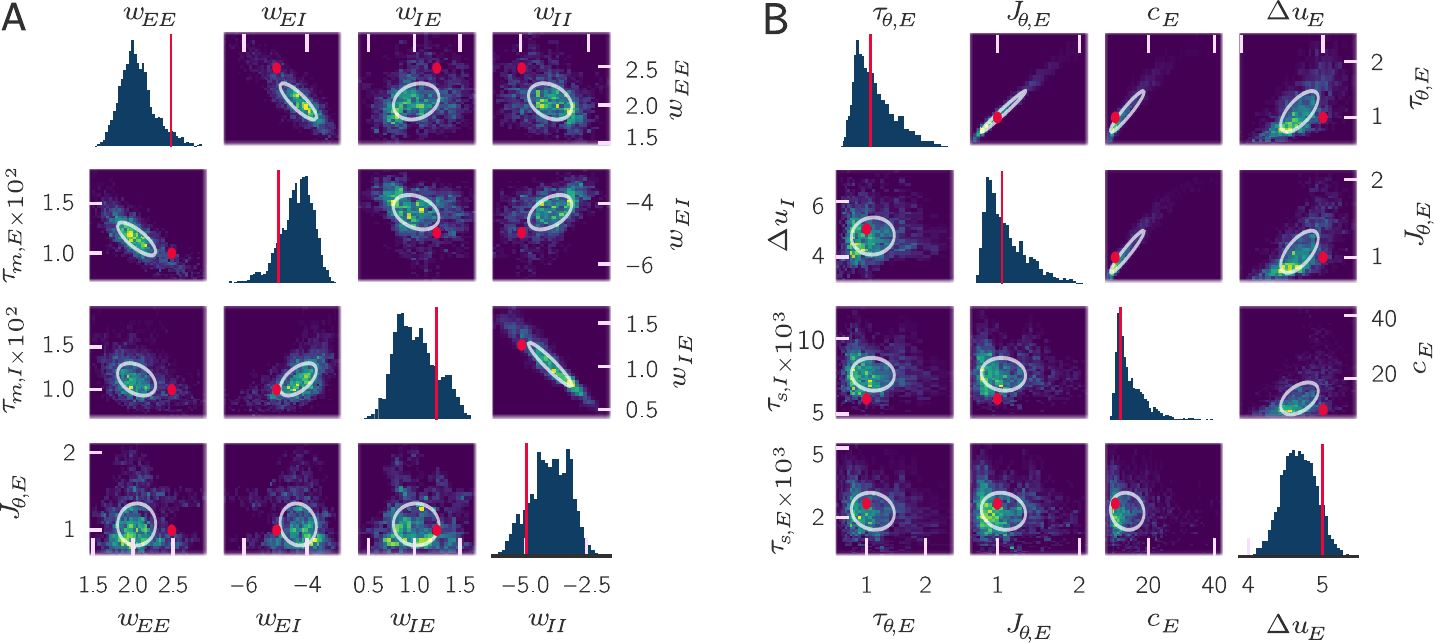}
  {\phantomsubcaption\label{fig:posteriorA}}
  {\phantomsubcaption\label{fig:posteriorB}}
  \caption{\captitle{Posterior probability highlights dependencies between model parameters.} Panels show one and two-parameter marginals; all panels within a column use the same parameter for their abscissa. \subreflabel{fig:posteriorA}
  Above diagonal: Full posterior over the connectivities $w$.
  Strongest (anti)correlation is between pairs impinging on the same population (i.e.\ $w_{EI}$--$w_{EE}$ and $w_{IE}$--$w_{II}$.)
  Below diagonal: Membrane time constants and adaptation strength show correlations with connectivity.
  Panels on the diagonal show the marginal for that column's parameters. Red dot or line shows the parameters' ground truth values. Ellipse is centered on the mean and corresponds to two standard deviations under a Gaussian model.
  The full posterior over all 14 parameters is shown in \creffloat{fig:full-posterior} and was obtained with HMC sampling using data generated with the two-population homogeneous microscopic model.
  \subreflabel{fig:posteriorB}
  Above diagonal: Tight correlation between $τ_{θ,E}$, $J_{θ,E}$ and $c_E$ suggests their ratios are most important to determining model dynamics.
  Below diagonal: There is little correlation between adaptation and synaptic parameters.
  Diagonal panels, red marks and ellipses are as in \subref{fig:posteriorA}.
  \label{fig:posterior}
  }
\end{figure}

It can often be desirable to know which parameters, or combinations of parameters, are constrained by the data. Bayesian inference, i.e. estimation of the posterior distribution over parameters given the data, can be used to not only identify the ‘best-fitting’ parameters, but also to characterize the uncertainty about these estimates. Notably, these uncertainties may be highly correlated across parameters:
For instance, one expects an increase in E connectivity to cancel a decrease in (negative) I connectivity to the same population, and this is confirmed by the correlation in the marginals shown in \creffloat{fig:posteriorA}. Interestingly, this correlation is in fact stronger for connectivities sharing the same \emph{target} than those sharing the same \emph{source}. More novel structure can be learned from \creffloat{fig:posteriorB}, such as the strong correlation between the adaptation parameters, or the complete absence of correlation between them and the synaptic parameters. In particular, the tight relationship between $J_{θ,E}$, $τ_{θ,E}$ and $c_E$ suggests that for determining model dynamics, the ratios ${J_{θ,E}}/{τ_{θ,E}}$ and ${J_{θ,E}}/{c_E}$ may be more important than any of those three quantities individually.

Since there are \num{14}~unknown parameters, the posterior is also \num{14}-dimensional; we represent it by displaying the joint distributions between pairs, obtained by marginalizing out the other \num{12} parameters (c.f.\ \cref{sec:methods:posteriors}). Training data here were generated in the same way as in \cref{sec:recovering-parameters}, from a homogeneous microscopic model with the parameters listed in \creffloat{tbl:parameters}.
To provide a sense of scale, we have drawn ellipses in \creffloat{fig:posterior} to indicate the volume corresponding to two standard deviations from the mean under a Gaussian model. In a number of cases it highlights how the true distribution is non-Gaussian – for example the distributions of $c_E$, $J_{θ,E}$ and $τ_{θ,E}$ are noticeably skewed.

A naive way to compute these 2D~marginals would be to numerically integrate the likelihood; however, given that that leaves \num{12}~dimensions to integrate, such an approach would be computationally unfeasible. Instead we used Hamiltonian Monte Carlo (HMC) sampling \citep{nealMCMCUsingHamiltonian2012,betancourtHamiltonianMonteCarlo2013}. Monte Carlo methods are guaranteed to asymptotically converge to the true posterior – a valuable feature when one wishes to deduce interactions between parameters from its structure.
Nevertheless, due to the complexity of mesoGIF{}'s likelihood,
memory and computational cost still required special consideration (c.f.\ \cref{sec:methods:posteriors}).

We note that the 2$σ$~ellipses in \creffloat{fig:posterior}, while informative, are imperfect indicators of the probability mass distribution. If the posterior is Gaussian, then each projection to a 2D~marginal places 86.5\% of the probability mass within the ellipse; however for non-Gaussian posteriors this number can vary substantially. Moreover, the markers for ground truth parameters shown in \creffloat{fig:posterior} may differ from the effective parameters found by the model (c.f.\ \cref{sec:inference-hetero-model}).

\subsection{Pushing the limits of generalization}
\label{sec:4pop-fit}

\begin{figure}
  \includegraphics{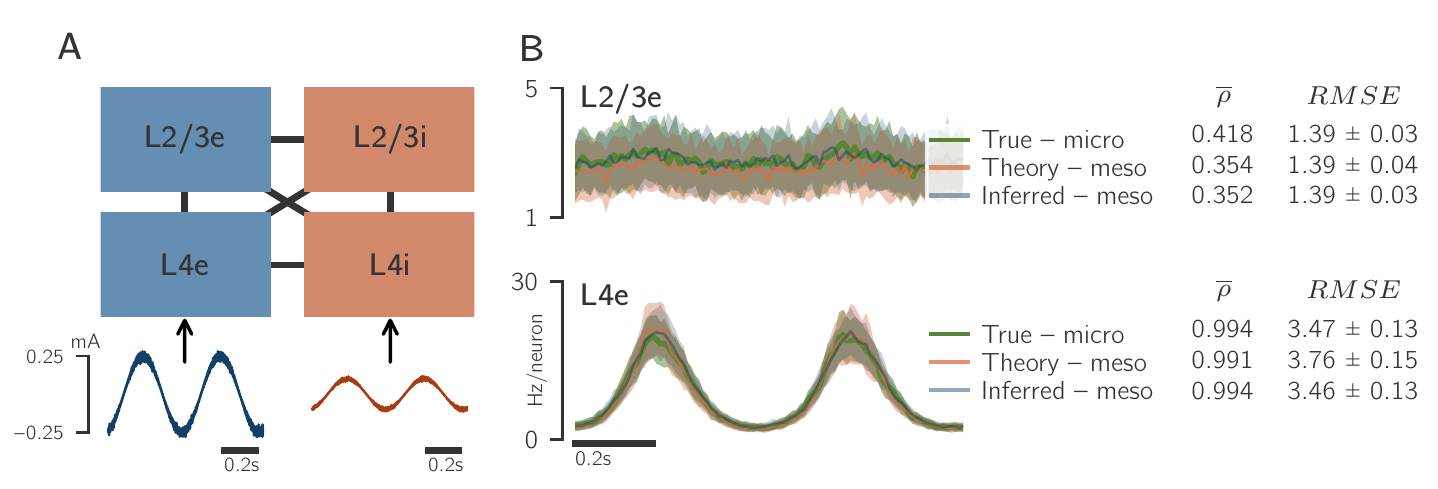}
  {\phantomsubcaption\label{fig:4pop-sketch}}
  {\phantomsubcaption\label{fig:4pop-output-L4e}}
  \caption{
  \captitle{Inference of a four population model with \num{36} free parameters.}
  \subreflabel{fig:4pop-sketch} Model represented E (blue) and I (red) populations from layers L2/3 and L4 of a cortical column. During training, only L4 populations received external sinusoidal input. The homogeneous microscopic model was used to generate data.
  \subreflabel{fig:4pop-output-L4e}
  The mesoscopic model matches aggregate microscopic dynamics (``True\,--\,micro''), both when using theoretical (``Theory\,--\,meso'') and inferred parameters (``Inferred\,--\,meso'').
  In contrast to the previous \lcnamecref{sec:4pop-fit}, correlation and RMSE scores are reported separately for each population; they are computed from \num{60} realizations of each models.
  \label{fig:4pop-setup}
  }
\end{figure}

The previous sections have shown that we can recover \num{14}~parameters of the two population model mesoGIF{} model. A natural question is whether this approach scales well to larger models.
We investigated this by considering four neuron populations representing the {L2/3} and {L4} layers of the Potjans-Diesmann micro-circuit \citep{potjansCelltypeSpecificCortical2014}. The associated higher-dimensional set of mesoscopic equations follow the same form as in previous sections \citep{schwalgerTheoryCorticalColumns2017}. There are \num{36}~free parameters in this model, of which \num{16}~are connectivities; they are listed in \cref{tbl:parameters}. Similar to previous sections, we trained mesoGIF{} on output from the microscopic model with sinusoidal drive (\creffloat{fig:4pop-sketch}).

\begin{figure}
 \includegraphics{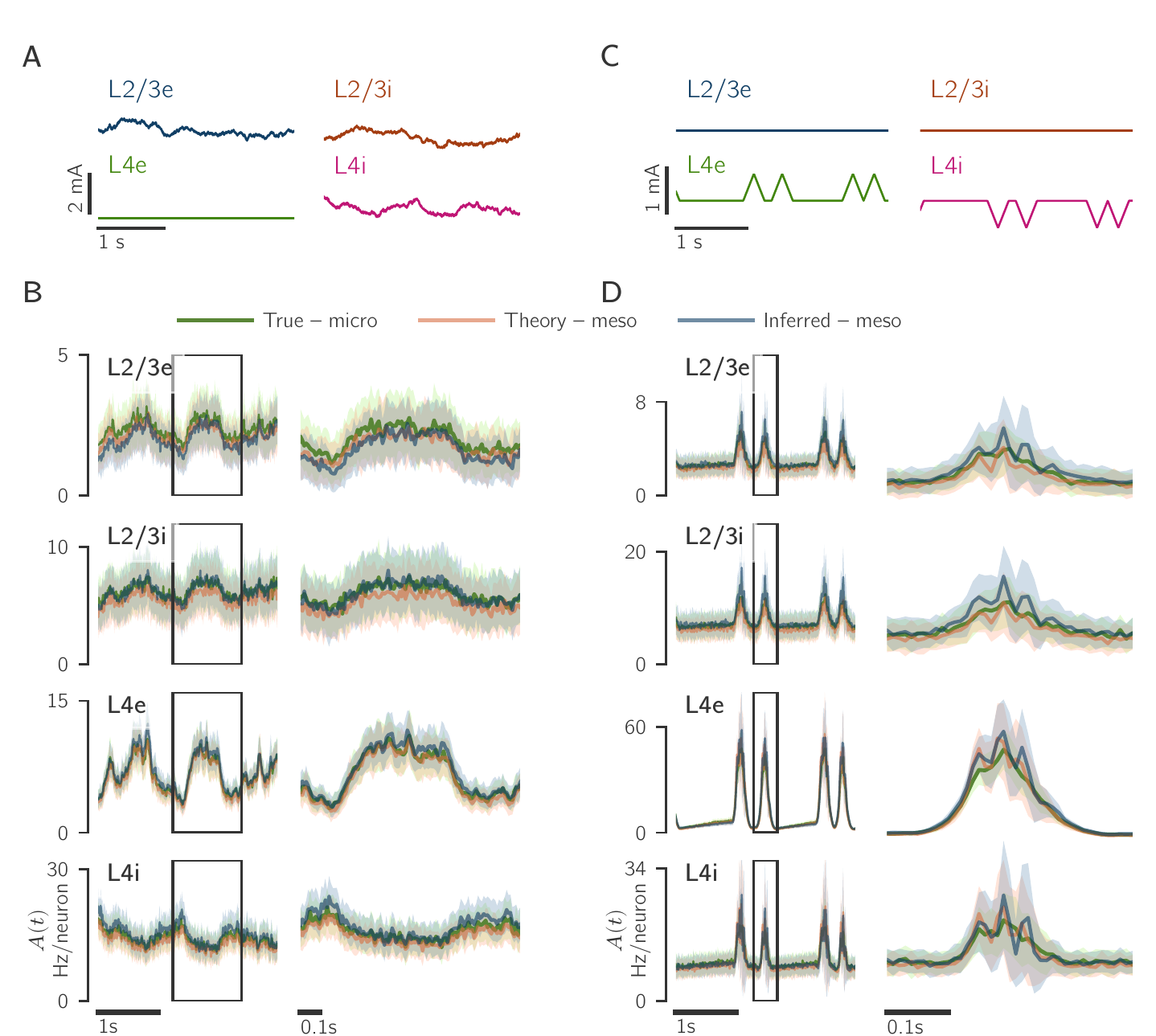}
  {\phantomsubcaption\label{fig:4pop-OUtest-input}}
  {\phantomsubcaption\label{fig:4pop-OUtest-output}}
  {\phantomsubcaption\label{fig:4pop-ramptest-input}}
  {\phantomsubcaption\label{fig:4pop-ramptest-output}}
 \caption{
 \captitle{Generalization errors appear with large deviations from the training input.}
 We test the \SI{36}{parameter} model inferred in \creffloat{fig:4pop-setup} under two different stimulation protocols. Lines and shading show mean and standard deviation over \num{60} realizations, computed as in \cref{sec:recovering-parameters}.
 \subrefformat{\subref{fig:4pop-OUtest-input},}\subreflabel{fig:4pop-OUtest-output} After completely removing external inputs to {L4e} (compare \subref{fig:4pop-OUtest-input} with the training input in \creffloat{fig:4pop-sketch}), predictions of the inferred and theoretical models are still indistinguishable.
 \subrefformat{\subref{fig:4pop-ramptest-input},}\subreflabel{fig:4pop-ramptest-output} To obtain visible deviations between inferred and theoretical models, we used inputs (\subref{fig:4pop-ramptest-input}) which stretch the mesoGIF{} assumptions. Oscillations are present in both the microscopic and mesoscopic models, but in the latter have much larger amplitudes: compare the blue and red traces to the thicker green trace in \subref{fig:4pop-ramptest-output}.
\label{fig:sim-compare-4pop}}
\end{figure}

The {L4} populations tend to drive the activity in this model, and we found that we do not need to provide any input to the {L2/3} neurons to get parameter estimates which accurately predict population activity (\creffloat{fig:sim-compare-4pop}, left): the small fluctuations in {L2/3} (\creffloat{fig:4pop-output-L4e}) suffice to provide constraints on those population parameters.
Those constraints of course are somewhat looser, and in particular connection strengths onto {L4} are not as well estimated when compared to ground truth (\creffloat{tbl:4pop-fit-values}).

Pushing the mesoscopic approximation beyond its validity limits using inputs with abrupt transitions understandably increases the discrepancy between ground truth and model (\creffloat{fig:sim-compare-4pop}, right). Indeed, such a strong input may cause neurons to fire in bursts, thereby breaking the quasi-renewal approximation (c.f.\ \cref{sec:model-summary}). During an input spike, the true model shows small oscillations; the theoretical mesoGIF{} reproduces these oscillations but with an exaggerated amplitude and higher variance between realizations, and in contrast to \cref{sec:inference-hetero-model}, the inferred model does no better. This larger discrepancy with the true model is reflected in the performance measures (c.f.\ \creffloats{tbl:performance-4pop-rmse}{tbl:performance-4pop-rho}), and is consistent with the observation that the mesoGIF{} has higher variance during bursts \citep[p.\ 15]{schwalgerTheoryCorticalColumns2017}. Slower time-scale dynamics are still accurately captured by both the theoretical and inferred models.

The capacity of the inferred model to generalize to unseen inputs is thus quite robust, with discrepancies between inferred and ground truth models only occurring when the test and training input were very different.
Of course this is in part due to mesoGIF{} being a good representation of the activity of homogeneous GIF neurons: while inference may compensate for some discrepancies between the model and the data, it still can only work within the freedom afforded by the model.

\section{Discussion}
\label{sec:discussion}

Population models play a key role in neuroscience: they may describe experimental data at the scale they are recorded and serve to simplify the dynamics of large numbers of neurons into a human-understandable form. These dynamics may occur on a range of scales, from the mesoscopic, limited to a single cortical column, to the macroscopic, describing interactions between regions across the entire brain. Mechanistic models allow us to bridge those scales, relating micro-scale interactions to meso- or macro-scale dynamics; of these, the model chosen for this study allows for rich dynamics at the single level by including synaptic, refractory and adaptation dynamics.

We have demonstrated that it is possible to fit a mechanistic population model to simulated data by maximizing the likelihood of its parameters, in much the same way as is already done with phenomenological models \citep{pillowSpatiotemporalCorrelationsVisual2008,mackeEmpiricalModelsSpiking2011,zhaoInterpretableNonlinearDynamic2016}. Since mechanistic models describe concrete, albeit idealized, biophysical processes, they have the additional benefit that their parameters can be understood in terms of those processes. Moreover, those parameters are typically not dependent on the applied input, and thus we can expect the inferred model to generalize to novel stimulus conditions.

We also found that after making a few parameters heterogeneous, averaging did not recover the most representative parameters. In general, when there is discrepancy between model and data, the effective parameters are difficult to recover analytically – data-driven methods then provide a valuable supplement to theoretical analysis, in order to ensure that a model actually represents the intended biological process. Nevertheless, since the inference procedure is agnostic to the model, it is up to the modeler to choose one for which the effective parameters remain interpretable.

\medskip

The approach we have presented requires only that a differentiable likelihood function be available, and thus is not limited to neuron population models. Stochastic models of neuron membrane potentials \citep{goldwynWhatWhereAdding2011}, of animal populations \citep{woodStatisticalInferenceNoisy2010} and of transition phenomena in physics and chemistry \citep[\S 7]{horsthemkeNoiseinducedTransitionsTheory2006} are examples for which parameters could be inferred using this approach.

In practice we expect some models to be more challenging than others. For instance, evaluating the likelihood of a spiking model typically involves integrating over all time courses of the subthreshold membrane potential compatible with the observed spike train \citep{paninskiMaximumLikelihoodEstimation2004}. This integral can be difficult to evaluate accurately, especially for models incorporating adaptation and refractoriness \citep{menaQuadratureMethodsRefractory2014,ramirezFastInferenceGeneralized2014}.
If evaluation of the likelihood is prohibitively expensive, likelihood-free approaches might be more appropriate \citep{lueckmannFlexibleStatisticalInference2017,papamakariosFastEpsilonFree2016}.

Of note also is that we required the dynamics to be formulated as a Markov process to express the likelihood (c.f.\ \cref{sec:methods:meso-logL}). We achieved this by constructing a state vector, but the size of this vector adds substantial computational cost and in practice there is a trade-off between the length of the integration time window and the number of units (here neuron populations) we can infer. Since neural field models are also computationally represented by long state vectors, inference on these models would be subject to a similar trade-off.
Finally, our current implementation assumes that the state ${S}$ (c.f.\ \cref{sec:details-meso-model}) can be fully reconstructed from observations. If only a partial reconstruction of ${S}$ is possible, undetermined components of ${S}$ form a latent state which must be inferred along with the parameters. This type of problem has already been studied in the context of dimensionality reduction \citep{mackeEmpiricalModelsSpiking2011,cunninghamDimensionalityReductionLargescale2014,ruleNeuralFieldModels2019}, and it is conceivable that such methods could be adapted to our framework. Such an approach would allow one to perform dimensionality reduction with mechanistic models of temporal dynamics.

The work of \citet{ruleNeuralFieldModels2019} presents an interesting complement to ours. The authors therein consider a neural field model where activities are observed only indirectly via a point-process, thus adressing the problem of inferring latent states. They infer both these states and the point-process parameters, but assume known parameters and neglect finite-size effects for the mesoscopic model; in contrast, here we inferred the mesoscopic model parameters while assuming that population states are observed. Inferring both mesoscopic model parameters and latent states remains a challenge for both of these approaches.

\medskip
To obtain posteriors, we employed a Hamiltonian Monte Carlo algorithm with minimal automatic tuning. We found this to work better than a more automatically tuned variant (c.f.~\cref{sec:methods:posteriors}), but it is beyond the scope of this work to provide a complete survey of sampling methods. The applicability of more recently developed algorithms such as Riemann manifold Monte Carlo \citep{girolamiRiemannManifoldLangevin2011}, sequential Monte Carlo \citep{moralSequentialMonteCarlo2006} and nested sampling \citep{skillingNestedSamplingGeneral2006} would be worth exploring in future work.
Variational methods such as that described by \citet{kucukelbirAutomaticDifferentiationVariational2017a} are another alternative to estimating posteriors which do not require sampling at all. They generally scale to large parameter spaces but do not provide the asymptotic guarantees of MCMC and may artifically smooth the resulting posterior.

\medskip

Important obstacles to using inference on complex models are the implementation and computational costs. Software tools developed for this work have helped limit the former, but the latter remains a challenge, with many of the figures shown requiring multiple days of computation on a personal workstation. While manageable for studying fixed networks, this would become an impediment for scaling to larger models, or tracking the evolution of parameter values by inferring them on successive time windows. For such tasks further work would be required to reduce the inference time, for example by investigating how large the integration time step for the mesoGIF{} model can be made, or by optimizing the current implementation. One might also attempt to derive a computationally simpler model, or make better use of parallelization and/or graphical processing units.

As noted by \citet[][\S~3.2]{ruleNeuralFieldModels2019}, an additional complication to inferring mechanistic model parameters is that they may be under-constrained. In our case, since mesoGIF{} is a rate model, the voltage scale can be chosen freely by setting the resting ($u_{\mathrm{rest}}$) and threshold ($u_{\mathrm{th}}$) potentials -- if we nonetheless attempt to infer them along with the noise scale (${Δ_{\mathrm{u}}}$), fits are unable to converge (c.f.\ \cref{sec:recovering-parameters,sec:sgd}). We avoided this problem by identifying the problematic parameters and fixing them to their known values. However, the development of a more systematic approach to dealing with under-constrained parameters is left for future investigations.

Since inference time is highly dependent on computational complexity, there is a trade-off between bottom-up models which attempt to match dynamics as closely as possible, and simpler top-down models which aim for computational efficiency; while the latter tend to provide better scalability, the former are likely to be more interpretable and allow for extrapolation to new dynamical regimes (c.f.\ \cref{sec:4pop-fit}). Choosing the right model thus remains a key component of data analysis and modelling.

Inference methods based on machine learning allow for flexible model design, using known biophysical parameter values when they are available, and inference to determine the others which are consistent with data. We hope this work further motivates the use of richer models in neuroscience, by providing tools to fit and validate them.


\section{Methods}
\label{sec:methods}

\subsection{Microscopic model}
\label{sec:details-micro-model}

We consider an ensemble of neurons grouped into $M$ populations; the symbols $i$, $j$ are used to label neurons, and $α$, $β$ to label populations. The neuron indices $i$, $j$ run across populations and are thus unique to each neuron.

Each neuron $i$ produces a spike train represented as a sum of Dirac delta functions,
\begin{equation}
 s_{i}(t) = \sum_{k} δ(t - t_{i,k}) \,,
\end{equation}
where $t_{i,k}$ is the time of its $k$-th spike. We denote $Γ_i^{β}$ the set of neuron indices from population $β$ which are presynaptic to neuron $i$, $w_{αβ}$ the strength of the connection from a neuron in population $β$ to another in population $α$, and $Δ_{αβ}$ the transmission delay between the two populations. As in \citet{schwalgerTheoryCorticalColumns2017}, we assume that intrinsic neural parameters are homogeneous across a given population. We further assume that  connection strengths depend only on the source and target populations; for a connection between neurons of population $\beta$ to those of population $\alpha$ , the strength is either $w_{\alpha\beta}$ with probability $p_{\alpha\beta}$ or zero with probability $1-p_{\alpha\beta}$.  Each spike elicits a post-synaptic current, which we sum linearly to obtain the synaptic inputs to neuron $i$ from $M$ populations,
\begin{equation}
 R_{α}I_{\mathrm{syn},i}(t) = τ_m^{α} \sum_{β=1}^M w_{αβ} \sum_{j \in Γ_i^{β}} \bigl(ε_{αβ} * s_j \bigr)(t) \,.
\end{equation}
The transmission delay is captured by shifting the synaptic kernel with a Heaviside function $Θ$:
\begin{equation}
 ε_{αβ}(t) = Θ(t - Δ_{αβ}) \frac{e^{-(t-Δ)/τ_{s,β}}}{τ_{s,β}} \,.
\end{equation}

Spike generation is modeled by a \emph{generalized integrate-and-fire} mechanism: leaky integration with adapting threshold, followed by an escape rate process. For each neuron $i$, the membrane potential $u_{i}$ and firing threshold $\vartheta_{i}$ evolve according to
\begin{align}
 τ_{m,α} \frac{d u_i}{d t} & = - u_i + u_{\mathrm{rest}, α} + R_{α}I_{\mathrm{ext}, α}(t) + R_{α}I_{\mathrm{syn},i}(t) \,; \label{eq:u-dynamics-cts}      \\
 \vartheta_i(t)         & = u_{\mathrm{th}, α} + \int_{-\infty}^{t} θ_α(t - t') s_i(t') \, \mathrm{d} t' \,. \label{eq:threshold-dynamics-cts-1}
\end{align}
Here, $θ_{α}$ is the adaptation kernel for population $α$ and $I_{\mathrm{ext}, α}$ the external input to that population. For this work we used an exponential adaptation kernel,
\begin{equation}
 θ_{α} (t) = \frac{J_{θ,α}}{τ_{θ,α}} e^{-t/τ_{θ}} \,,
\end{equation}
which allows us to rewrite \cref{eq:threshold-dynamics-cts-1} as
\begin{equation}
 \label{eq:threshold-dynamics-cts}
 τ_{θ} \frac{d \vartheta_{i}}{d t}(t) = - \vartheta_i(t) + u_{\mathrm{th}, α} + J_{θ,α} s_i(t) \,.
\end{equation}
Spikes are generated stochastically with an escape rate (also called conditional intensity or hazard rate), calculated with the inverse link function $f$:
\begin{align}
 λ_i(t) &= f( u_i(t) - \vartheta_i(t) ) \,. \label{eq:firing-rate-general}\\
\intertext{For this work we used}
  λ_i(t) & = c_{α} \exp( (u_i(t) - \vartheta_i(t)) / {Δ_{\mathrm{u},α}} ) \,, \label{eq:firing-rate-specific}
\end{align}
where ${Δ_{\mathrm{u}}}$ parameterizes the amount of noise (or equivalently, the softness of the threshold) and $c$ is the firing rate when $u(t) = \vartheta(t)$.

Once a spike is emitted, a neuron's potential is reset to $u_r$ and clamped to this value for a time $t_{\mathrm{ref}}$ corresponding to its absolute refractory period. It then evolves again according to \cref{eq:u-dynamics-cts}. All model parameters are summarized in \creffloat{tbl:parameters}.

\robustify\bfseries
\sisetup{detect-weight=true,detect-inline-weight=math}
\begin{table}
  \centering
  \caption{\captitle{Parameters for the micro- and mesoscopic models.} For the mesoscopic populations, the ensemble of neuron parameters is replaced by a single effective value for that population. For each parameter, we indicate the number of components in the two and four population models; adaptation parameters have fewer components because the model assumes no adaptation for inhibitory neurons. Boldface is used to indicate inferred parameters; the remainder are fixed to the known ground truth values listed in \creffloat{tbl:parameter-values}. This results in respectively \num{14} and \num{36} free parameters for the two- and four-population models. A brief discussion of how we chose which parameters to infer is given at the end of \cref{sec:sgd}.
  \label{tbl:parameters}
  }
 \begin{tabular}{lSSl}
  \toprule
  & \multicolumn{2}{c}{No. of components} & \\
  \cmidrule(rl){2-3}
  Parameter       & {2 pop.} & {4 pop.} & Description                    \tabularnewline\midrule
  $p$          & 4 & 16 & connection probability            \tabularnewline
  $\bm{w}$ &  \bfseries 4 & \bfseries 16 & \textbf{connection weight}            \tabularnewline
  $Δ$          & 4 & 16 & transmission delay                         \tabularnewline
  $N$          & 2 & 4 & no. of neurons in pop.            \tabularnewline
  $R$          & 2 & 4 & membrane resistance               \tabularnewline
  $u_{\mathrm{rest}}$     & 2 & 4 & membrane resting potential        \tabularnewline
  $\bm{τ_m}$   & \bfseries 2 & \bfseries 4 & \textbf{membrane time constant}   \tabularnewline
  $t_{\mathrm{ref}}$      & 2 & 4 & absolute refractory period        \tabularnewline
  $u_{\mathrm{th}}$  & 2 & 4 & non-adapting threshold   \tabularnewline
  $u_{\mathrm{r}}$   & 2 & 4 & reset potential          \tabularnewline
  $\bm{c}$     & \bfseries 2 & \bfseries 4 & \textbf{escape rate at threshold} \tabularnewline
  $\bm{{Δ_{\mathrm{u}}}}$   & \bfseries 2 & \bfseries 4 &\textbf{noise level}              \tabularnewline
  $\bm{τ_s}$ & \bfseries 2 & \bfseries 4 & \textbf{synaptic time constant}     \tabularnewline
  $\bm{J_θ}$  & \bfseries 1 & \bfseries 2 & \textbf{adaptation strength}      \tabularnewline
  $\bm{τ_θ}$ & \bfseries 1 & \bfseries 2 & \textbf{adpatation time constant} \tabularnewline
  \bottomrule
 \end{tabular}
\end{table}

\subsection{Mesoscopic model}
\label{sec:details-meso-model}

The mesoscopic equations describe the interaction of population activities (total number of spikes per second per neuron) in closed form: they can be integrated without the need to simulate indiviual neurons. This is achieved by identifying each neuron $i$ by its age $τ_i$ and making the assumptions stated in \cref{sec:model-summary}: that each population is homogeneous, that neurons are all-to-all connected with effective weights $p_{αβ} w_{αβ}$, and that dynamics are well approximated as a quasi-renewal process. Under these conditions it is possible to rewrite the dynamical equations in terms of the refractory densities $ρ_α(t,τ)$ – the proportion of neurons with age $τ_i \in [τ, τ+dτ)$ in each population $α$. With very large populations $N_α$, we can neglect finite-size fluctuations and $ρ$ satisfies the transport equation \citep{wilsonExcitatoryInhibitoryInteractions1972,gerstnerPopulationDynamicsSpiking2000,chizhovEfficientEvaluationNeuron2008,gerstnerNeuronalDynamicsSingle2014}:
\begin{align}
  \label{eq:RDM-transport-equation}
  \frac{\partial ρ_α}{\partial t} + \frac{\partial ρ_α}{\partial τ} &= -λ_α(t,τ) ρ \,, & ρ_α(0, t) &= A_α(t)\,.
\end{align}
Neuronal dynamics and synaptic interactions are captured within the functional form of the hazard rate $λ_α(t,τ)$, which depends only on $τ$ and on the history of population activities. In the limit $N_α \to \infty$, the evolution of $A(t)$ matches its expectation $a(t)$ and is obtained by integrating over all neurons:
\begin{align}
  \label{eq:RDM-activity-infinite-N}
  N_α \to \infty: \quad A_α(t) &= a_α(t) = \int_{0}^\infty λ_α(t,τ) ρ_α(t,τ) \,\mathrm{d}τ \,.
\end{align}
For finite $N$, the expression for the expected activity becomes \citep{schwalgerMindLastSpike2019,schwalgerTheoryCorticalColumns2017}
\begin{equation}
\label{eq:RDM-expected-activity} a_α(t) = \int_{0}^\infty λ_α(t,τ) ρ_α(t,τ) dτ + Λ_α(t)\left(1 - \int_0^\infty ρ_α(t,τ)\right) \,,
\end{equation}
where $Λ(t)$ is a rate function that accounts for finite-size effects in the refractory density. The activity then follows a stochastic process described by
\begin{align}
  \label{eq:cts-A-dynamics}
  A_α(t) &= \frac{n_α(t)}{Ndt} \,, & n_α(t) &\sim \Binom(Na(t)dt; N_α) \,.
\end{align}

For this work we discretize time into steps of length $Δt$, and instead of the refractory density work with the vector $m^{(k)}_α$, where $m^{(k)}_{α,l}$ is formally defined as the expected number of neurons of age $τ \in [lΔt, (l+1))$:
\begin{align}
  m^{(k)}_{α,l} &= \int_τ^{τ+Δt^-} \hspace{-1em} N_α ρ_α(t_k,l\,Δt) \, dτ \,, & (l = 1,\dotsc,K < \infty)\,.
\end{align}
Here the superscript $(k)$ indicates the simulation time step and $l$ the age bin.
Since refractory effects are negligible for sufficiently old neurons, $m^{(k)}$ only needs to be computed for a finite number of age bins $K$ (c.f.\ \cref{sec:meso-equations}, as well as \eqname~(86) from \citet{schwalgerTheoryCorticalColumns2017}).

We similarly compute the firing rates at time $t_k$ as a vector $λ^{(k)}_{α,l}$, $l=1,\dotsc, K$. The expected number of spikes in a time bin,
\begin{equation}
  \bar{n}^{(k)}_{α} = \Expect\left[n^{(k)}_α\right] \,,
\end{equation}
can then be computed in analogy with \cref{eq:RDM-expected-activity}, by summing the products $λ^{(k)}_{α,l} m^{(k)}_{α,l}$ over $l$ and adding a finite-size correction; the precise equations used to evaluate $m^{(k)}_{α,l}$, $λ^{(k)}_{α,l}$ and $\bar{n}^{(k)}_{α}$ are listed in \cref{sec:meso-equations}.
We can convert spike counts to activities by dividing by $N_α Δt$:
\begin{align}
  \label{eq:A-a-def}
  a^{(k)}_{α} &\coloneqq \frac{\bar{n}_{α}^{(k)}}{N_α Δt} \,, &
  A^{(k)}_{α} &\coloneqq \frac{n_{α}^{(k)}}{N_α Δt} \,.
\end{align}

For the following, it will be convenient to define the single-neuron firing probability,
\begin{equation}
  \label{eq:meso-fire-prob}
  p_{α,η}^{(k)} \coloneqq \frac{\bar{n}_{α,η}^{(k)}}{N_α} \,,
\end{equation}
where the subscript $η$ makes explicit the dependence on the model parameters. This allows us to rewrite \cref{eq:A-a-relation} as
\begin{equation}
  \label{eq:meso-binomial}
  n_{α}^{(k)} \sim \Binom\left( p_{α,η}^{(k)} ; N_α  \right) \,,
\end{equation}
where $p_{α,η}^{(k)} = p_{α,η}(t_k | \mathcal{H}_{t_k})$ depends on the activity history $\mathcal{H}_{t_k}$ (\cref{eq:hist-meso}).
Because $K$ is finite, we can replace $\mathcal{H}_{t_k}$ by a finite state-vector ${S}^{(k)}$, obtained by concatenating all variables required to update $n^{(k)}$ (c.f.\ \cref{sec:meso-equations}, especially \cref{eq:state-vector}):
\begin{equation}
  \label{eq:state-vector-abbrev}
{S}^{(k)} = \left(n^{(k)}, {m}^{({k})}, {λ}^{({k})}, \dotsc \right) \,.
\end{equation}
The update equations for ${S}^{(k)}$ are Markovian by construction, which simplifies the expression of the model's likelihood presented in the next \lcnamecref{sec:methods:meso-logL}.

\subsection{Likelihood for the mesoscopic model}
\label{sec:methods:meso-logL}

As stated in \cref{sec:details-meso-model}, the mesoGIF{} model can be cast in a Markovian form, which allows us to expand the probability of observing a sequence of spike counts as a recursive product. If that sequence has length $L$ and an initial time point $k_0$, then that probability is
\begin{equation}
  p\biggl(\bigl\{n_α^{(k)}\bigr\}_{\substack{k=k_0\dots k_0+L-1\\α=\mathrlap{1\dots M}\hphantom{k_0\dots k_0+L-1}}} \biggr) = \prod_{α=1}^M \prod_{k=k_0}^{L+k_0-1} p\left(n_α^{(k)} | S^{(k)}\right) \,.
\end{equation}
The likelihood of this sequence then follows directly from the probability mass function of a binomial, using the definitions for $n_α^{(k)}$ and $p^{(k)}_{α,η}$ defined above;
\begin{equation}
  L_{k_0;L}  = \prod_{α=1}^{M} \prod_{k = k_0}^{k_0+L-1} \binom{N_α}{n_α^{(k)}} \left(p^{(k)}_{α,η}\right)^{n_α^{(k)}} \left(1 - p^{(k)}_{α,η}\right)^{N_{\alpha}n_α^{(k)}} \,.
\end{equation}
We note that the $n_α^{(k)}$ are observed data points, and are thus constant when maximizing the likelihood.

Expanding the binomial coefficient, the log-likelihood becomes
\begin{multline}
\label{eq:log-likelihood-batch}
\log L_{k_0;L}(η) = \sum_{α=1}^{M} \sum_{k = k_0}^{k_0+L-1} \log\left(N_α! \right)- \log \left(n_α^{(k)}!\right) - \log \left((N_α - n_α^{(k)})!\right) \\
  + n_α^{(k)} \log \left(\tilde{p}_{α,η}^{(k)} \right) + \left(N_α - n_α^{(k)}\right) \log \left(1 - \tilde{p}_{α,η}^{(k)}\right) \,,
\end{multline}
where we clipped the probability $\tilde{p}_α^{(k)}$ to avoid writing separate expressions for $p^{(k)}_{α,η} \in 0, 1$,
\begin{align}
  \tilde{p}_{α,η}^{(k)} &= \begin{cases}
  ε & \text{if $p_{α,η}^{(k)} \leq ε$} \,,\\
  p_α^{(k)} & \text{if $ε \leq p_{α,η}^{(k)} \leq 1-ε$} \,,\\
  1-ε & \text{if $p_{α,η}^{(k)} \geq 1-ε$} \,.
\end{cases}
\end{align}
Clipping also avoids issues where the firing probability $p_α^{(k)}$ exceeds 1, which occurs when one explores the parameter space. (This can happen when parameters are such that the chosen $Δt$ is no longer small enough for the underlying Poisson assumption to be valid, although it should \emph{not} occur around the true parameters. See the discussion by \citet[p.~48]{schwalgerTheoryCorticalColumns2017}.) We found that with double precision, a tolerance $ε = 1\times 10^{-8}$ worked well.

For numerical stability, logarithms of factorials are computed with a dedicated function such as SciPy's \texttt{gammaln} \citep{jonesSciPyOpenSource2001}. For optimization, the term $\log \left(N_α!\right)$ can be omitted from the sum since it is constant.

\subsection{Initializing the model}
\label{sec:initialization}

Although the updates to the state ${S}$ are deterministic (c.f.\ \cref{sec:details-meso-model}), only the components $n_α^{(k_0)}$ of the initial state ${S}^{(k_0)}$ is known – unobserved components can easily number in the thousands. We get around this problem in the same manner as in \citet{schwalgerTheoryCorticalColumns2017}: by making an initial guess that is consistent with model assumptions (survival counts sum to $N_α$, etc.) and letting the system evolve until it has forgotten its initial condition. We note that the same problem is encountered when training recurrent neural networks, whereby the first data points are used to ``burn-in'' unit activations before training can begin. For the results we presented, we used a variation of the initialization scheme used by
\citet{schwalgerTheoryCorticalColumns2017} which we call the ''silent initialization``.

\paragraph{Silent initialization}  Neurons are assumed to have never fired, and thus they are all ``free''. This results in large spiking activity in the first few time bins, which then relaxes to realistic levels.
\begin{algorithm}
  \caption{Silent initialization scheme. \label{alg:init-scheme-silent}}
  \begin{algorithmic}[1]
    \State $n_α \gets 0$
    \State $h_α, u_{α,i} \gets u_{\mathrm{rest}, i}$
    \State $x_α \gets N_α$
    \State $λ_{α,i}, λ_{\mathrm{free},α}, g_α, m_{α,i}, v_{α,i}, y_{αβ}, z_{α} \gets 0$
  \end{algorithmic}
\end{algorithm}

This initialization scheme has the advantage of being simple and needing no extra computation, but with the high-dimensional internal state ${S}$, also requires a large burn-in time of around \SI{10}{s}. This can be largely mitigated by using sequential batches (\cref{alg:sequential-adam}).

We also experimented with intializing the model at a stationary point (\cref{sec:alternative-initialization}), but in the cases we considered it did not provide a notable improvement in computation time.

\subsection{Estimating parameters}
\label{sec:sgd}

To maximize the likelihood, we used adam{} \citep{kingmaAdamMethodStochastic2014}, a momentum-based stochastic gradient descent algorithm, for which gradients were computed automatically with Theano \citep{thetheanodevelopmentteamTheanoPythonFramework2016} (c.f.\ \cref{sec:software}). Training parameters are listed in \creffloat{tbl:fit-parameters}.

\begin{table}
  \centering
 \caption{Fitting parameters for adam{}. Learning rate, $β_1$ and $β_2$ are as defined in \citet{kingmaAdamMethodStochastic2014}.
  \label{tbl:fit-parameters}}
 \begin{tabular}{ccl}
  \toprule
  fit parameter      & value & comment\tabularnewline\midrule
  learning rate      & \num{0.01} & adam{} parameter \tabularnewline
  $β_1$             & \num{0.1} & adam{} parameter \tabularnewline
  $β_2$             & \num{0.001} & adam{} parameter \tabularnewline
  $g_{\mathrm{clip}}$ & \num{100} & clipping threshold \tabularnewline
  ${L_{\mathrm{burnin}}}$ & \SI{10}{s} & data burn-in \tabularnewline
  ${B_{\mathrm{burnin}}}$ & \SI{0.3}{s} & mini-batch burn-in  \tabularnewline
  ${γ_L}$ & \SI{1} & ${L_{\mathrm{burnin}}}$ noise factor \tabularnewline
  ${γ_B}$ & \SI{0.1} & ${B_{\mathrm{burnin}}}$ noise factor \tabularnewline
  \bottomrule
 \end{tabular}
\end{table}

Despite the similarities, there remain important practical differences between fitting the mesoscopic model and training a recurrent neural network (RNN). Notably, RNN weights are more freely rescaled, allowing the use of single precision floating point arithmetic. In the case of the mesoscopic model, the dynamic range is wider and we found it necessary to use double precision.

Compared to a neural network, the mesoscopic update equations (\namecrefs{eq:link-function}~(\ref{eq:link-function}–\ref{eq:spike-generation-appendix})) are also more expensive to compute, in our case slowing down parameter updates by at least an order of magnitude.

The subsequences of data (``mini-batches'') used to train an RNN are usually selected at random: at each iteration, a random time step $k_0$ is selected, from which the next ${B_{\mathrm{burnin}}}$ data points are used for burn-in and the following $B$ data points form the mini-batch. This becomes problematic when long burn-in times are required, not only because it requires long computation times, but also because it wastes a lot of data.
We addressed this problem by keeping the state across iterations (Alg.~\ref{alg:sequential-adam}): since this is a good guess of what it should be after updating the parameters, it reduces the required burn-in time by an order of magnitude. However this requires batches to follow one another, breaking the usual assumption that they are independently selected. In practice this seemed not to be a problem; in anecdotal comparisons, we found that training with either a) randomly selected batches and stationary initialization (\cref{alg:init-scheme-stationary}), or b) sequential batches and silent initialization (\cref{alg:init-scheme-silent}), required comparable numbers of iterations to converge to similar parameter values. Computation time in the case of random batches however was much longer.

We also found that bounding the gradient helped make inference more robust. We set maximum values for each gradient component and rescaled the gradient so that no component exceeded its maximum (Alg.~\ref{alg:sequential-adam}, lines \ref{alg:rescale-begin} to \ref{alg:rescale-end}).

Maximizing the posterior rather than the likelihood by multiplying the latter by parameter priors (to obtain the MAP estimate rather than the MLE) helped prevent the fit from getting stuck in unphysical regions far from the true parameters, where the likelihood may not be informative. We used noninformative priors (c.f.\ \creffloat{tbl:parameter-values}) so as to ensure that they didn't artificially constrain the fits. Fits were also initialized by sampling from the prior.

Choosing adequate external inputs may also impact fit performance, as in general, sharp stimuli exciting transients on multiple timescales tend to be more informative than constant input \citep{iolovOptimalDesignEstimation2017}. That being said, even under constant input, the fluctuations in a finite-sized neuron population still carry some information, and anecdotal evidence suggests that these can be sufficient to infer approximate model parameters.
In this paper, we used a sinusoidal input with frozen white noise to train the mesoGIF{} model -- with only one dominant time scale, this input is more informative than constant input but far from optimal for the purpose of fitting. This made it a reasonable choice for computing baseline performance measures.

Finally, to allow fits to converge, it is essential to avoid fitting any ill-defined or degenerate parameters. For example, as explained in \cref{sec:recovering-parameters}, we fixed the parameters $u_{\mathrm{rest}}$ and $u_{\mathrm{th}}$ because the mesoGIF{} model is invariant under a rescaling of the voltage; for simplicity we also fixed $u_{\mathrm{r}}$ and $R$ even though this was not strictly necessary. The parameters $w$ and $p$ are similarly degenerate (c.f.\ \cref{eq:update-htot}) and we fixed $p$. The parameters $N$, $Δ$ and $t_{\mathrm{ref}}$ are effectively discrete (either in numbers of neurons or time bins), and they were also fixed to simplify the implementation. \Cref{tbl:parameters} summarizes the inferred and non-inferred parameters.

\begin{algorithm}
  \caption{Training with sequential mini-batches. The gradient is normalized before computing adam{} updates. Note that the state is not reinitialized within the inner loop.%
  \label{alg:sequential-adam}%
  }
  \begin{algorithmic}[1]
    \Repeat
      \State ${S} \gets \text{initialize state}$
      \State $k' \sim \Uniform(0, {γ_L} B)$  \Comment{Randomize initialization burn-in}
      \State $k_0 \gets {L_{\mathrm{burnin}}} + k'$
      \While{$k_0 < L - B$}  \Comment{Scan data sequentially}
        \State $g \gets \nabla \log L(η, A_{k_0:k_0+B})$ \Comment{Log-likelihood gradient on the mini-batch}
        \If {$\text{any}(|g| > g_{\mathrm{clip}})$}  \Comment{Normalize gradients with $L^\infty$ norm} \label{alg:rescale-begin}
          \State $g_{\mathrm{max}} \gets \text{max}(|Δη|)$
          \State $g \gets \frac{g_{\mathrm{clip}}}{g_{\mathrm{max}}} Δη$
        \EndIf \label{alg:rescale-end}
        \State $η \gets adam{}(g)$ \Comment{Update parameters updates with adam{}}
        \State $k' \sim \Uniform({B_{\mathrm{burnin}}}, (1+{γ_B}) {B_{\mathrm{burnin}}})$  \Comment{Randomize batch burn-in}
        \State $k_0 \gets k_0 + k'$
      \EndWhile
    \Until{converged.}
  \end{algorithmic}
\end{algorithm}

\subsection{Estimating the posterior}
\label{sec:methods:posteriors}

The posteriors in \cref{sec:posteriors} were obtained using Hamiltonian Monte Carlo \citep{nealMCMCUsingHamiltonian2012,betancourtHamiltonianMonteCarlo2013} (HMC). Having expressed the likelihood with Theano made it straightforward to use the implementation in PyMC3 \citep{salvatierProbabilisticProgrammingPython2016} -- HamiltonianMC -- to sample the likelihood; the sampling parameters we used are listed in \creffloat{tbl:MCMC-setup}.

Although straightforward, this approach pushes the limit of what can be achieved with currently implemented samplers: because the likelihood of this model is expensive to evaluate, even coarse distributions can take hours to obtain. In addition, the large state vector required sufficiently large amounts of memory to make the automatically tuned NUTS \citep{hoffmanNoUturnSamplerAdaptively2014} sampler impractical. (NUTS stores the most recent states in order to tune the sampling parameters.) In an application with experimental data, one would want to reserve sufficient computational resources to perform at least basic validation of the obtained that posterior, using for example the methods described in \citet{gelmanBayesianDataAnalysis2014} and \citet{taltsValidatingBayesianInference2018}.

In order for samplers to find the high probability density region in finite time, we found it necessary to initialize them with the MAP estimate. This also ensured that their mass matrix was tuned on an area of the posterior with appropriate curvature.
In applications where the posterior has multiple modes, one should be able to identify them from the collection of fits. The high probability density region around each mode should then be sampled separately, integrated, and combined with the others to obtain the full posterior. (See e.g.\ \citet{vanhaasterenMarginalLikelihoodCalculation2014} for integration methods for MCMC chains.)

Finally, as with parameter optimization, we found that the use of at least double precision floats was required in order to obtain consistent results.

\begin{table}
  \centering
  \caption{
  Specification the MCMC sampler.
  \label{tbl:MCMC-setup}
  }
  \begin{tabular}{ll}
    \toprule
    Algorithm  & \texttt{HamiltonianMC} (PyMC3\citep{salvatierProbabilisticProgrammingPython2016}) \tabularnewline
    \midrule
    step scale & \num{0.0025} \tabularnewline
    path length & \num{0.1}  \tabularnewline
    tuning steps & \num{20} \tabularnewline
    initialization & \texttt{jitter+adapt_diag}  \tabularnewline
    start      & ${η}_{\scriptscriptstyle \mathrm{MAP}}$ estimate  \tabularnewline
    no. of samples & \num{2000} \tabularnewline
    \midrule
    total run time & \SI{201}{h} \tabularnewline
    \bottomrule
  \end{tabular}
\end{table}

\subsection{Measuring performance}
\label{sec:performance-measures}

In order to assess the performance of our inference method, we quantified the discrepancy between a simulation using ground truth parameters and another using inferred parameters; the same input was used for both simulations, and was different from the one used for training. Following \citet{augustinLowdimensionalSpikeRate2017}, discrepancy was quantified using both correlation ($ρ$) and root mean square error ($\RMSE$); these are reported according to the amount of data $L$ used to train the model, which may be given either in time bins or seconds.

The correlation between activity traces from the ground truth and inferred models, respectively ${A^{\mathrm{\scriptscriptstyle true}}}(t)$ and ${\hat{A}^{(L)}}(t)$, was obtained by computing the \emph{per-trial} Pearson coefficient for each of the $M$ populations and averaging the results across populations to report a single value:
\begin{equation}
 \label{eq:corr-def}
 ρ({A^{\mathrm{\scriptscriptstyle true}}}, {\hat{A}^{(L)}}) = \frac{1}{M} \sum_{α=1}^M
 \frac { \Braket{ \bigl({A_{α}^{\vphantom{(L)}\mathrm{\scriptscriptstyle true}}} - \braket{{A_{α}^{\vphantom{(L)}\mathrm{\scriptscriptstyle true}}}}\bigr) \bigl({\hat{A}_{α}^{(L)}} - \braket{{\hat{A}_{α}^{(L)}}}_k\bigr) }_k }
 {\sqrt{ \Braket{ \bigl({A_{α}^{\vphantom{(L)}\mathrm{\scriptscriptstyle true}}} - \braket{{A_{α}^{\vphantom{(L)}\mathrm{\scriptscriptstyle true}}}}_k\bigr)^2 \bigl({\hat{A}_{α}^{(L)}} - \braket{{\hat{A}_{α}^{(L)}}}_k\bigr)^2 }_k }} \,.
\end{equation}
Here brackets indicate averages over time,
\begin{equation*}
 \braket{A}_k \coloneqq \tfrac{1}{L'} \sum_{k=k_0}^{k_0 + L'} A^{(k)} \,,
\end{equation*}
with $k$ a discretized time index. The initial time point $k_0$ sets the burn-in period; in all calculations below, we set it to correspond to \SI{10}{s} to ensure that any artifacts due to the initialization have washed away. The value of $L'$ need not be the same as $L$, and we set it to \num{9000} (corresponding to \SI{9}{s}) for all discrepancy estimates.

As with correlation, the per-trial RMSE was averaged across populations,
\begin{equation}
 \label{eq:rms-def}
 \RMSE({A^{\mathrm{\scriptscriptstyle true}}}, {\hat{A}^{(L)}}) \coloneqq \sqrt{ \frac{1}{M} \sum_{α=1}^M \Braket{\left({\hat{A}_{α}^{(L)}} - {A_{α}^{\vphantom{(L)}\mathrm{\scriptscriptstyle true}}}\right)^2}_k} \,.
\end{equation}

Because the models are stochastic, \cref{eq:corr-def,eq:rms-def} describe random variables. Thus, for each of our results, we generated ensembles of realizations $\{{A^{\mathrm{\scriptscriptstyle true}, r}}\}_{r=1}^{R_1}$, $\{{A^{\mathrm{\scriptscriptstyle true}, r}}'\}_{r=1}^{R_2}$ and $\{\hat{A}^{r}\}_{r=1}^{R_3}$, each with a different set of random seeds. We compute the $ρ$ and $\RMSE$ for the $R_1\times R_2$ pairs $({A^{\mathrm{\scriptscriptstyle true}, r}}, \hat{A}^{r})$, as well the $R_1 \times R_3$ combinations $({A^{\mathrm{\scriptscriptstyle true}, r}}, {A^{\mathrm{\scriptscriptstyle true}, r}}')$, from which we empirically estimate the mean and standard deviation of those measures. Values for the pairs $({A^{\mathrm{\scriptscriptstyle true}, r}}, {A^{\mathrm{\scriptscriptstyle true}, r}}')$ provide an estimate of the best achievable value for a given measure.

Another way to address the stochasticity of these measures is to use \emph{trial-averaged} traces:
\begin{align}
  \bar{ρ}(L) &= ρ({\bar{A}^{\mathrm{\scriptscriptstyle true}}}, \hat{\bar{A}}) \,, \label{eq:corrbar-def}\\
  \overline{\RMSE}(L) &= \RMSE({\bar{A}^{\mathrm{\scriptscriptstyle true}}}, \hat{\bar{A}}) \label{eq:rmsbar-def}\,;
\end{align}
where the trial-averaged activity,
\begin{equation*}
  \bar{A}_{α}^{(k)} \coloneqq \frac{1}{R} \sum_{r=1}^R A_α(t_k | \mathcal{H}_{t_k}^r) \,,
\end{equation*}
is as in \cref{eq:trial-averaged-A}.
Because trial-averaged measures only provide a point estimate, we used bootstrapping to estimate their variability. We resampled the ensemble of realizations with replacement to generate a new ensemble of same size $R$, and repeated this procedure \SI{100}{times}. This yielded a set of $R$ measures (either $\bar{ρ}$ or $\overline{\RMSE}$), for which we computed the sample standard deviation. Note that in contrast to per-trial measures, errors on trial-averaged measurements vanish in the limit of large number of trials $R$ and thus are not indicative of the variability between traces.

We found the pair of measures $(\bar{ρ}, \RMSE)$ (\cref{eq:corrbar-def,eq:rms-def}) to provide a good balance between information and conciseness (c.f. \cref{sec:data-requirements}). We generally used $R_1$ = $R_2$ = \num{50} and $R_3$ = \num{100} for the ensembles, with the exception of \cref{fig:param-error} where $R_1$ = $R_2$ = $R_3$ = \num{20}. We also ensured that sets of trial-averaged measures use the same number of trials, to ensure comparability.

\subsection{Stimulation and integration details}
\label{sec:stimulation}

All external inputs used in this paper are shared within populations and frozen across realizations. They are distinct from the escape noise (\cref{eq:s-lambda-relation,eq:A-a-relation}), which is \emph{not} frozen across realizations.

\paragraph{Sine-modulated white noise input}
\label{sec:stimulation-sine-wn}

For inferring parameters in all our work, we generated training data with a sine-modulated stimulus of the form
\begin{equation}
 \label{eq:sine-modulated-wn-input}
 I_{\mathrm{ext}}(t) = B \sin (ω t) \, \cdot (1 \, + \, q ξ(t)) \,,
\end{equation}
\noindent where $ξ(t)$ is the output of a white noise process with $\langle ξ(t) ξ(t') \rangle = δ(t - t')$. This input was chosen to be weakly informative, in order to provide a baseline for the inference procedure. The values of $B$, $ω$ and $q$ are listed in \creffloat{tbl:sin-wn-Iparams}. The integration time step was set to \SI{0.2}{ms} for microscopic simulations and \SI{1}{ms} for mesoscopic simulations. We then tested the fitted model with the inputs described below.

\begin{table}
  \centering
 \caption{Parameters for the sine-modulated input.
  \label{tbl:sin-wn-Iparams}}
 \begin{tabular}{lSSSSSSc}
  \toprule
    & \multicolumn{2}{c}{2 pop. model}  & \multicolumn{4}{c}{4 pop. model} & {Unit} \\
    \cmidrule(lr){2-3} \cmidrule(lr){4-7}
    & {E} & {I}  & {L2/3e} & {L2/3i} & {L4e} & {L4i} \\
  \midrule
  $B$ &  0.25  & 0.1  &    0.0 &    0.0 &  0.25 &  0.1 &  \si{mA} \\
  $ω$ &  2.0   & 2.0  &    2.0 &    2.0 &  2.0 &  2.0 & --\\
  $q$ &  4.0   & 4.0  &    4.0 &    4.0 &  4.0 &  4.0 & \si{mA} \\
  \bottomrule
 \end{tabular}
\end{table}

\paragraph{OU process input}

Fit performance in \cref{sec:data-requirements,sec:4pop-fit} was measured using an input produced by an Ornstein-Uhlenbeck (OU) process defined by
\begin{equation}
 \label{eq:OU-input}
 \frac{dI_{\mathrm{test}}}{dt} = -\frac{(I_{\mathrm{test}} - {μ_{\mathrm{\scriptscriptstyle OU}}})}{{τ_{\mathrm{\scriptscriptstyle OU}}}} \, dt \,+\, \sqrt{\frac{2}{{τ_{\mathrm{\scriptscriptstyle OU}}}}} q \, dW \,.
\end{equation}
Here ${μ_{\mathrm{\scriptscriptstyle OU}}}$, ${τ_{\mathrm{\scriptscriptstyle OU}}}$ and $q$ respectively set the mean, correlation time and noise amplitude of the input, while $dW$ denotes increments of a Wiener process. The parameter values and initial condition ($I_{\mathrm{test}}(0)$) are listed in \creffloat{tbl:OU-params}.

\begin{table}[!ht]
  \centering
 \caption{Parameters for the OU-process input (\cref{eq:OU-input}).
  \label{tbl:OU-params}}
 \begin{tabular}{cSSSSSSs}
  \toprule
  {} & \multicolumn{2}{c}{2 pop. model} & \multicolumn{4}{c}{4 pop. model} & unit \\
   \cmidrule(lr){2-3} \cmidrule(lr){4-7}
       & {E} & {I} & {L2/3e} & {L2/3i}    & {L4e} & {L4i} \\
  \midrule
  ${μ_{\mathrm{\scriptscriptstyle OU}}}$ &  0.1     & 0.05  &  1   & 1   & 0  & 1   & mA \\
  ${τ_{\mathrm{\scriptscriptstyle OU}}}$  & 1        & 1     &  2   & 2   & {–}  & 2   & s \\
  $q$  &  0.125     & 0.125  &  0.5 & 0.5 & 0  & 0.5 & mA \\
  $I_{\mathrm{test}}(0)$ & 0.1 & 0.05   &  0.5 & 0.5 & 0  & 0.5 & mA \\
  \bottomrule
 \end{tabular}
\end{table}

\paragraph{Impulse input}

We further tested the generalizability of the four population model using an input composed of sharp synchronous ramps. As the transient response is qualitatively different from the sinusoidal oscillations used to fit the model, this is a way of testing the robustness of the inferred parameters to extrapolation. The input had the following form:

\newcommand{\tset}{\mathcal{T}}
\newcommand{\impulse}{\mathcal{J}}
\begin{align}
  I_α(t) &= \sum_{t_0 \in \tset} \impulse_{t_0}(t) \, \\
  \impulse_{t_0}(t) &=
  \begin{cases}
    B \left(1-\frac{|t-t_0|}{d}\right) & \text{if $|t-t_0| \leq d$,} \\
    0 & \text{otherwise.}
  \end{cases}
\end{align}
\noindent The input was generated with $d = \SI{0.15}{s}$, $B = (0, 0, 0.6, -0.6)\,\si{mA}$. Impulses were placed at
\begin{equation*}
 \tset = \{11.0,  11.7,  12.2,  12.9,  14.1,  14.5,  15.5,  15.8,  16.2,  16.8\}\,\si{s} \,.
\end{equation*}

\paragraph{Numerical integration} For all simulations of the mesoGIF{} model, we used a time step of \SI{1}{ms}. We also used a \SI{1}{ms} time step when inferring parameters. Simulations of the microscopic GIF model require finer temporal resolution, and for those we used time steps of \SI{0.2}{ms}.
In order to have the same inputs at both temporal resolutions, they were generated using the finer time step, and coarse-grained by averaging.

We used the Euler-Maruyama scheme to integrate inputs; the GIF and mesoGIF{} models are given as update equations of the form $A(t+Δt) = F(A(t))$, and thus already define an integration scheme.

\subsection{Software}
\label{sec:software}

We developed software for expressing likelihoods of dynamical systems by building on general purpose machine learning libraries:
\texttt{Theano_shim} (\texttt{https://\allowbreak github\allowbreak .com/m\allowbreak ackelab/\allowbreak theano\_shim}) is a thin layer over the numerical backend, allowing one to execute the same code either using Theano \citep{thetheanodevelopmentteamTheanoPythonFramework2016} or Numpy \citep{jonesSciPyOpenSource2001}.
\texttt{Sinn} (\texttt{https://\allowbreak github.com/\allowbreak mackelab/\allowbreak sinn}) makes use of \texttt{theano_shim} to provide a backend-agnostic set of high-level abstractions to build dynamical models.
Finally, a separate repository (\texttt{https://\allowbreak github.com/\allowbreak mackelab/\allowbreak fsGIF}) provides the code specific to this paper.

\subsection*{Acknowledgments}

We thank Pedro Gonçalves, Giacomo Bassetto, Tilo Schwalger and David Dahmen for discussions and comments on the manuscript.
AR and AL were supported by NSERC (Canada); AL also acknowledges support from the Humboldt Foundation. JHM was supported by  the German Research Foundation (DFG) through  SFB 1089, and the German Federal Ministry of Education and Research (BMBF, project `ADMIMEM', FKZ 01IS18052 A-D).

\appendix
\section*{Appendix}
\nopagebreak[4]

\makeatletter
\@secpenalty 0  
\typeout{Section penalty: \the\@secpenalty}
\makeatother

\section{Priors and parameter values}
\label{sec:priors-parameters}

For both microscopic and mesoscopic models, unless otherwise specified in the text, we used the same parameters values as our ground truth values. Values are listed in \creffloat{tbl:parameter-values} and are based on those given in \citet{schwalgerTheoryCorticalColumns2017}, and we follow the recommendation therein of adjusting resting potentials $u_{\mathrm{rest}}$ to maintain realistic firing rates. To facilitate simulations, we also reduced the population sizes by a factor of 50 and correspondingly up-scaled the connectivity weights by a factor of $\sqrt{50}$, to maintain a balanced E-I network \citep{vogelsNeuralNetworkDynamics2005}.

Prior distributions on inferred parameters were set sufficiently broad to be considered noninformative. Prior distributions are independent of the population, so as to ensure that any inferred feature (e.g. excitatory vs inhibitory connections) is due to the data.

The two population heteregeneous model was obtained by sampling similar but tighter distributions as the prior (\creffloat{tbl:hetero-model-dist-params}). Only membrane and adaptation time constants were sampled; other parameters were as in \creffloat{tbl:parameter-values}.

\begin{inlinetable}
  \centering
  \small
  \renewcommand{\arraystretch}{1.2}
  \setlength{\tabcolsep}{0.9\oldtabcolsep}
  \centering
  \caption{Default parameter values and priors; symbols are the same as in \creffloat{tbl:parameters}. Most values are those given in  Tables~1 and 2 of \citet{schwalgerTheoryCorticalColumns2017}.
  Priors are given as scalar distributions because they are the same for all components.
  The p.d.f.\ of $Γ(α,θ)$ is $\frac{x^{α-1} e^{-x/θ}}{θ^α Γ(α)}$.
  \label{tbl:parameter-values}}
  \begin{tabular}{lllll<{\,}@{}l}
  \toprule
  & \multicolumn{2}{c}{Value} \tabularnewline \cmidrule{2-3}
  Parameter    & 2 pop. model & 4 pop. model & unit  & \multicolumn{2}{l}{Prior distribution} \tabularnewline
  \midrule
  pop. labels\! & E, I & L2/3e, L2/3i, L4e, L4i \tabularnewline
  $N$          & $(438, 109)$
    & $(413, 116, 438, 109)$
    & \tabularnewline
  $R$          & $(19, 11.964)$
    & $(0, 0, 19, 11.964)$
    & $Ω$\tabularnewline
  $u_{\mathrm{rest}}$     & $(20, 19.5)$
    & $(18, 18, 25, 20)$
    & mV\tabularnewline[0.5ex]
  $p$
    &  $\left(\begin{smallmatrix}
        0.0497 & 0.1350 \\
        0.0794 & 0.1597
      \end{smallmatrix}\right)$
    & $\left(\begin{smallmatrix}
        0.1009 & 0.1689 & 0.0437 & 0.0818 \\
        0.1346 & 0.1371 & 0.0316 & 0.0515 \\
        0.0077 & 0.0059 & 0.0497 & 0.1350 \\
        0.0691 & 0.0029 & 0.0794 & 0.1597
      \end{smallmatrix}\right)$
    &\tabularnewline[3.5ex]
  ${w}$
    &  $\left(\begin{smallmatrix} 2.482 & -4.964 \\ 1.245 & -4.964 \end{smallmatrix}\right)$
    & $\left(\begin{smallmatrix}
      1.245 & -4.964 & 1.245 & -4.964 \\
          1.245 & -4.964 & 1.245 & -4.964 \\
          1.245 & -4.964 & 2.482 & -4.964 \\
          1.245 & -4.964 & 1.245 & -4.964
      \end{smallmatrix}\right)$
    & {mV}   &  \footnotesize $w$ & \footnotesize $\sim \mathcal{N}\left(0, 4^2\right)$  \tabularnewline[0.5ex]
  ${τ_m}$
    & (0.01, 0.01)
    & (0.01, 0.01, 0.01, 0.01)
    & {s} & \footnotesize $\log_{10} τ_m$ & \footnotesize $\sim \mathcal{N}(-2, 2^2)$ \tabularnewline
  $t_{\mathrm{ref}}$
    & (0.002, 0.002)
    & (0.002, 0.002, 0.002, 0.002)
    & s\tabularnewline
  ${u_{\mathrm{th}}}$
    & (15, 15)
    & (15, 15, 15, 15)
    & {mV}  & \footnotesize $u_{\mathrm{th}}$ & \footnotesize $\sim \mathcal{N}(15, 10^2)$ \tabularnewline
  ${u_{\mathrm{r}}}$
    & (0, 0)
    & (0, 0, 0, 0)
    & {mV} & \footnotesize $u_{\mathrm{r}}$ & \footnotesize $\sim \mathcal{N}(0, 10^2)$ \tabularnewline
  ${c}$
    & (10, 10)
    & (10, 10, 10, 10)
    & {Hz} & \footnotesize $c$ & \footnotesize $\sim Γ(2, 5)$ \tabularnewline
  ${{Δ_{\mathrm{u}}}}$
    & (5, 5)
    & (5, 5, 5, 5)
    & {mV} & \footnotesize ${Δ_{\mathrm{u}}}$ & \footnotesize $\sim Γ(3, 1.5)$ \tabularnewline
  $Δ$
    & 0.001 
    & 0.001
    & s\tabularnewline
  ${τ_s}$
    & (0.003, 0.006)
    & (0.003, 0.006, 0.003, 0.006)
    & {s} & \footnotesize $\log_{10} τ_s$ & \footnotesize $\sim \mathcal{N}(-3, 3^2)$ \tabularnewline
  ${J_θ}$
    & (1.0, 0)
    & (1.0, 0, 1.0, 0)
    & {mV} & \footnotesize $J_θ$ & $\sim Γ(2, 0.5)$ \tabularnewline
  ${τ_θ}$
    & (1.0, --)
    & (1.0, --, 1.0, --)
    & {s} & \footnotesize $\log_{10} τ_θ$ & \footnotesize $\sim \mathcal{N}(-1, 5^2)$ \tabularnewline
  \bottomrule
 \end{tabular}
  \setlength{\tabcolsep}{0.95\oldtabcolsep}
\end{inlinetable}

\begin{inlinetable}
  \centering
  \caption{Distribution parameters for the heterogeneous model. Each parameter was sampled from a log-normal distribution $\log_{10} \mathcal{N}(μ, σ^2)$ with mean $μ$ and variance $σ^2$. No adaptation was modeled in the inhibitory population, so $τ_{θ,I}$ was not sampled.
  \label{tbl:hetero-model-dist-params}
  }
  \begin{tabular}{lcc}
    \toprule
  Heterogeneous model & \multicolumn{2}{c}{Distribution parameter} \\\cmidrule{2-3}
  \hfill parameter \hfill &  $μ$ &  $σ$ \\
  \midrule
  $\log_{10} τ_{m,E}$ & -1.6 &  0.5 \\
  $\log_{10} τ_{m,I}$ & -1.8 &  0.5 \\
  $\log_{10} τ_{θ,E}$ & -0.7 &  0.5 \\
  \bottomrule
  \end{tabular}
\end{inlinetable}

\FloatBarrier

\section{Inferred parameters}

\begin{inlinetable}
  \centering
  \caption{\captitle{Inferred parameters for a heterogeneous population} (\cref{sec:inference-hetero-model}); values are given in vector format, as $(η_E, η_I)$. Corresponding average values for the heterogeneous microscopic model are given for comparison. (The heterogeneous model was homogeneous in all parameters except $τ_m$ and $τ_θ$.)
 \label{tbl:hetero-fit-result}
 }
 \begin{tabular}{llll}
  \toprule
  Parameter  & Inferred value   & \parbox{9em}{Average\\ heterogeneous value} & Unit\tabularnewline[.4ex]
  \midrule
  \Tstrut $w$
  & $\left(\begin{smallmatrix} 1.59 & -5.05 \\ 0.73 & -3.43 \end{smallmatrix}\right)$
  & $\left(\begin{smallmatrix} 2.482 & -4.964 \\ 1.245 & -4.964 \end{smallmatrix}\right)$
  & mV\tabularnewline[1ex]
  $τ_m$   & (0.011, 0.008)     &  (0.056, 0.046)          & s\tabularnewline
  $c$     & (5.05, 5.22)       &  (10, 10)                & Hz\tabularnewline
  ${Δ_{\mathrm{u}}}$    & (5.09, 4.09)       &  (5, 5)                  & mV\tabularnewline
  $τ_s$   & (0.0046, 0.0109)   &  (0.003, 0.006)          & s\tabularnewline
  $J_θ$   & (0.538, 0)         &  (1.0, 0)                & mV\tabularnewline
  $τ_θ$   & (0.131, --)        &  (0.380, --)          & s\tabularnewline
  \bottomrule
 \end{tabular}
\end{inlinetable}

\begin{inlinetable}
  \centering
  \caption{
  \captitle{Inferred values for the \num{4} population model.} The values for the homogeneous microscopic model used in \creffloats{fig:4pop-setup}{fig:sim-compare-4pop} are listed on the right. Theory predicts these to be the best parameterization for the mesoscopic model, and thus should be recovered by maximizing the posterior (MAP values). Since {L2/3} receives no external input in the training data, the inferred parameters for those populations are understandably further from theory.
  \label{tbl:4pop-fit-values}
  }
  \setlength{\tabcolsep}{0.85\oldtabcolsep}
  \sisetup{table-format = +1.3}
  \begin{tabular}{lSSSSc@{\hspace{0.75em}}SSSS}
    \toprule
       & \multicolumn{ 4 }{c}{MAP} && \multicolumn{ 4 }{c}{Theory}\\
          & {$L2/3e$} & {$L2/3i$} & {$L4e$} & {$L4i$} && {$L2/3e$} & {$L2/3i$} & {$L4e$} & {$L4i$} \\ \cmidrule(lr){2-5} \cmidrule(lr){7-10}
          $w_{ L2/3e\leftarrow \cdot}$ & 0.734 & -5.629 & 1.546 & -5.292 && 1.245 & -4.964 & 1.245 & -4.964 \\
      $w_{ L2/3i\leftarrow \cdot}$ & 1.181 & -5.406 & 1.419 & -4.294 && 1.245 & -4.964 & 1.245 & -4.964 \\
      $w_{ L4e\leftarrow \cdot}$ & 1.528 & -0.637 & 2.058 & -4.213 && 1.245 & -4.964 & 2.482 & -4.964 \\
      $w_{ L4i\leftarrow \cdot}$ & 0.174 & 1.112 & 1.046 & -3.994 && 1.245 & -4.964 & 1.245 & -4.964 \\
      $τ_m$ & 0.016 & 0.015 & 0.008 & 0.009 && 0.010 & 0.010 & 0.010 & 0.010 \\
      $c$ & 16.717 & 18.170 & 9.020 & 9.680 && 10.000 & 10.000 & 10.000 & 10.000 \\
      $Δu$ & 7.435 & 6.453 & 4.750 & 4.420 && 5.000 & 5.000 & 5.000 & 5.000 \\
      $τ_s$ & 0.001 & 0.006 & 0.002 & 0.009 && 0.003 & 0.006 & 0.003 & 0.006 \\
      $J_θ$ & 0.232 &  {—}  & 0.967 &  {—}  && 1.000 &  {—}  & 1.000 &  {—}  \\
      $τ_θ$ & 0.425 &  {—}  & 1.596 &  {—}  && 1.000 &  {—}  & 1.000 &  {—} \\
    \bottomrule
  \end{tabular}
  \setlength{\tabcolsep}{\oldtabcolsep}
\end{inlinetable}

\FloatBarrier
\section{Alternative initialization scheme}
\label{sec:alternative-initialization}

Compared to the silent initialization (\cref{sec:initialization}), the ''stationary initialization`` finds a more realistic initial state, which reduces the burn-in time required by about an order of magnitude. This makes it more practical when minibatches are selected random, and we used this scheme to validate \cref{alg:sequential-adam} (c.f.\ \cref{sec:sgd}). However in general we found the computational gain to be offset by the added cost of solving a self-consistent equation for each batch.

\paragraph{Stationary initialization} Assuming zero external input, we find a self-consistent equation for the stationary activity $A^*$ (c.f.\ \cref{sec:stationary-state-derivation}). After solving numerically for $A^*$, the other state variables are then easily computed.

\begin{algorithm}
  \caption{Stationary initialization scheme. \label{alg:init-scheme-stationary}}
  \begin{algorithmic}[1]
    \State $A_α^* \gets $ \text{Solve} \cref{eq:self-consistent-equation}
    \State $n_α \gets A_α^* \,N_α\, Δt$
    \State $h_α, y_{αβ}, u_{α,i}, g_α, λ_{\mathrm{free},α}, λ_{α,i}, x_α, z_{α}, m_{α,i}, v_{α,i} \gets$ Evaluate
    Eqs.~(\ref{eq:update-y}–\ref{eq:update-v})
    with $A_α^*$
  \end{algorithmic}
\end{algorithm}

\FloatBarrier

\section{Data requirements for different parameter sets}
\label{sec:app:fit-statistics}

In \cref{sec:data-requirements}, we showed that less than $\num{10}{s}$ of data were sufficient to infer the parameters of the two-population mesoGIF{} model. Of course, the exact data requirements will depend on how many parameters we need to infer and which they are (e.g.\ $w$ vs $τ_m$).

\begin{figure}
  \includegraphics{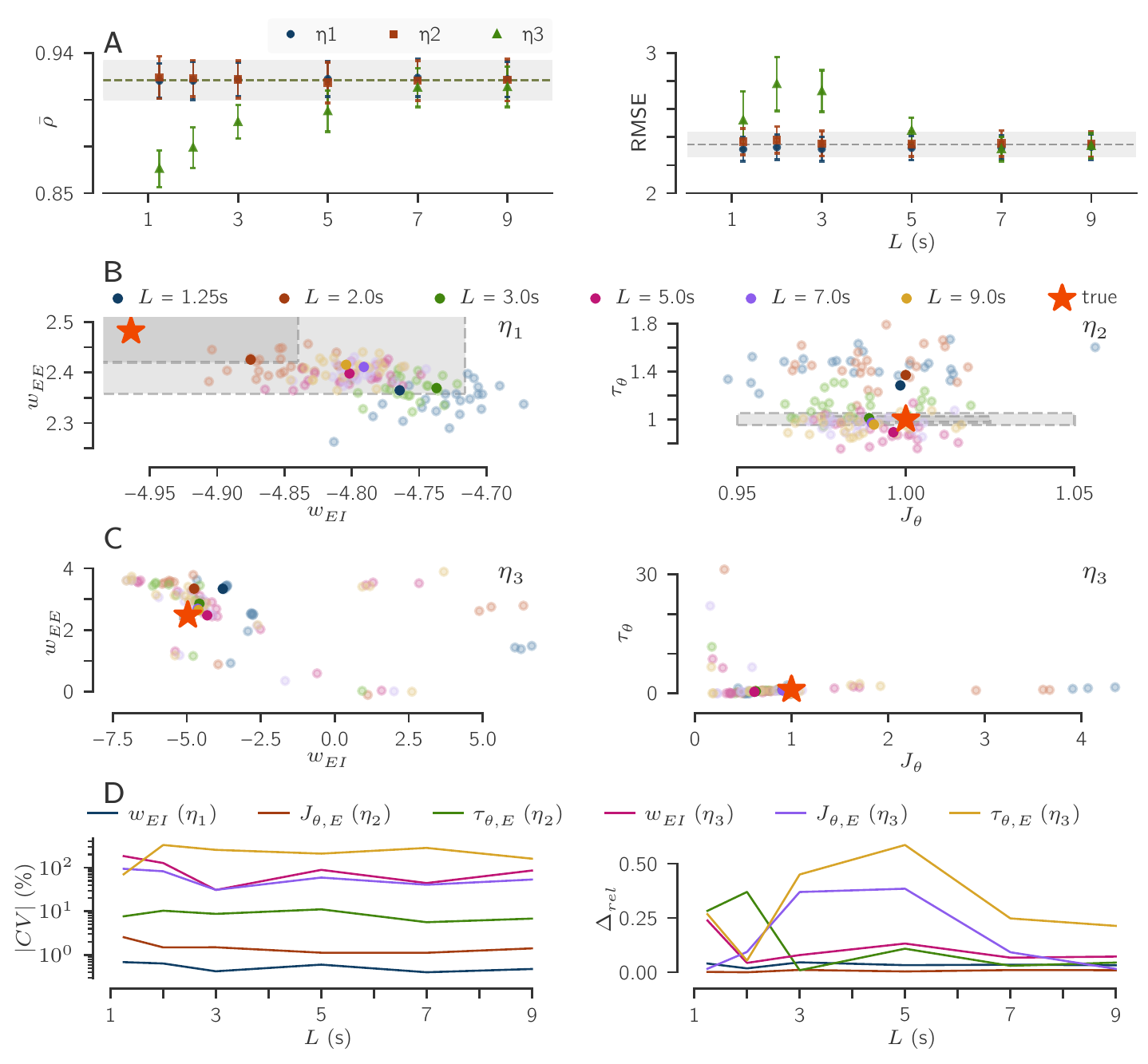}
  {\phantomsubcaption\label{fig:data-reqs-corr-rms}}
  {\phantomsubcaption\label{fig:data-reqs-scatter}}
  {\phantomsubcaption\label{fig:data-reqs-scatter-eta5}}
  {\phantomsubcaption\label{fig:data-reqs-CV}}
  \caption{
  \captitle{Fits of many parameters are less consistent.}
  \subreflabel{fig:data-reqs-corr-rms}~As the number of inferred parameters is increased, more data is required to estimate them. ($η_1$, $η_2$ are parameter sets corresponding respectively to connectivity and adaptation.  $η_3 \supset (η_1 \cup η_2$) is the set of all parameters. Definitions in \creffloat{tbl:param-subsets}.)
  \subreflabel{fig:data-reqs-scatter}~Results from 24 fits for subsets ${η_1}$ (left) and ${η_2}$ (right) for different amounts $L$ of data. Star indicates the true parameters, and gray boxes the 5 and 10\% relative errors $Δ_{\mathrm{rel}}$. Fits cluster around the MAP, which for finite amounts of data will not exactly coincide with the ground truth values. Darker dots indicate the fit with the highest likelihood. The consistency of estimates for the adaptation parameters, with $τ_{θ,E} = \SI{1}{s}$, is particularly improved with longer data traces.
  \subreflabel{fig:data-reqs-scatter-eta5}~Same as in (\subref{fig:data-reqs-scatter}) but all parameters were simultaneously inferred. The reduced consistency is noticeable by the change of scale, at which the 5\% and 10\% relative error boxes are not visible.
  \subreflabel{fig:data-reqs-CV}~When going from inferring smaller (${η_1}$, ${η_2}$) to larger (${η_3}$) subsets of parameters, the increase in relative error for the same number of fits is relatively modest (right) compared to the wider area in parameter space to which fits converge (left). Figure traces statistics for different parameters as a function of the amount of data ($L$) and the subset of parameters which were fit simultaneously (${η_1}$--${η_3}$).
  Values for all data/subset combinations are given in \creffloats{tbl:all-fits-deltas}{tbl:all-fits-CVs}.
  \label{fig:data-reqs}
  }
\end{figure}

To explore this issue, we repeated the inference procedure for the parameter subsets listed in \creffloat{tbl:param-subsets}, performing \num{24}~fits for each subset using different amounts of data. Subsets ${η_1}$ and ${η_2}$ parameterize respectively the connectivity and the adaptation, while ${η_3}$ is the full set used for \creffloat{fig:param-error}. A similar figure to \creffloat{fig:param-error} with all three subsets is shown in \creffloat{fig:data-reqs-corr-rms}.

\begin{inlinetable}
  \centering
  \caption{Definition of parameter subsets for the two population model. There are only two adaptation parameters because inhibitory populations have no adaptation in this model.
  \label{tbl:param-subsets}}
  \begin{tabular}{ll}
    \toprule
    Subset label & Included parameters \tabularnewline
    \midrule
    \Tstrut ${η_1}$ & $\{w_{EE}, w_{EI}, w_{IE}, w_{II}\}$ \\
    ${η_2}$ & $\{τ_{θ,E}, J_{θ,E}\}$ \\
    ${η_3}$ & $η_1 \cup η_3 \cup \{c_E, c_I, {Δ_{\mathrm{u},E}}, {Δ_{\mathrm{u},I}}, τ_{m,E}, τ_{m,I}, τ_{s,E}, τ_{s,I}\}$ \\
    \bottomrule
  \end{tabular}
\end{inlinetable}

With the smaller subsets (${η_1}$, ${η_2}$), \SI{1.25}{s} of data was sufficient to get good accuracy of the inferred dynamics (\creffloat{fig:data-reqs-corr-rms}). However working with such small amounts of data incurs a substantial computational cost. Firstly because the fits converge less consistently, thus requiring more fits to find a good estimate of the MAP (\creffloat{fig:data-reqs}, \subref{fig:data-reqs-scatter}, \subref{fig:data-reqs-scatter-eta5} and \subref{fig:data-reqs-CV} left). And secondly because the algorithm optimizations making use of the longer traces (c.f.\ \cref{sec:sgd}) are no longer as effective, making each iteration slower on average.

Since we know the ground truth parameters, we can further estimate the expected error by computing the relative difference between true and inferred parameter values. For a parameter $η$ and its estimate $η^{(L)}$ obtained by using $L$ seconds of data, this is calculated as
\begin{equation}
  \label{eq:rel-error-def}
  Δ_{\mathrm{rel}}\left(\hat{η}^{(L)}\right) \coloneqq \left\lvert\frac{\hat{η}^{(L)} - η}{η}\right\rvert  \,.
\end{equation}
The number of fits required to achieve this performance will vary according to the nature and number of parameters; indeed with more parameters to infer, we found that fits terminated further from the true values. A simple way then to quantify the uncertainty of any one particular fit is the sample standard deviation $σ_{η}$ of the set of found optima from a collection of fits. In order to make the $σ_{η}$ comparable between parameters, we normalized by the parameter mean $μ_{η}$ to obtain the coefficient of variation:
\begin{equation}
  \label{eq:CV-def}
  \left\lvert CV(η^{(L)})\right\rvert \stackrel{\text{\tiny def}}{=} \left\lvert σ_{η^{(L)}} / μ_{η^{(L)}}\right\rvert
\end{equation}
Relative error and CV values for all parameter subsets are listed in \creffloats{tbl:all-fits-deltas}{tbl:all-fits-CVs}.

\begin{inlinetable}[!htb]
  \centering
\caption{
  Relative error for the fits shown in \cref{sec:data-requirements}.
  \label{tbl:all-fits-deltas}
}
  \sisetup{table-format = 1.3}
  {\footnotesize\begin{tabular}{llSSSSSS}
\toprule
& & \multicolumn{6}{c}{L} \\
\cmidrule(rl){3-8}
Subset & Parameter  &    {1.25} &    {2.00} &     {3.00} &     {5.00} &     {7.00} &     {9.00} \\
\midrule
${η_1}$ & $w_{EE}$ & 0.047 & 0.023 & 0.045 & 0.034 & 0.029 & 0.027 \\
   & $w_{EI}$ & 0.040 & 0.018 & 0.046 & 0.033 & 0.035 & 0.032 \\
   & $w_{IE}$ & 0.013 & 0.038 & 0.000 & 0.001 & 0.005 & 0.024 \\
   & $w_{II}$ & 0.018 & 0.005 & 0.022 & 0.018 & 0.012 & 0.005 \\
${η_2}$ & $J_{θ,E}$ & 0.002 & 0.000 & 0.011 & 0.004 & 0.010 &   0.009 \\
   & $τ_{θ,E}$ & 0.283 & 0.370 & 0.009 & 0.108 & 0.030 &   0.045 \\
${η_3}$
   & $w_{EE}$ & 0.345 & 0.348 & 0.151 & 0.001 & 0.084 & 0.067 \\
   & $w_{EI}$ & 0.238 & 0.043 & 0.079 & 0.132 & 0.067 & 0.072 \\
   & $w_{IE}$ & 0.017 & 0.556 & 0.630 & 0.427 & 0.244 & 0.178 \\
   & $w_{II}$ & 0.070 & 0.495 & 0.503 & 0.326 & 0.180 & 0.136 \\
   & $J_{θ,E}$ & 0.016 & 0.094 & 0.369 & 0.385 & 0.092 & 0.016 \\
   & $τ_{θ,E}$ & 0.267 & 0.054 & 0.450 & 0.586 & 0.248 & 0.213 \\
   & $c_{E}$ & 0.420 & 0.376 & 0.469 & 0.382 & 0.239 & 0.160 \\
   & $c_{I}$ & 2.825 & 0.006 & 0.133 & 0.142 & 0.128 & 0.161 \\
   & $\Delta u_E$ & 0.215 & 0.182 & 0.090 & 0.086 & 0.092 & 0.052 \\
   & $\Delta u_I$ & 0.520 & 0.338 & 0.466 & 0.302 & 0.129 & 0.058 \\
   & $τ_{m,E}$ & 0.190 & 0.117 & 0.057 & 0.177 & 0.052 & 0.037 \\
   & $τ_{m,I}$ & 2.590 & 0.619 & 0.430 & 0.393 & 0.235 & 0.219 \\
   & $τ_{s,E}$ & 0.744 & 0.101 & 0.038 & 0.101 & 0.039 & 0.142 \\
   & $τ_{s,I}$ & 0.271 & 0.119 & 0.138 & 0.132 & 0.081 & 0.081 \\
\bottomrule
\end{tabular}}
\end{inlinetable}
\begin{inlinetable}[!ht]
  \centering
\caption{
  Coefficients of variation for the collections of fits shown in \cref{sec:data-requirements}.
  \label{tbl:all-fits-CVs}
}
\sisetup{table-format = +4.2}
\setlength{\tabcolsep}{1.0\oldtabcolsep}
{\footnotesize\begin{tabular}{ll
                 S[table-format=+3.2]
                 S[table-format=+3.2]
                 SSSS}
\toprule
  & & \multicolumn{6}{c}{L} \\
  \cmidrule(rl){3-8}
 Subset & Parameter  &    {1.25} &    {2.00} &     {3.00} &     {5.00} &     {7.00} &     {9.00} \\
\midrule
${η_1}$ & $w_{EE}$ &    1.33 &    1.05 &     0.95 &     0.72 &     0.61 &     0.82 \\
   & $w_{EI}$ &    0.69 &    0.63 &     0.43 &     0.62 &     0.40 &     0.48 \\
   & $w_{IE}$ &    1.38 &    1.48 &     1.53 &     1.75 &     1.90 &     1.36 \\
   & $w_{II}$ &    0.40 &    0.46 &     0.56 &     0.39 &     0.61 &     0.65 \\
${η_2}$ & $J_{θ,E}$ &    2.56 &    1.49 &     1.50 &     1.13 &     1.13 &      1.42 \\
   & $τ_{θ,E}$ &    7.70 &   10.33 &     8.75 &    11.11 &     5.64 &      6.82 \\
${η_3}$
   & $w_{EE}$ &   31.63 &   31.73 &    29.26 &    37.84 &    38.29 &    31.49 \\
   & $w_{EI}$ &  183.77 &  127.48 &    30.84 &    88.90 &    44.35 &    85.80 \\
   & $w_{IE}$ &   64.08 &  100.98 &  1833.28 &  1143.47 &  1489.30 &  2405.89 \\
   & $w_{II}$ &   53.90 &   65.92 &    36.25 &    78.14 &    37.70 &    55.10 \\
   & $J_{θ,E}$ &   94.03 &   82.51 &    30.99 &    59.25 &    40.58 &    53.40 \\
   & $τ_{θ,E}$ &   70.53 &  329.84 &   255.21 &   209.10 &   282.68 &   160.36 \\
   & $c_{E}$ &   83.08 &   65.81 &    37.16 &    39.42 &    48.50 &    47.80 \\
   & $c_{I}$ &   29.31 &   28.84 &    34.10 &    52.49 &    49.66 &    52.50 \\
   & $\Delta u_E$ &   14.98 &   19.16 &     6.11 &    11.37 &     7.58 &     8.84 \\
   & $\Delta u_I$ &   32.84 &   30.33 &    13.28 &    41.22 &    21.86 &    34.30 \\
   & $τ_{m,E}$ &  429.43 &  420.71 &   428.10 &   431.68 &   441.05 &   444.64 \\
   & $τ_{m,I}$ &  256.36 &  269.36 &   346.42 &   337.09 &   314.64 &   288.72 \\
   & $τ_{s,E}$ &  427.66 &  441.61 &   447.11 &   446.85 &   446.84 &   447.20 \\
   & $τ_{s,I}$ &  238.26 &  240.74 &    65.92 &   262.20 &    71.13 &   295.76 \\
\bottomrule
\end{tabular}}
\setlength{\tabcolsep}{\oldtabcolsep}
\end{inlinetable}

\FloatBarrier

\section{Mesoscopic update equations}
\label{sec:meso-equations}

This \lcnamecref{sec:meso-equations} first describes the quantities composing the state vector for the mesoGIF{} model, then lists the equations used for this paper. All equations are for discretized time, and we use a superscript ${(k)}$ to indicate the $k$-th time step.  For derivations and a more complete discussion of the variables involved, see \citet{schwalgerTheoryCorticalColumns2017}.

\subsection{Construction of the state vector}
\label{sec:construction-state-vector}

In order to obtain a finite state-vector (c.f.\ \cref{sec:details-meso-model}), neurons are divided into two categories: ``free'`and ``refractory''; the assignment of neurons to either category changes over time, following a discretized form of the transport equation~\eqref{eq:RDM-transport-equation}.

\emph{Refractory neurons} are still in the absolute or relative refractory period caused by their last spike, and thus have a higher firing threshold. Since the height of the threshold is dependent on that spike's time, we track a vector $m_{\alpha}^{(k)}$, indexed by the age $l$.  We define the scalar $m_{\alpha,l}^{(k)}$ as our estimate of the number of neurons at time $t_k$ which last fired at time $t_{k-l}$. A vector $v_{α}^{(k)}$ similarly tracks the variance of that estimate. The adaptation of the neurons depends on their age, such that their firing rate is also given by a vector, $λ_{α}^{(k)}$. With an adaptation time scale $τ_θ$ of \SI{1}{s} and time steps of \SI{1}{ms}, these vectors each comprise around $K = \num{1000}$ age bins. For a more detailed discussion on properly choosing $K$, see \eqname~(86) in \citet{schwalgerTheoryCorticalColumns2017}.

\emph{Free neurons}, meanwhile, have essentially forgotten their last spike: their firing threshold has relaxed back to its resting state, and so they can be treated as identical, independent of when that last spike was. One scalar per population, $λ_{\mathrm{free},α}^{(k)}$, suffices to describe their firing rate. Scalars $x_α^{(k)}$ and $z_α^{(k)}$ respectively track the estimated mean and variance of the number of free neurons.

In the case of an infinite number of neurons, the firing rates $λ_{α}^{(k)}$ and $λ_{\mathrm{free},α}^{(k)}$ would be exact, but for finite populations a further correction $P_{Λ,\alpha}^{(k)}$ must be made to account for statistical fluctuations.
Combining $λ_{\mathrm{free},\alpha}^{(k)}$, ${λ}_{α}^{(k)}$ and $P_{Λ,\alpha}^{(k)}$, one can compute $\bar{n}_{\alpha}^{(k)}$, the expected number of spikes at $t_k$. The definition of $n^{(k)}$ then follows as described in \cref{sec:details-meso-model}.

\begin{figure}
 \includegraphics[width=1.0\fullwidth]{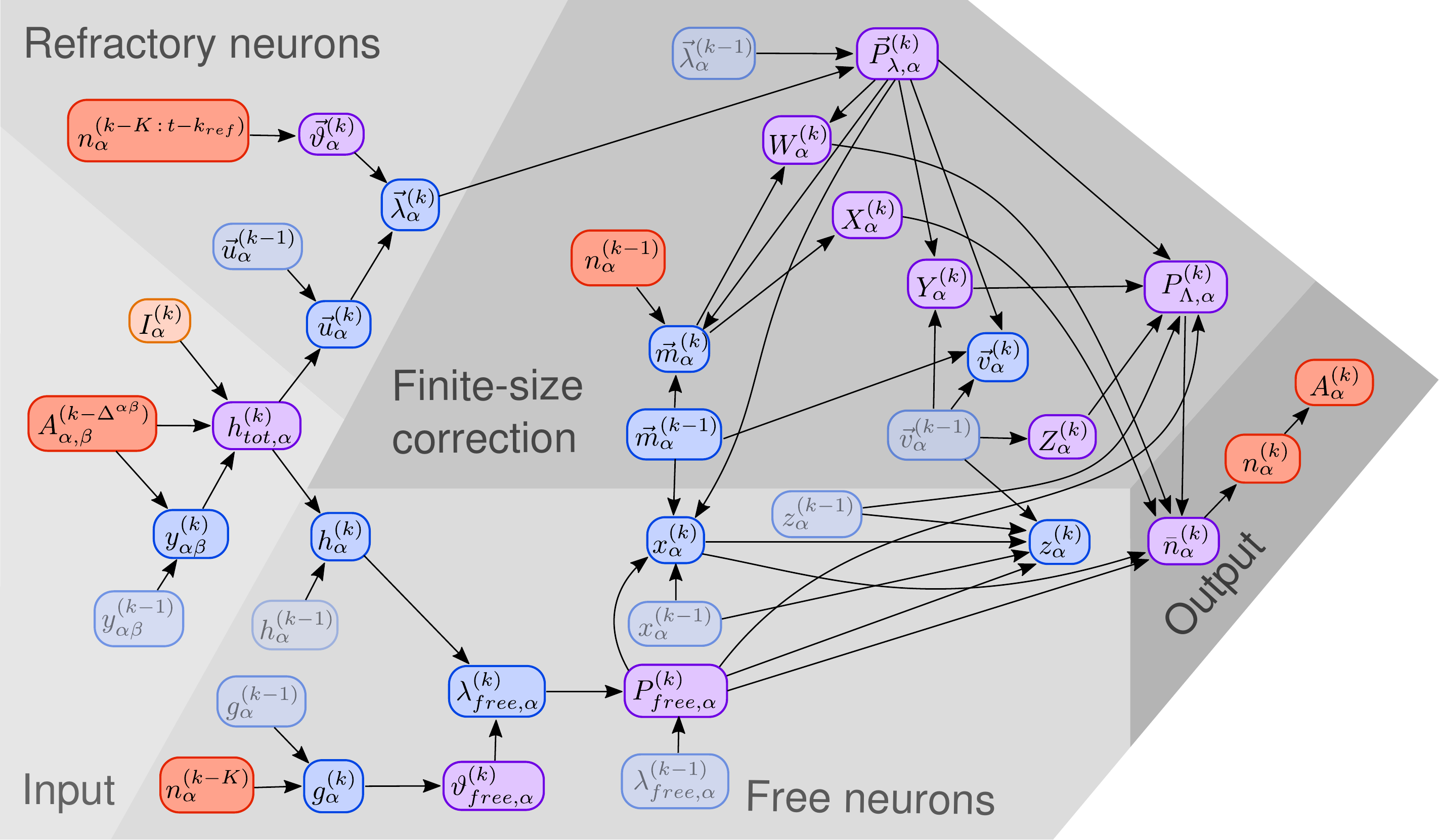}
 \caption{\captitle{Graphical representation of the mesoscopic model.} An arrow $x \rightarrow y$ indicates that $x$ is required in order to compute $y$. Red (orange) boxes indicate observed variables (input). Variables in blue boxes must be stored until the next iteration, and along with the activity $A$, form the model's state. Intermediate variables shown in purple do not need to be stored. Indices in parentheses indicate the time step, greek letters the population index. During simulation, mesoscopic model parameters (not shown, but determine the computations along arrows) are fixed, and mesoscopic output variables $A_{\alpha}^{(k)}$ are generated in a given time step; these values form the input for the next time step. During inference, the input is obtained from the training data, and is used to  compute the sequence of binomial means $\bar{n}_{\alpha}^{(k)}$. These outputs, along with the observed outputs in the training data, are used to compute the likelihood. The gradient descent algorithm then changes the model parameters after each batch of training data to maximize the likelihood. See also \citet[\figname~12]{schwalgerTheoryCorticalColumns2017}.
\label{fig:unrolled-fsGIF}}
\end{figure}

For both refractory and free neurons, the dependency of their time evolution on the spiking history of the network is taken into account by convolving the population activities (one per population) with synaptic, membrane and adaptation kernels.
Following \citet{schwalgerTheoryCorticalColumns2017}, we express these as exponential filters; this allows the associated convolutions to be respectively replaced by three additional dynamic variables $y$, $h$ and $g$, making forward simulations more efficient. Replacing temporal filters by dynamic variables has the additional important benefit of making the dynamics Markovian when we consider them as updates on a state ${S}^{(k)}$, composed of the concatenation of the blue variables in \cref{fig:unrolled-fsGIF},
\pagebreak[0]\begin{samepage}
\begin{equation}
  \label{eq:state-vector}
{S}^{(k)} \coloneqq \left(n^{(k)}, y^{({k})}, g^{({k})}, h^{({k})}, \bm{u^{({k})}}, \bm{λ^{({k})}}, λ^{({k})}_{\mathrm{free}}, \bm{m^{({k})}}, \bm{v^{({k})}}, x^{({k})}, z^{({k})}\right) \,.
\end{equation}
For clarity, we have here typeset in bold the components of ${S}^{(k)}$ with both population and age dimensions.
\end{samepage}


\subsection{Update equations}

The equations below follow from \citet{schwalgerTheoryCorticalColumns2017} after setting the synaptic filter to an exponential:
$ε_{αβ}(s) = Θ(s-Δ_{αβ})e^{-(s-Δ)/τ_s,β} / τ_s(β)$. They depend on the inverse link function $f$, relating membrane potential to spiking probability, and a refractory/adaptation kernel $θ$. Throughout this work we used
\begin{align}
  \label{eq:link-function}
  f_α(u') &= c_α \exp(u'/{Δ_{\mathrm{u},α}})
\intertext{and}
  \label{eq:adaptation-kernel}
  θ_α(t) &= \begin{cases}
    \infty & \text{if $t < t_{\mathrm{ref,α}}$,} \\
    \frac{J_{θ,α}}{τ_{θ,α}} e^{(t-t_{\mathrm{ref,α}})/τ_{θ,α}} & \text{otherwise.}
  \end{cases}
\end{align}
The quasi-renewal kernel \citep{naudCodingDecodingAdapting2012} used below is defined as
\begin{equation}
  \tilde{θ}_{α}(t) = {Δ_{\mathrm{u},α}} \left[1 - e^{-θ_α(t)/{Δ_{\mathrm{u},α}}}\right] \,.
\end{equation}
\noindent State vectors assign the index 0 to the time $Δt$, such that they run from $θ_0 = θ(Δt)$ to $θ_K = θ((K+1)Δt)$, with $K \in {\mathbb N}$. We define ${k_{\mathrm{ref}}}$ to be the lengths of the absolute refractory periods in time bins, i.e.\ $t_{\mathrm{ref,α}} = {k_{\mathrm{ref}, α}} Δt$ for each population $α$.

\paragraph{Total input}\label{total-input}

\begin{align}
 h^{(k+1)}        & = u_{rest} + (u^{(k)} - u_{rest})e^{-Δt/τ_m} + h_{tot}                                       \label{eq:update-h}     \,, \\
 y_{αβ}^{(k+1)} & = A_{β}(t_k - Δ_{αβ}) + \left[y_{αβ}^{(k)} - A_{β}(t_k - Δ_{αβ})\right] e^{-Δt/τ_{s,β}}  \label{eq:update-y} \,.
\end{align}

\begin{equation}
  \label{eq:update-htot}
 \begin{split}
  h_{tot,α} &= RI_{ext}^{(k)} \left(1 - e^{-Δt/τ_{m,α}}\right) + τ_{m,α} \sum_{β=1}^M p_{αβ} N_β w_{αβ} \Biggl\{ A_β(t \!-\! Δ_{αβ}) \\
  &\quad + \frac{τ_{s,β} e^{-\tfrac{Δt}{τ_{s,β}}} \! \left[y_{αβ}^{(k)} - A_β(t_k \!-\! Δ_{αβ})\right]
   - e^{-\tfrac{Δt}{τ_{m,α}}} \! \left[τ_{s,β} y_{αβ}^{(k)} - τ_{m,α} A_β(t_k \!-\! Δ_{αβ})\right]}
  {τ_{s,β} - τ_{m,α}} \Biggr\}
 \end{split}
\end{equation}

\paragraph{Membrane potential, refractory neurons}\label{membrane-potential-refractory-neurons}

\begin{equation}
  \label{eq:update-u}
 u_{α,i}^{(k+1)} =
 \begin{cases}
  u_{r,α}                                                                          & \text{$0 \leq i < {k_{\mathrm{ref}, α}}$,} \\
  u_{rest,α} + (u_{α,i-1}^{(k)} - u_{rest,α}) e^{-Δt/τ_{m,α}} + h_{tot,α}^{(k)} & \text{$i \geq {k_{\mathrm{ref}, α}}$.}
 \end{cases}
\end{equation}

\paragraph{Firing threshold}\label{firing-threshold}

\begin{align}
 \vartheta_{αi}^{(k+1)}      & = \vartheta_{\mathrm{free},α}^{(k+1)} + θ_{αi} + \frac{1}{N} \sum_{j=i+1}^K \tilde{θ}_{α,j} Δn_α^{(k-j-1)} \,, \\
 \vartheta_{\mathrm{free},α}^{(k+1)} & = u_{th,α} + J_{θ,α} e^{-T/τ_{θ,α}} g^{(k+1)}     \,, \\
 \label{eq:update-g}
 g^{(k+1)}_α                & = e^{-Δt/τ_{θ,α}} g_α^{(k)} + (1 - e^{-Δt/τ_{θ,α}}) A_α^{(k+1-K)} \,.
\end{align}

\paragraph{Firing probabilities}\label{firing-probabilities}

\begin{align}
  \label{eq:update-l}
 λ_{\mathrm{free},α}^{(k)} &= f(h_α^{(k)} - \vartheta_{\mathrm{free},α}^{(k)}) \,,
   & λ_{αi}^{(k)} & =
 \begin{cases}
 0 & \text{$0 \leq i < {k_{\mathrm{ref}, α}}$,} \\
 f(u_{αi}^{(k)} - \vartheta_{αi}^{(k)}) & \text{${k_{\mathrm{ref}, α}} \leq i < K$.}
 \end{cases} \\
 P_{\mathrm{free},α}^{(k)} & = 1 - e^{-\bar{λ}_{\mathrm{free},α}^{(k)} Δt} \,, & P_{λ,αi}^{(k)} & = 1 - e^{-\bar{λ}_{αi}^{(k)} Δt} \,,
\end{align}

where

\begin{align}
 \bar{λ}_{\mathrm{free},α}^{(k)} & = [λ_{\mathrm{free},α}^{(k-1)} + λ_{\mathrm{free},α}^{(k)}]/2 \,, \\
 \bar{λ}_{αi}^{(k)}    & = [λ_{α,i-1}^{(k-1)} + λ_{αi}^{(k)}]/2 \,.
\end{align}

\paragraph{Survival counts}\label{population-dynamics}

\begin{align}
 \bar{n}_α^{(k)} &= \sum_{i=0}^{K-1} P_{λ,αi}^{(k)} \bar{m}_{αi}^{(k)} + P_{\mathrm{free},α}^{(k)} x_α^{(k)}
 + P_{Λ,α}^{(k)} \left( N_α - \sum_{i=0}^{K-1} \bar{m}_{αi}^{(k)} - x_α^{(k)} \right) \,, \\
 a_α^{(k)} &= \frac{\bar{n}_α^{(k)}}{N_α Δt}  \,, \label{eq:exp-a-meso}
\end{align}

where

\begin{equation}
 P_{Λ,α}^{(k)} = \frac{\sum_{i=0}^{K-1} P_{λ,αi}^{(k)} v_{αi}^{(k)} + P_{\mathrm{free}}^{(k)} z_{α}^{(k)}}{\sum_{i=0}^{K-1} v_{αi}^{(k)} + z_{α}^{(k)}} \,,
\end{equation}

\begin{alignat}{2}
  \label{eq:update-x}
 x_{α}^{(k)} \; & =\; \sum_{i=K}^{\infty} \bar{m}_{αi}^{(k)} \; &   & =\; (1 - P_{\mathrm{free},α}^{(k)}) x_{α}^{(k-1)} + m_{αK}^{(k)}    \,,                                     \\
  \label{eq:update-z}
 z_{α}^{(k)} \; & =\; \sum_{i=K}^{\infty} v_{αi}^{(k)} \;       &   & =\; (1 - P_{\mathrm{free},α}^{(k)})^{2} z_{α}^{(k-1)} + P_{\mathrm{free},α}^{(k)} x_{α}^{(k-1)} + v_{αK}^{(k)} \,,
\end{alignat}

\begin{align}
  \label{eq:update-m}
 \bar{m}_{αi}^{(k)} & =
 \begin{cases}
 n_{α}^{(k-1)}     & \text{if $i = 0$,} \\
 \mathrlap{[1 - P_{λ,αi}^{(k)}] \bar{m}_{α,i-1}^{(k-1)}}
   \hphantom{[1 - P_{λ,αi}^{(k)}]^2 v_{α,i-1}^{(k-1)} + P_{λ,αi}^{(k)} m_{α,i-1}^{(k-1)} }
   & \text{otherwise;}
 \end{cases} \\
 \label{eq:update-v}
 v_{αi}^{(k)}       & =
 \begin{cases}
 0                    & \text{if $i = 0$,} \\
 [1 - P_{λ,αi}^{(k)}]^2 v_{α,i-1}^{(k-1)} + P_{λ,αi}^{(k)} \bar{m}_{αi-1}^{(k-1)} & \text{otherwise.}
 \end{cases}
\end{align}

\paragraph{Spike generation}

\begin{equation}
  \label{eq:spike-generation-appendix}
 n_{α}^{(k)} \sim \Binom(\bar{n}_{α}^{(k)} / N_{α} ; N_{α} ) \,.
\end{equation}
\noindent This last equation is the one identified as \cref{eq:meso-binomial} in the main text.
\FloatBarrier
\section{Performance of four population models}
\newcommand{\cc}[1]{\rlap{#1}\hphantom{Theory}}
\begin{inlinetable}
  \centering
  \caption{
    {Performance of four population models -- Per-trial $\RMSE$ (\cref{eq:rms-def}). Measures computed from \num{60}~realizations of each model.}
    \label{tbl:performance-4pop-rmse}
    }
   \sisetup{
            table-number-alignment=center,
            table-figures-integer = 1,
            table-figures-decimal = 2,
            table-figures-uncertainty = 2,
            tight-spacing = false}
\begin{tabular}{llSSSSSSSS}
\toprule
        &                  & \multicolumn{4}{c}{$\RMSE$} \\
  Input & Model &  {L2/3e} & {L2/3i} & {L4e} & {L4i} \\
\cmidrule(rl){1-2} \cmidrule(lr){3-6}
Sine & \cc{True} -- micro &  1.39 \pm 0.03 &  3.46 \pm 0.10 &  3.47 \pm 0.13 &  4.59 \pm 0.13 \\
        & \cc{Theory} -- meso &  1.39 \pm 0.04 &  3.37 \pm 0.10 &  3.76 \pm 0.15 &  4.51 \pm 0.13 \\
        & \cc{MAP} -- meso &  1.39 \pm 0.03 &  3.49 \pm 0.09 &  3.46 \pm 0.13 &  4.53 \pm 0.13 \\
OU & \cc{True} -- micro &  1.22 \pm 0.03 &  3.14 \pm 0.08 &  2.26 \pm 0.07 &  5.13 \pm 0.15 \\
        & \cc{Theory} -- meso &  1.21 \pm 0.03 &  3.06 \pm 0.09 &  2.26 \pm 0.07 &  4.95 \pm 0.15 \\
        & \cc{MAP} -- meso &  1.22 \pm 0.03 &  3.11 \pm 0.08 &  2.25 \pm 0.06 &  5.30 \pm 0.14 \\
Impulse & \cc{True} -- micro &  1.54 \pm 0.05 &  3.64 \pm 0.11 &  5.32 \pm 0.46 &  5.11 \pm 0.23 \\
        & \cc{Theory} -- meso &  1.59 \pm 0.06 &  3.63 \pm 0.11 &  7.99 \pm 0.66 &  5.88 \pm 0.36 \\
        & \cc{MAP} -- meso &  1.70 \pm 0.07 &  3.96 \pm 0.13 &  7.74 \pm 0.59 &  6.00 \pm 0.36 \\
\bottomrule
\end{tabular}
\end{inlinetable}
\begin{inlinetable}
  \centering
  \caption{
    {Performance of four population models -- Trial-averaged correlation (\cref{eq:corrbar-def}). Measures computed from \num{60}~realizations of each model.}
    \label{tbl:performance-4pop-rho}
    }
   \sisetup{
            table-number-alignment=center,
            table-figures-integer = 1,
            table-figures-decimal = 3,
            table-figures-uncertainty = 0,
            }
  \begin{tabular}{llSSSSSSSS}
  \toprule
          &                  & \multicolumn{4}{c}{$\bar{ρ}$} \\
  Input & Model &  {L2/3e} & {L2/3i} & {L4e} & {L4i} \\
  \cmidrule(rl){1-2} \cmidrule(lr){3-6}
  Sine & \cc{True} -- micro &     0.418 &   0.386 & 0.994 & 0.948 \\
        & \cc{Theory} -- meso &     0.354 &   0.348 & 0.991 & 0.945 \\
        & \cc{MAP} -- meso &     0.352 &   0.455 & 0.994 & 0.951 \\
OU & \cc{True} -- micro &     0.829 &   0.694 & 0.977 & 0.905 \\
        & \cc{Theory} -- meso &     0.815 &   0.717 & 0.978 & 0.914 \\
        & \cc{MAP} -- meso &     0.855 &   0.756 & 0.977 & 0.916 \\
Impulse & \cc{True} -- micro &     0.914 &   0.879 & 0.996 & 0.927 \\
        & \cc{Theory} -- meso &     0.880 &   0.858 & 0.979 & 0.870 \\
        & \cc{MAP} -- meso &     0.912 &   0.896 & 0.988 & 0.887 \\
  \bottomrule
  \end{tabular}
\end{inlinetable}
\FloatBarrier
\pagebreak[4]
\section{Posterior for the 2 population mesoscopic model}
\begin{inlinefigure}[!ht]
  \includegraphics[width=\textwidth,height=0.85\textheight,keepaspectratio=True]{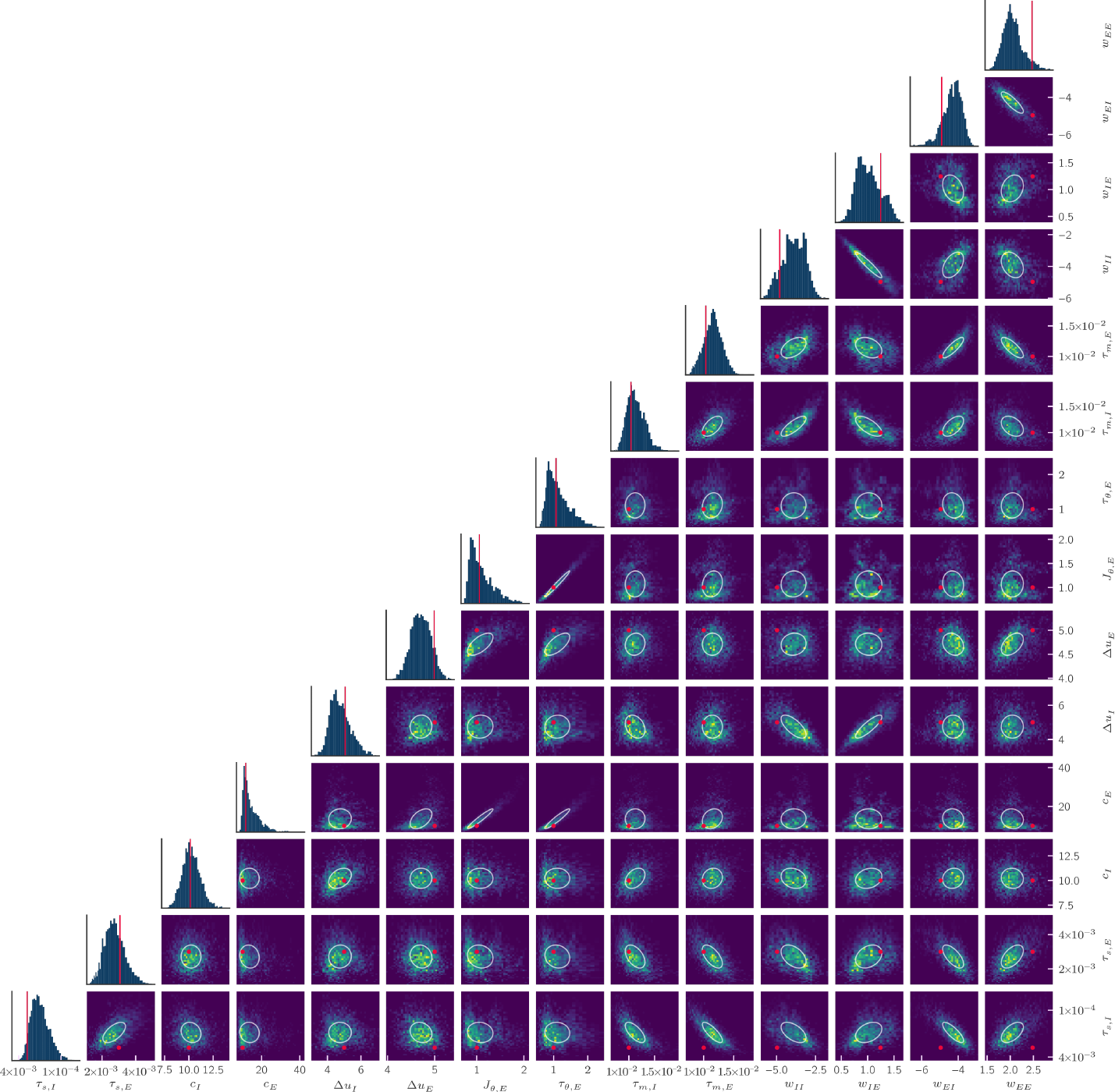}
  \caption{Full posterior for the two population mesoscopic model. Red point indicates the true values and ellipses trace the two standard-deviation isoline assuming a Gaussian model. Many parameter pairs show noticeable correlation, such as $τ_{θ,E}$ and $J_{θ,E}$, or $w_{IE}$ and ${Δ_{\mathrm{u},I}}$.
  \label{fig:full-posterior}
  }
\end{inlinefigure}

\FloatBarrier

\pagebreak[4]
\section{Fit dynamics}
\label{sec:fit-dynamics-4pop}

When fitting to data produced with a homogeneous microscopic model, inferred parameters are consistent with those predicted by the mesoscopic theory.

\begin{inlinefigure}[H]
  \includegraphics{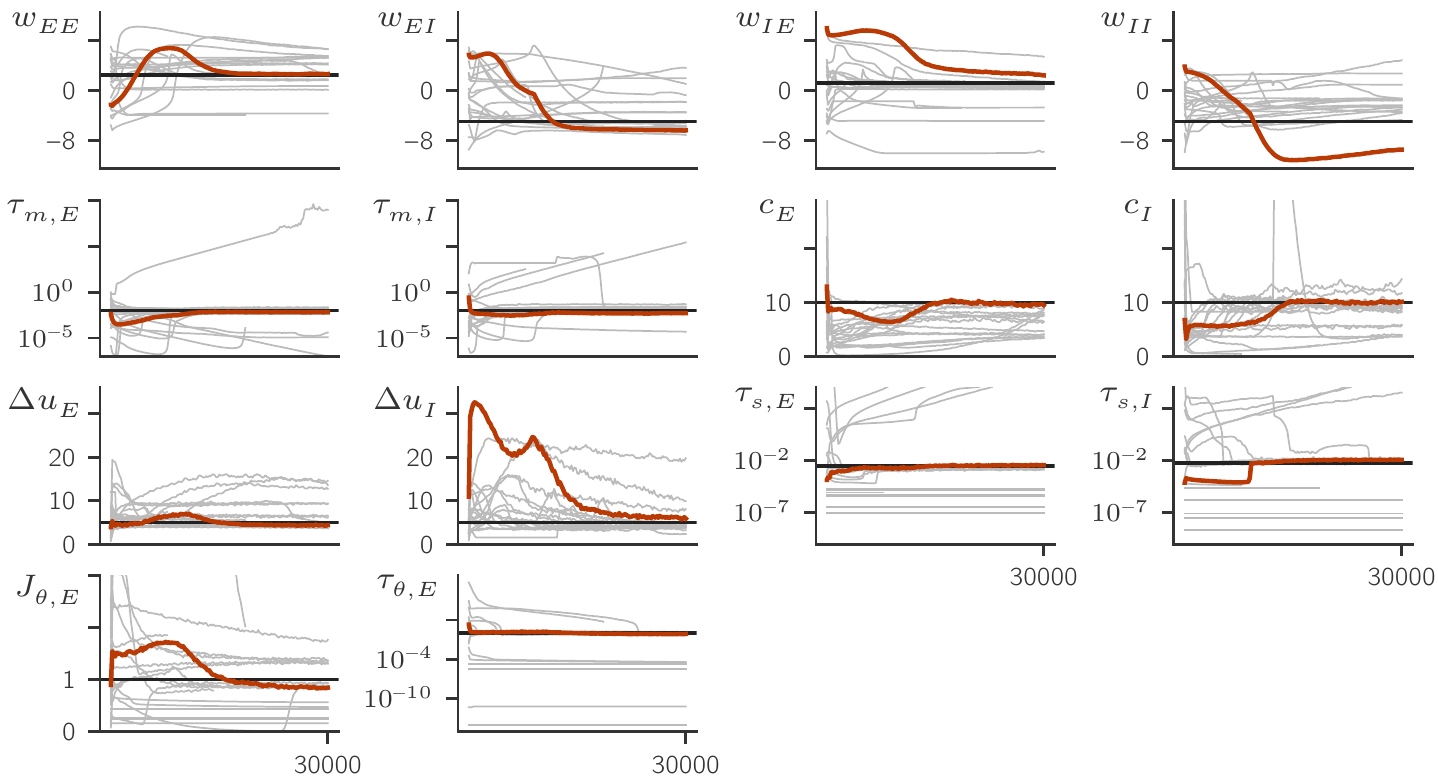}
  \caption{\captitle{Fit dynamics for the two population model.} The \SI{25}{fits} used to infer parameters for the two population model in \cref{sec:recovering-parameters}. Fits in red are those that resulted in a likelihood within \num{5} orders of magnitude of the maximum, with brighter red indicating closer to the maximum. A total of \SI{14}{parameters} were inferred; black lines indicate theoretical values.
  \label{fig:2pop-fits}
  }
\end{inlinefigure}

\pagebreak[3]
\begin{inlinefigure}[!ht]
  \includegraphics[width=\maxwidth,height=\maxheight,keepaspectratio=true]{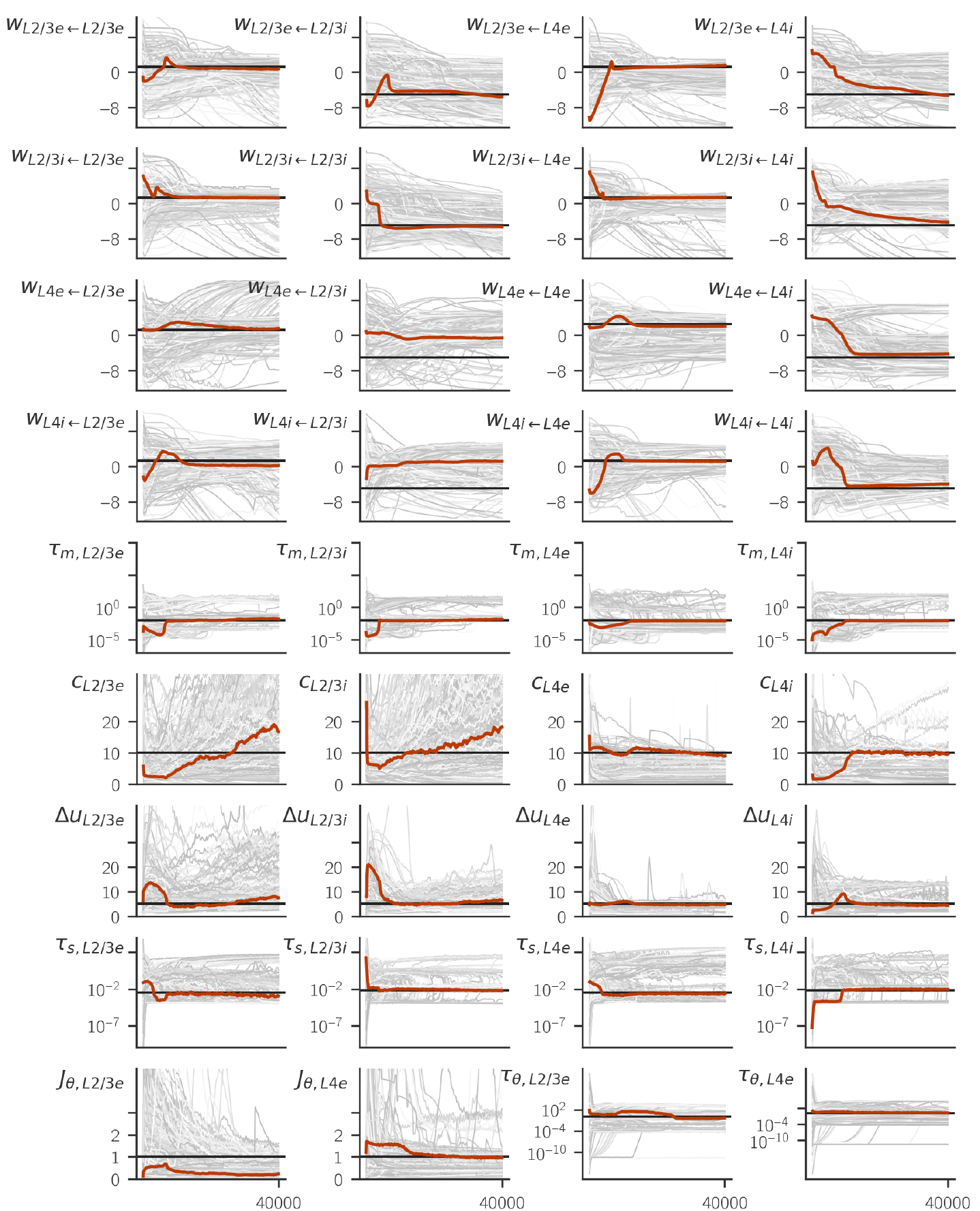}
  \caption{\captitle{Fit dynamics for the four population model.}
  The \SI{622}{fits} used to infer parameters for the four population model in \cref{sec:4pop-fit}. Although certain parameters would benefit from more iterations (e.g.\ $c$), most have converged within \num[scientific-notation=true]{4e4} iterations. A total of \SI{36}{parameters} were inferred; black lines indicate theoretical values.
  \label{fig:4pop-fits}
  }
\end{inlinefigure}

\FloatBarrier

\section{Self-consistent equation for the mesoscopic stationary state}
\label{sec:stationary-state-derivation}
We derive the stationary state for the case where \(I_{ext} \equiv 0\). For analytical tractability, we assume that finite-size fluctuations are negligible (effectively, that $N_α$ is very large), such that in the stationary state the activity is constant. We denote this activity $A^{*}$.

Having no fluctuations means that expected and observed spikes are interchangeable and equal to a constant;
\begin{equation}
  n_α^{(k)} = \bar{n}_α^{(k)} = A^* \, N_α \, Δt \,.
\end{equation}
This means that the number of spikes never overshoots nor undershoots $\bar{n}_α$, and the correction factor $P_Λ$ is zero. Equivalently, we can state that
\begin{equation}
\label{eq:conservation-condition}
   N = \sum_{i=0}^{K-1} \bar{m}_i + x \,.
\end{equation}

Substituting the stationary values $A^*, h^*,\dotsc$ into the equations of the previous \lcnamecref{sec:meso-equations}, we obtain equations for the stationary state. For instance,
\begin{align}
  h^{*} &= u_{rest,α} + τ_m \sum_{β=1}^M p_{αβ}N_{β}w_{αβ}A_{αβ}^{*} \,, \\
  y_{αβ}^{*} &= A_{αβ}^{*} \,, \\
  &\;\,\vdots \notag
\end{align}
and so on. Combining these with \cref{eq:conservation-condition}, we obtain a self-consistency relation,
\begin{multline}
   \label{eq:self-consistent-equation}
   1 = A_{α}^{*} Δt \left\{
           {k_{\mathrm{ref}, α}} + 1 + \sum_{i={k_{\mathrm{ref}, α}}}^{K-1} \exp \left[ - \sum_{j={k_{\mathrm{ref}, α}}+1}^{i-1} f(a_{αj} + b_{αj}^{β} A_{β}^{*} - c_{αj} A_{α}^{*}) Δt \right] \right. \\
           \left. + \frac {\exp \left[ - \sum_{j={k_{\mathrm{ref}, α}}}^{K-1} f(a_{αj} + b_{αj}^{β} A_{β}^{*} - c_{αj} A_α^*) Δt \right]}
                   {1 - \exp \left[ - f({a'}_α + {b'}_α^β - {c'}_α A_α^*) Δt \right]}
       \right\} \,,
\end{multline}
\noindent where ${k_{\mathrm{ref}, α}}$ is the number of bins corresponding to the absolute refractory period of that population. The terms therein are given by
\begin{align}
   a_{αj} &=
      e^{-(j-{k_{\mathrm{ref}, α}}+1) Δt / τ_{m,α}} (u_{r,α} - u_{rest,α}) + u_{rest,α} - u_{th,α} - θ_{αj} \,, \\
   b_{αj}^β &=
      (1 - e^{-(j-{k_{\mathrm{ref}, α}}+1)Δt/τ_{m,α}}) \frac{1 - e^{-Δt/τ_{m,β}}}{1 - e^{-Δt/τ_{m,α}}}τ_m p_{α}^{β} N^{β} w_{α}^{β} \,, \\
   c_{αj} &=
      J_{θ,α} e^{-T/τ_{θ,α}} + Δt \sum_{j'=j+1}^K \tilde{θ}_{αj'} \,, \\
   a_{α}' &=
       u_{rest,α} - u_{th,α} \,, \\
   b_{α}^{\prime β} &=
       (1 - e^{-Δt/τ_m}) τ_m p_{α}^{β} N^{β} w_{α}^{β} \,, \\
   c_{α}' &=
       J_{θ,α} e^{-T/τ_{θ,α}}  \,;
\end{align}
\noindent and the inverse link function $f$ and the kernels $θ$ and $\tilde{θ}$ are as in \cref{sec:meso-equations}.

\Cref{eq:self-consistent-equation} can be solved numerically for $A^*$, after which the other state variables are easily calculated from the expressions in \cref{sec:meso-equations}. We used SciPy's \citep{jonesSciPyOpenSource2001} \texttt{root} function with an initial guess of $A^*_α = 1$ to solve for $A^*$. Since the stationary initialization was ultimately only used this to validate \cref{alg:init-scheme-silent} (c.f.\ \cref{sec:alternative-initialization}), we did no further analysis of \cref{eq:self-consistent-equation}, and in particular leave the determination of conditions for which its solutions are unique to future work.


\begin{thebibliography}{}

\bibitem [\protect \citeauthoryear {%
Augustin%
, Ladenbauer%
, Baumann%
\BCBL {}\ \BBA {} Obermayer%
}{%
Augustin%
\ \protect \BOthers {.}}{%
{\protect \APACyear {2017}}%
}]{%
augustinLowdimensionalSpikeRate2017}
\APACinsertmetastar {%
augustinLowdimensionalSpikeRate2017}%
\begin{APACrefauthors}%
Augustin, M.%
, Ladenbauer, J.%
, Baumann, F.%
\BCBL {}\ \BBA {} Obermayer, K.%
\end{APACrefauthors}%
\unskip\
\newblock
\APACrefYearMonthDay{2017}{}{}.
\newblock
{\BBOQ}\APACrefatitle {Low-Dimensional Spike Rate Models Derived from Networks
  of Adaptive Integrate-and-Fire Neurons: {{Comparison}} and Implementation}
  {Low-dimensional spike rate models derived from networks of adaptive
  integrate-and-fire neurons: {{Comparison}} and implementation}.{\BBCQ}
\newblock
\APACjournalVolNumPages{PLOS Computational Biology}{13}{6}{e1005545}.
\PrintBackRefs{\CurrentBib}

\bibitem [\protect \citeauthoryear {%
Barak%
}{%
Barak%
}{%
{\protect \APACyear {2017}}%
}]{%
barakRecurrentNeuralNetworks2017}
\APACinsertmetastar {%
barakRecurrentNeuralNetworks2017}%
\begin{APACrefauthors}%
Barak, O.%
\end{APACrefauthors}%
\unskip\
\newblock
\APACrefYearMonthDay{2017}{}{}.
\newblock
{\BBOQ}\APACrefatitle {Recurrent Neural Networks as Versatile Tools of
  Neuroscience Research} {Recurrent neural networks as versatile tools of
  neuroscience research}.{\BBCQ}
\newblock
\APACjournalVolNumPages{Current Opinion in Neurobiology}{46}{}{1-6}.
\PrintBackRefs{\CurrentBib}

\bibitem [\protect \citeauthoryear {%
Betancourt%
\ \BBA {} Girolami%
}{%
Betancourt%
\ \BBA {} Girolami%
}{%
{\protect \APACyear {2013}}%
}]{%
betancourtHamiltonianMonteCarlo2013}
\APACinsertmetastar {%
betancourtHamiltonianMonteCarlo2013}%
\begin{APACrefauthors}%
Betancourt, M\BPBI J.%
\BCBT {}\ \BBA {} Girolami, M.%
\end{APACrefauthors}%
\unskip\
\newblock
\APACrefYearMonthDay{2013}{}{}.
\newblock
{\BBOQ}\APACrefatitle {Hamiltonian {{Monte Carlo}} for {{hierarchical models}}}
  {Hamiltonian {{Monte Carlo}} for {{hierarchical models}}}.{\BBCQ}
\newblock
\APACjournalVolNumPages{arXiv:1312.0906 [stat]}{}{}{}.
\PrintBackRefs{\CurrentBib}

\bibitem [\protect \citeauthoryear {%
Chizhov%
\ \BBA {} Graham%
}{%
Chizhov%
\ \BBA {} Graham%
}{%
{\protect \APACyear {2008}}%
}]{%
chizhovEfficientEvaluationNeuron2008}
\APACinsertmetastar {%
chizhovEfficientEvaluationNeuron2008}%
\begin{APACrefauthors}%
Chizhov, A\BPBI V.%
\BCBT {}\ \BBA {} Graham, L\BPBI J.%
\end{APACrefauthors}%
\unskip\
\newblock
\APACrefYearMonthDay{2008}{{\APACmonth{01}}}{}.
\newblock
{\BBOQ}\APACrefatitle {Efficient Evaluation of Neuron Populations Receiving
  Colored-Noise Current Based on a Refractory Density Method} {Efficient
  evaluation of neuron populations receiving colored-noise current based on a
  refractory density method}.{\BBCQ}
\newblock
\APACjournalVolNumPages{Physical Review E}{77}{1}{011910}.
\PrintBackRefs{\CurrentBib}

\bibitem [\protect \citeauthoryear {%
Cunningham%
\ \BBA {} Yu%
}{%
Cunningham%
\ \BBA {} Yu%
}{%
{\protect \APACyear {2014}}%
}]{%
cunninghamDimensionalityReductionLargescale2014}
\APACinsertmetastar {%
cunninghamDimensionalityReductionLargescale2014}%
\begin{APACrefauthors}%
Cunningham, J\BPBI P.%
\BCBT {}\ \BBA {} Yu, B\BPBI M.%
\end{APACrefauthors}%
\unskip\
\newblock
\APACrefYearMonthDay{2014}{}{}.
\newblock
{\BBOQ}\APACrefatitle {Dimensionality Reduction for Large-Scale Neural
  Recordings} {Dimensionality reduction for large-scale neural
  recordings}.{\BBCQ}
\newblock
\APACjournalVolNumPages{Nature Neuroscience}{17}{11}{1500-1509}.
\PrintBackRefs{\CurrentBib}

\bibitem [\protect \citeauthoryear {%
Doiron%
, {Litwin-Kumar}%
, Rosenbaum%
, Ocker%
\BCBL {}\ \BBA {} Josic%
}{%
Doiron%
\ \protect \BOthers {.}}{%
{\protect \APACyear {2016}}%
}]{%
doironMechanicsStatedependentNeural2016}
\APACinsertmetastar {%
doironMechanicsStatedependentNeural2016}%
\begin{APACrefauthors}%
Doiron, B.%
, {Litwin-Kumar}, A.%
, Rosenbaum, R.%
, Ocker, G\BPBI K.%
\BCBL {}\ \BBA {} Josic, K.%
\end{APACrefauthors}%
\unskip\
\newblock
\APACrefYearMonthDay{2016}{}{}.
\newblock
{\BBOQ}\APACrefatitle {The Mechanics of State-Dependent Neural Correlations}
  {The mechanics of state-dependent neural correlations}.{\BBCQ}
\newblock
\APACjournalVolNumPages{Nature Neuroscience}{19}{3}{383-393}.
\PrintBackRefs{\CurrentBib}

\bibitem [\protect \citeauthoryear {%
Dumont%
, Payeur%
\BCBL {}\ \BBA {} Longtin%
}{%
Dumont%
\ \protect \BOthers {.}}{%
{\protect \APACyear {2017}}%
}]{%
dumontStochasticfieldDescriptionFinitesize2017}
\APACinsertmetastar {%
dumontStochasticfieldDescriptionFinitesize2017}%
\begin{APACrefauthors}%
Dumont, G.%
, Payeur, A.%
\BCBL {}\ \BBA {} Longtin, A.%
\end{APACrefauthors}%
\unskip\
\newblock
\APACrefYearMonthDay{2017}{}{}.
\newblock
{\BBOQ}\APACrefatitle {A Stochastic-Field Description of Finite-Size Spiking
  Neural Networks} {A stochastic-field description of finite-size spiking
  neural networks}.{\BBCQ}
\newblock
\APACjournalVolNumPages{PLOS Computational Biology}{13}{8}{e1005691}.
\PrintBackRefs{\CurrentBib}

\bibitem [\protect \citeauthoryear {%
Gelman%
\ \protect \BOthers {.}}{%
Gelman%
\ \protect \BOthers {.}}{%
{\protect \APACyear {2014}}%
}]{%
gelmanBayesianDataAnalysis2014}
\APACinsertmetastar {%
gelmanBayesianDataAnalysis2014}%
\begin{APACrefauthors}%
Gelman, A.%
, Carlin, J\BPBI B.%
, Stern, H\BPBI S.%
, Dunson, D\BPBI B.%
, Vehtari, A.%
\BCBL {}\ \BBA {} Rubin, D\BPBI B.%
\end{APACrefauthors}%
\unskip\
\newblock
\APACrefYear{2014}.
\newblock
\APACrefbtitle {Bayesian Data Analysis} {Bayesian data analysis}.
\newblock
\APACaddressPublisher{{Boca Raton}}{{CRC Press}}.
\PrintBackRefs{\CurrentBib}

\bibitem [\protect \citeauthoryear {%
Gerstner%
}{%
Gerstner%
}{%
{\protect \APACyear {2000}}%
}]{%
gerstnerPopulationDynamicsSpiking2000}
\APACinsertmetastar {%
gerstnerPopulationDynamicsSpiking2000}%
\begin{APACrefauthors}%
Gerstner, W.%
\end{APACrefauthors}%
\unskip\
\newblock
\APACrefYearMonthDay{2000}{}{}.
\newblock
{\BBOQ}\APACrefatitle {Population {{dynamics}} of {{spiking Neurons}}: {{Fast
  transients}}, {{asynchronous states}}, and {{Locking}}} {Population
  {{dynamics}} of {{spiking Neurons}}: {{Fast transients}}, {{asynchronous
  states}}, and {{Locking}}}.{\BBCQ}
\newblock
\APACjournalVolNumPages{Neural Computation}{12}{1}{43-89}.
\PrintBackRefs{\CurrentBib}

\bibitem [\protect \citeauthoryear {%
Gerstner%
, Paninski%
, Naud%
\BCBL {}\ \BBA {} Kistler%
}{%
Gerstner%
\ \protect \BOthers {.}}{%
{\protect \APACyear {2014}}%
}]{%
gerstnerNeuronalDynamicsSingle2014}
\APACinsertmetastar {%
gerstnerNeuronalDynamicsSingle2014}%
\begin{APACrefauthors}%
Gerstner, W.%
, Paninski, L.%
, Naud, R.%
\BCBL {}\ \BBA {} Kistler, W\BPBI M.%
\end{APACrefauthors}%
\unskip\
\newblock
\APACrefYear{2014}.
\newblock
\APACrefbtitle {Neuronal Dynamics from Single Neurons to Networks and Models of
  Cognition} {Neuronal dynamics from single neurons to networks and models of
  cognition}.
\newblock
\APACaddressPublisher{{Cambridge}}{{Cambridge University Press}}.
\newblock
\APACrefnote{OCLC: 945459025}
\PrintBackRefs{\CurrentBib}

\bibitem [\protect \citeauthoryear {%
Girolami%
\ \BBA {} Calderhead%
}{%
Girolami%
\ \BBA {} Calderhead%
}{%
{\protect \APACyear {2011}}%
}]{%
girolamiRiemannManifoldLangevin2011}
\APACinsertmetastar {%
girolamiRiemannManifoldLangevin2011}%
\begin{APACrefauthors}%
Girolami, M.%
\BCBT {}\ \BBA {} Calderhead, B.%
\end{APACrefauthors}%
\unskip\
\newblock
\APACrefYearMonthDay{2011}{}{}.
\newblock
{\BBOQ}\APACrefatitle {Riemann Manifold {{Langevin}} and {{Hamiltonian Monte
  Carlo}} Methods} {Riemann manifold {{Langevin}} and {{Hamiltonian Monte
  Carlo}} methods}.{\BBCQ}
\newblock
\APACjournalVolNumPages{Journal of the Royal Statistical Society: Series B
  (Statistical Methodology)}{73}{2}{123-214}.
\PrintBackRefs{\CurrentBib}

\bibitem [\protect \citeauthoryear {%
Goldwyn%
\ \BBA {} {Shea-Brown}%
}{%
Goldwyn%
\ \BBA {} {Shea-Brown}%
}{%
{\protect \APACyear {2011}}%
}]{%
goldwynWhatWhereAdding2011}
\APACinsertmetastar {%
goldwynWhatWhereAdding2011}%
\begin{APACrefauthors}%
Goldwyn, J\BPBI H.%
\BCBT {}\ \BBA {} {Shea-Brown}, E.%
\end{APACrefauthors}%
\unskip\
\newblock
\APACrefYearMonthDay{2011}{}{}.
\newblock
{\BBOQ}\APACrefatitle {The {{what}} and {{where}} of {{adding channel noise}}
  to the {{Hodgkin}}-{{Huxley equations}}} {The {{what}} and {{where}} of
  {{adding channel noise}} to the {{Hodgkin}}-{{Huxley equations}}}.{\BBCQ}
\newblock
\APACjournalVolNumPages{PLOS Computational Biology}{7}{11}{e1002247}.
\PrintBackRefs{\CurrentBib}

\bibitem [\protect \citeauthoryear {%
Greenberg%
, Nonnenmacher%
\BCBL {}\ \BBA {} Macke%
}{%
Greenberg%
\ \protect \BOthers {.}}{%
{\protect \APACyear {2019}}%
}]{%
greenbergAutomaticPosteriorTransformation2019}
\APACinsertmetastar {%
greenbergAutomaticPosteriorTransformation2019}%
\begin{APACrefauthors}%
Greenberg, D.%
, Nonnenmacher, M.%
\BCBL {}\ \BBA {} Macke, J.%
\end{APACrefauthors}%
\unskip\
\newblock
\APACrefYearMonthDay{2019}{}{}.
\newblock
{\BBOQ}\APACrefatitle {Automatic posterior transformation for likelihood-free
  inference} {Automatic posterior transformation for likelihood-free
  inference}.{\BBCQ}
\newblock
\BIn{} \APACrefbtitle {International {{Conference}} on {{Machine Learning}}}
  {International {{Conference}} on {{Machine Learning}}}\ (\BPG~2404-2414).
\PrintBackRefs{\CurrentBib}

\bibitem [\protect \citeauthoryear {%
Haviv%
, Rivkind%
\BCBL {}\ \BBA {} Barak%
}{%
Haviv%
\ \protect \BOthers {.}}{%
{\protect \APACyear {2019}}%
}]{%
havivUnderstandingControllingMemory2019}
\APACinsertmetastar {%
havivUnderstandingControllingMemory2019}%
\begin{APACrefauthors}%
Haviv, D.%
, Rivkind, A.%
\BCBL {}\ \BBA {} Barak, O.%
\end{APACrefauthors}%
\unskip\
\newblock
\APACrefYearMonthDay{2019}{}{}.
\newblock
{\BBOQ}\APACrefatitle {Understanding and controlling memory in recurrent neural
  networks} {Understanding and controlling memory in recurrent neural
  networks}.{\BBCQ}
\newblock
\APACjournalVolNumPages{arXiv:1902.07275 [cs, stat]}{}{}{}.
\PrintBackRefs{\CurrentBib}

\bibitem [\protect \citeauthoryear {%
Hawrylycz%
\ \protect \BOthers {.}}{%
Hawrylycz%
\ \protect \BOthers {.}}{%
{\protect \APACyear {2016}}%
}]{%
hawrylyczInferringCorticalFunction2016}
\APACinsertmetastar {%
hawrylyczInferringCorticalFunction2016}%
\begin{APACrefauthors}%
Hawrylycz, M.%
, Anastassiou, C.%
, Arkhipov, A.%
, Berg, J.%
, Buice, M.%
, Cain, N.%
\BDBL {}{MindScope}%
\end{APACrefauthors}%
\unskip\
\newblock
\APACrefYearMonthDay{2016}{}{}.
\newblock
{\BBOQ}\APACrefatitle {Inferring Cortical Function in the Mouse Visual System
  through Large-Scale Systems Neuroscience} {Inferring cortical function in the
  mouse visual system through large-scale systems neuroscience}.{\BBCQ}
\newblock
\APACjournalVolNumPages{Proceedings of the National Academy of
  Sciences}{113}{27}{7337-7344}.
\PrintBackRefs{\CurrentBib}

\bibitem [\protect \citeauthoryear {%
Hoffman%
\ \BBA {} Gelman%
}{%
Hoffman%
\ \BBA {} Gelman%
}{%
{\protect \APACyear {2014}}%
}]{%
hoffmanNoUturnSamplerAdaptively2014}
\APACinsertmetastar {%
hoffmanNoUturnSamplerAdaptively2014}%
\begin{APACrefauthors}%
Hoffman, M\BPBI D.%
\BCBT {}\ \BBA {} Gelman, A.%
\end{APACrefauthors}%
\unskip\
\newblock
\APACrefYearMonthDay{2014}{}{}.
\newblock
{\BBOQ}\APACrefatitle {The {{No}}-{{U}}-Turn Sampler: Adaptively Setting Path
  Lengths in {{Hamiltonian Monte Carlo}}.} {The {{No}}-{{U}}-turn sampler:
  Adaptively setting path lengths in {{Hamiltonian Monte Carlo}}.}{\BBCQ}
\newblock
\APACjournalVolNumPages{Journal of Machine Learning
  Research}{15}{1}{1593--1623}.
\PrintBackRefs{\CurrentBib}

\bibitem [\protect \citeauthoryear {%
Horsthemke%
\ \BBA {} Lefever%
}{%
Horsthemke%
\ \BBA {} Lefever%
}{%
{\protect \APACyear {2006}}%
}]{%
horsthemkeNoiseinducedTransitionsTheory2006}
\APACinsertmetastar {%
horsthemkeNoiseinducedTransitionsTheory2006}%
\begin{APACrefauthors}%
Horsthemke, W.%
\BCBT {}\ \BBA {} Lefever, R.%
\end{APACrefauthors}%
\unskip\
\newblock
\APACrefYear{2006}.
\newblock
\APACrefbtitle {Noise-Induced Transitions: Theory and Applications in Physics,
  Chemistry, and Biology} {Noise-induced transitions: Theory and applications
  in physics, chemistry, and biology}\ (\PrintOrdinal{2. print}\ \BEd)\
  (\BNUM~15).
\newblock
\APACaddressPublisher{{Berlin}}{{Springer}}.
\newblock
\APACrefnote{OCLC: 255634759}
\PrintBackRefs{\CurrentBib}

\bibitem [\protect \citeauthoryear {%
{Ian Goodfellow}%
, {Yoshua Bengio}%
\BCBL {}\ \BBA {} {Aaron Courville}%
}{%
{Ian Goodfellow}%
\ \protect \BOthers {.}}{%
{\protect \APACyear {2016}}%
}]{%
iangoodfellowDeepLearning2016}
\APACinsertmetastar {%
iangoodfellowDeepLearning2016}%
\begin{APACrefauthors}%
{Ian Goodfellow}%
, {Yoshua Bengio}%
\BCBL {}\ \BBA {} {Aaron Courville}.%
\end{APACrefauthors}%
\unskip\
\newblock
\APACrefYear{2016}.
\newblock
\APACrefbtitle {Deep {{Learning}}} {Deep {{Learning}}}.
\newblock
\APACaddressPublisher{}{{MIT Press}}.
\PrintBackRefs{\CurrentBib}

\bibitem [\protect \citeauthoryear {%
Iolov%
, Ditlevsen%
\BCBL {}\ \BBA {} Longtin%
}{%
Iolov%
\ \protect \BOthers {.}}{%
{\protect \APACyear {2017}}%
}]{%
iolovOptimalDesignEstimation2017}
\APACinsertmetastar {%
iolovOptimalDesignEstimation2017}%
\begin{APACrefauthors}%
Iolov, A.%
, Ditlevsen, S.%
\BCBL {}\ \BBA {} Longtin, A.%
\end{APACrefauthors}%
\unskip\
\newblock
\APACrefYearMonthDay{2017}{}{}.
\newblock
{\BBOQ}\APACrefatitle {Optimal {{design}} for {{estimation}} in {{diffusion
  processes}} from {{first hitting times}}} {Optimal {{design}} for
  {{estimation}} in {{diffusion processes}} from {{first hitting
  times}}}.{\BBCQ}
\newblock
\APACjournalVolNumPages{SIAM/ASA Journal on Uncertainty
  Quantification}{5}{}{88-110}.
\PrintBackRefs{\CurrentBib}

\bibitem [\protect \citeauthoryear {%
Jones%
, Oliphant%
, {Pearu Peterson}%
\BCBL {}\ \protect \BOthers {.}}{%
Jones%
\ \protect \BOthers {.}}{%
{\protect \APACyear {2001\textendash{}}}%
}]{%
jonesSciPyOpenSource2001}
\APACinsertmetastar {%
jonesSciPyOpenSource2001}%
\begin{APACrefauthors}%
Jones, E.%
, Oliphant, T.%
, {Pearu Peterson}%
\BCBL {}\ \BOthersPeriod {.}\end{APACrefauthors}%
\unskip\
\newblock
\APACrefYearMonthDay{2001\textendash{}}{}{}.
\newblock
\APACrefbtitle {{{SciPy}}: {{Open}} Source Scientific Tools for {{Python}}.}
  {{{SciPy}}: {{Open}} source scientific tools for {{Python}}.}
\newblock
\APACrefnote{[Online; accessed 2019-06-03]}
\PrintBackRefs{\CurrentBib}

\bibitem [\protect \citeauthoryear {%
Kingma%
\ \BBA {} Ba%
}{%
Kingma%
\ \BBA {} Ba%
}{%
{\protect \APACyear {2014}}%
}]{%
kingmaAdamMethodStochastic2014}
\APACinsertmetastar {%
kingmaAdamMethodStochastic2014}%
\begin{APACrefauthors}%
Kingma, D\BPBI P.%
\BCBT {}\ \BBA {} Ba, J.%
\end{APACrefauthors}%
\unskip\
\newblock
\APACrefYearMonthDay{2014}{}{}.
\newblock
{\BBOQ}\APACrefatitle {Adam: {{A method}} for {{stochastic optimization}}}
  {Adam: {{A method}} for {{stochastic optimization}}}.{\BBCQ}
\newblock
\APACjournalVolNumPages{arXiv:1412.6980 [cs]}{}{}{}.
\PrintBackRefs{\CurrentBib}

\bibitem [\protect \citeauthoryear {%
Kucukelbir%
, Tran%
, Ranganath%
, Gelman%
\BCBL {}\ \BBA {} Blei%
}{%
Kucukelbir%
\ \protect \BOthers {.}}{%
{\protect \APACyear {2017}}%
}]{%
kucukelbirAutomaticDifferentiationVariational2017a}
\APACinsertmetastar {%
kucukelbirAutomaticDifferentiationVariational2017a}%
\begin{APACrefauthors}%
Kucukelbir, A.%
, Tran, D.%
, Ranganath, R.%
, Gelman, A.%
\BCBL {}\ \BBA {} Blei, D\BPBI M.%
\end{APACrefauthors}%
\unskip\
\newblock
\APACrefYearMonthDay{2017}{}{}.
\newblock
{\BBOQ}\APACrefatitle {Automatic {{differentiation variational inference}}}
  {Automatic {{differentiation variational inference}}}.{\BBCQ}
\newblock
\APACjournalVolNumPages{Journal of Machine Learning Research}{18}{14}{1-45}.
\PrintBackRefs{\CurrentBib}

\bibitem [\protect \citeauthoryear {%
Lueckmann%
\ \protect \BOthers {.}}{%
Lueckmann%
\ \protect \BOthers {.}}{%
{\protect \APACyear {2017}}%
}]{%
lueckmannFlexibleStatisticalInference2017}
\APACinsertmetastar {%
lueckmannFlexibleStatisticalInference2017}%
\begin{APACrefauthors}%
Lueckmann, J\BHBI M.%
, Goncalves, P\BPBI J.%
, Bassetto, G.%
, {\"O}cal, K.%
, Nonnenmacher, M.%
\BCBL {}\ \BBA {} Macke, J\BPBI H.%
\end{APACrefauthors}%
\unskip\
\newblock
\APACrefYearMonthDay{2017}{}{}.
\newblock
{\BBOQ}\APACrefatitle {Flexible Statistical Inference for Mechanistic Models of
  Neural Dynamics} {Flexible statistical inference for mechanistic models of
  neural dynamics}.{\BBCQ}
\newblock
\BIn{} I.~Guyon\ \BOthers {.}\ (\BEDS), \APACrefbtitle {Advances in {{Neural
  Information Processing Systems}} 30} {Advances in {{Neural Information
  Processing Systems}} 30}\ (\BPGS\ 1289--1299).
\newblock
\APACaddressPublisher{}{{Curran Associates, Inc.}}
\PrintBackRefs{\CurrentBib}

\bibitem [\protect \citeauthoryear {%
Macke%
\ \protect \BOthers {.}}{%
Macke%
\ \protect \BOthers {.}}{%
{\protect \APACyear {2011}}%
}]{%
mackeEmpiricalModelsSpiking2011}
\APACinsertmetastar {%
mackeEmpiricalModelsSpiking2011}%
\begin{APACrefauthors}%
Macke, J\BPBI H.%
, Buesing, L.%
, Cunningham, J\BPBI P.%
, Yu, B\BPBI M.%
, Shenoy, K\BPBI V.%
\BCBL {}\ \BBA {} Sahani, M.%
\end{APACrefauthors}%
\unskip\
\newblock
\APACrefYearMonthDay{2011}{}{}.
\newblock
{\BBOQ}\APACrefatitle {Empirical Models of Spiking in Neural Populations}
  {Empirical models of spiking in neural populations}.{\BBCQ}
\newblock
\BIn{} J.~{Shawe-Taylor}, R\BPBI S.~Zemel, P\BPBI L.~Bartlett, F.~Pereira\BCBL
  {}\ \BBA {} K\BPBI Q.~Weinberger\ (\BEDS), \APACrefbtitle {Advances in
  {{Neural Information Processing Systems}} 24} {Advances in {{Neural
  Information Processing Systems}} 24}\ (\BPGS\ 1350--1358).
\newblock
\APACaddressPublisher{}{{Curran Associates, Inc.}}
\PrintBackRefs{\CurrentBib}

\bibitem [\protect \citeauthoryear {%
Mart{\'i}%
, Brunel%
\BCBL {}\ \BBA {} Ostojic%
}{%
Mart{\'i}%
\ \protect \BOthers {.}}{%
{\protect \APACyear {2018}}%
}]{%
martiCorrelationsSynapsesPairs2018}
\APACinsertmetastar {%
martiCorrelationsSynapsesPairs2018}%
\begin{APACrefauthors}%
Mart{\'i}, D.%
, Brunel, N.%
\BCBL {}\ \BBA {} Ostojic, S.%
\end{APACrefauthors}%
\unskip\
\newblock
\APACrefYearMonthDay{2018}{}{}.
\newblock
{\BBOQ}\APACrefatitle {Correlations between Synapses in Pairs of Neurons Slow
  down Dynamics in Randomly Connected Neural Networks} {Correlations between
  synapses in pairs of neurons slow down dynamics in randomly connected neural
  networks}.{\BBCQ}
\newblock
\APACjournalVolNumPages{Physical Review E}{97}{6}{062314}.
\PrintBackRefs{\CurrentBib}

\bibitem [\protect \citeauthoryear {%
Mena%
\ \BBA {} Paninski%
}{%
Mena%
\ \BBA {} Paninski%
}{%
{\protect \APACyear {2014}}%
}]{%
menaQuadratureMethodsRefractory2014}
\APACinsertmetastar {%
menaQuadratureMethodsRefractory2014}%
\begin{APACrefauthors}%
Mena, G.%
\BCBT {}\ \BBA {} Paninski, L.%
\end{APACrefauthors}%
\unskip\
\newblock
\APACrefYearMonthDay{2014}{}{}.
\newblock
{\BBOQ}\APACrefatitle {On {{quadrature methods}} for {{refractory point process
  likelihoods}}} {On {{quadrature methods}} for {{refractory point process
  likelihoods}}}.{\BBCQ}
\newblock
\APACjournalVolNumPages{Neural Computation}{26}{12}{2790--2797}.
\PrintBackRefs{\CurrentBib}

\bibitem [\protect \citeauthoryear {%
Mensi%
\ \protect \BOthers {.}}{%
Mensi%
\ \protect \BOthers {.}}{%
{\protect \APACyear {2012}}%
}]{%
mensiParameterExtractionClassification2012}
\APACinsertmetastar {%
mensiParameterExtractionClassification2012}%
\begin{APACrefauthors}%
Mensi, S.%
, Naud, R.%
, Pozzorini, C.%
, Avermann, M.%
, Petersen, C\BPBI C\BPBI H.%
\BCBL {}\ \BBA {} Gerstner, W.%
\end{APACrefauthors}%
\unskip\
\newblock
\APACrefYearMonthDay{2012}{}{}.
\newblock
{\BBOQ}\APACrefatitle {Parameter Extraction and Classification of Three
  Cortical Neuron Types Reveals Two Distinct Adaptation Mechanisms} {Parameter
  extraction and classification of three cortical neuron types reveals two
  distinct adaptation mechanisms}.{\BBCQ}
\newblock
\APACjournalVolNumPages{Journal of Neurophysiology}{107}{6}{1756-1775}.
\PrintBackRefs{\CurrentBib}

\bibitem [\protect \citeauthoryear {%
Meyer%
, Williamson%
, Linden%
\BCBL {}\ \BBA {} Sahani%
}{%
Meyer%
\ \protect \BOthers {.}}{%
{\protect \APACyear {2017}}%
}]{%
meyerModelsNeuronalStimulusResponse2017}
\APACinsertmetastar {%
meyerModelsNeuronalStimulusResponse2017}%
\begin{APACrefauthors}%
Meyer, A\BPBI F.%
, Williamson, R\BPBI S.%
, Linden, J\BPBI F.%
\BCBL {}\ \BBA {} Sahani, M.%
\end{APACrefauthors}%
\unskip\
\newblock
\APACrefYearMonthDay{2017}{}{}.
\newblock
{\BBOQ}\APACrefatitle {Models of {{Neuronal Stimulus}}-{{Response Functions}}:
  {{Elaboration}}, {{Estimation}}, and {{Evaluation}}} {Models of {{Neuronal
  Stimulus}}-{{Response Functions}}: {{Elaboration}}, {{Estimation}}, and
  {{Evaluation}}}.{\BBCQ}
\newblock
\APACjournalVolNumPages{Frontiers in Systems Neuroscience}{10}{}{109}.
\PrintBackRefs{\CurrentBib}

\bibitem [\protect \citeauthoryear {%
Moral%
, Doucet%
\BCBL {}\ \BBA {} Jasra%
}{%
Moral%
\ \protect \BOthers {.}}{%
{\protect \APACyear {2006}}%
}]{%
moralSequentialMonteCarlo2006}
\APACinsertmetastar {%
moralSequentialMonteCarlo2006}%
\begin{APACrefauthors}%
Moral, P\BPBI D.%
, Doucet, A.%
\BCBL {}\ \BBA {} Jasra, A.%
\end{APACrefauthors}%
\unskip\
\newblock
\APACrefYearMonthDay{2006}{}{}.
\newblock
{\BBOQ}\APACrefatitle {Sequential {{Monte Carlo}} Samplers} {Sequential {{Monte
  Carlo}} samplers}.{\BBCQ}
\newblock
\APACjournalVolNumPages{Journal of the Royal Statistical Society: Series B
  (Statistical Methodology)}{68}{3}{411-436}.
\PrintBackRefs{\CurrentBib}

\bibitem [\protect \citeauthoryear {%
Naud%
\ \BBA {} Gerstner%
}{%
Naud%
\ \BBA {} Gerstner%
}{%
{\protect \APACyear {2012}}%
}]{%
naudCodingDecodingAdapting2012}
\APACinsertmetastar {%
naudCodingDecodingAdapting2012}%
\begin{APACrefauthors}%
Naud, R.%
\BCBT {}\ \BBA {} Gerstner, W.%
\end{APACrefauthors}%
\unskip\
\newblock
\APACrefYearMonthDay{2012}{}{}.
\newblock
{\BBOQ}\APACrefatitle {Coding and {{decoding}} with {{adapting neurons}}: {{A
  population approach}} to the {{Peri}}-{{Stimulus Time Histogram}}} {Coding
  and {{decoding}} with {{adapting neurons}}: {{A population approach}} to the
  {{Peri}}-{{Stimulus Time Histogram}}}.{\BBCQ}
\newblock
\APACjournalVolNumPages{PLOS Computational Biology}{8}{10}{e1002711}.
\PrintBackRefs{\CurrentBib}

\bibitem [\protect \citeauthoryear {%
Neal%
}{%
Neal%
}{%
{\protect \APACyear {2012}}%
}]{%
nealMCMCUsingHamiltonian2012}
\APACinsertmetastar {%
nealMCMCUsingHamiltonian2012}%
\begin{APACrefauthors}%
Neal, R\BPBI M.%
\end{APACrefauthors}%
\unskip\
\newblock
\APACrefYearMonthDay{2012}{}{}.
\newblock
{\BBOQ}\APACrefatitle {{{MCMC}} Using {{Hamiltonian}} Dynamics} {{{MCMC}} using
  {{Hamiltonian}} dynamics}.{\BBCQ}
\newblock
\APACjournalVolNumPages{arXiv:1206.1901 [physics, stat]}{}{}{}.
\PrintBackRefs{\CurrentBib}

\bibitem [\protect \citeauthoryear {%
Nykamp%
\ \BBA {} Tranchina%
}{%
Nykamp%
\ \BBA {} Tranchina%
}{%
{\protect \APACyear {2000}}%
}]{%
nykampPopulationDensityApproach2000}
\APACinsertmetastar {%
nykampPopulationDensityApproach2000}%
\begin{APACrefauthors}%
Nykamp, D\BPBI Q.%
\BCBT {}\ \BBA {} Tranchina, D.%
\end{APACrefauthors}%
\unskip\
\newblock
\APACrefYearMonthDay{2000}{}{}.
\newblock
{\BBOQ}\APACrefatitle {A {{population density approach that facilitates
  large}}-{{scale modeling}} of {{neural networks}}: {{analysis}} and an
  {{application}} to {{orientation tuning}}} {A {{population density approach
  that facilitates large}}-{{scale modeling}} of {{neural networks}}:
  {{analysis}} and an {{application}} to {{orientation tuning}}}.{\BBCQ}
\newblock
\APACjournalVolNumPages{Journal of Computational Neuroscience}{8}{1}{19-50}.
\PrintBackRefs{\CurrentBib}

\bibitem [\protect \citeauthoryear {%
Pandarinath%
\ \protect \BOthers {.}}{%
Pandarinath%
\ \protect \BOthers {.}}{%
{\protect \APACyear {2018}}%
}]{%
pandarinathInferringSingletrialNeural2018}
\APACinsertmetastar {%
pandarinathInferringSingletrialNeural2018}%
\begin{APACrefauthors}%
Pandarinath, C.%
, O'Shea, D\BPBI J.%
, Collins, J.%
, Jozefowicz, R.%
, Stavisky, S\BPBI D.%
, Kao, J\BPBI C.%
\BDBL {}Sussillo, D.%
\end{APACrefauthors}%
\unskip\
\newblock
\APACrefYearMonthDay{2018}{}{}.
\newblock
{\BBOQ}\APACrefatitle {Inferring Single-Trial Neural Population Dynamics Using
  Sequential Auto-Encoders} {Inferring single-trial neural population dynamics
  using sequential auto-encoders}.{\BBCQ}
\newblock
\APACjournalVolNumPages{Nature Methods}{15}{10}{805}.
\PrintBackRefs{\CurrentBib}

\bibitem [\protect \citeauthoryear {%
Paninski%
, Pillow%
\BCBL {}\ \BBA {} Simoncelli%
}{%
Paninski%
\ \protect \BOthers {.}}{%
{\protect \APACyear {2004}}%
}]{%
paninskiMaximumLikelihoodEstimation2004}
\APACinsertmetastar {%
paninskiMaximumLikelihoodEstimation2004}%
\begin{APACrefauthors}%
Paninski, L.%
, Pillow, J\BPBI W.%
\BCBL {}\ \BBA {} Simoncelli, E\BPBI P.%
\end{APACrefauthors}%
\unskip\
\newblock
\APACrefYearMonthDay{2004}{}{}.
\newblock
{\BBOQ}\APACrefatitle {Maximum {{likelihood estimation}} of a {{stochastic
  integrate}}-and-{{fire neural encoding model}}} {Maximum {{likelihood
  estimation}} of a {{stochastic integrate}}-and-{{fire neural encoding
  model}}}.{\BBCQ}
\newblock
\APACjournalVolNumPages{Neural Computation}{16}{12}{2533-2561}.
\PrintBackRefs{\CurrentBib}

\bibitem [\protect \citeauthoryear {%
Papamakarios%
\ \BBA {} Murray%
}{%
Papamakarios%
\ \BBA {} Murray%
}{%
{\protect \APACyear {2016}}%
}]{%
papamakariosFastEpsilonFree2016}
\APACinsertmetastar {%
papamakariosFastEpsilonFree2016}%
\begin{APACrefauthors}%
Papamakarios, G.%
\BCBT {}\ \BBA {} Murray, I.%
\end{APACrefauthors}%
\unskip\
\newblock
\APACrefYearMonthDay{2016}{}{}.
\newblock
{\BBOQ}\APACrefatitle {Fast $\epsilon$-free {{inference}} of {{simulation
  models}} with {{Bayesian conditional density estimation}}} {Fast
  $\epsilon$-free {{inference}} of {{simulation models}} with {{Bayesian
  conditional density estimation}}}.{\BBCQ}
\newblock
\BIn{} D\BPBI D.~Lee, M.~Sugiyama, U\BPBI V.~Luxburg, I.~Guyon\BCBL {}\ \BBA {}
  R.~Garnett\ (\BEDS), \APACrefbtitle {Advances in {{Neural Information
  Processing Systems}} 29} {Advances in {{Neural Information Processing
  Systems}} 29}\ (\BPGS\ 1028--1036).
\newblock
\APACaddressPublisher{}{{Curran Associates, Inc.}}
\PrintBackRefs{\CurrentBib}

\bibitem [\protect \citeauthoryear {%
Papamakarios%
, Sterratt%
\BCBL {}\ \BBA {} Murray%
}{%
Papamakarios%
\ \protect \BOthers {.}}{%
{\protect \APACyear {2018}}%
}]{%
papamakariosSequentialNeuralLikelihood2018}
\APACinsertmetastar {%
papamakariosSequentialNeuralLikelihood2018}%
\begin{APACrefauthors}%
Papamakarios, G.%
, Sterratt, D\BPBI C.%
\BCBL {}\ \BBA {} Murray, I.%
\end{APACrefauthors}%
\unskip\
\newblock
\APACrefYearMonthDay{2018}{}{}.
\newblock
{\BBOQ}\APACrefatitle {Sequential {{Neural likelihood}}: {{fast
  likelihood}}-free {{inference}} with {{autoregressive flows}}} {Sequential
  {{Neural likelihood}}: {{fast likelihood}}-free {{inference}} with
  {{autoregressive flows}}}.{\BBCQ}
\newblock
\APACjournalVolNumPages{arXiv:1805.07226 [cs, stat]}{}{}{}.
\PrintBackRefs{\CurrentBib}

\bibitem [\protect \citeauthoryear {%
Pillow%
\ \protect \BOthers {.}}{%
Pillow%
\ \protect \BOthers {.}}{%
{\protect \APACyear {2008}}%
}]{%
pillowSpatiotemporalCorrelationsVisual2008}
\APACinsertmetastar {%
pillowSpatiotemporalCorrelationsVisual2008}%
\begin{APACrefauthors}%
Pillow, J\BPBI W.%
, Shlens, J.%
, Paninski, L.%
, Sher, A.%
, Litke, A\BPBI M.%
, Chichilnisky, E\BPBI J.%
\BCBL {}\ \BBA {} Simoncelli, E\BPBI P.%
\end{APACrefauthors}%
\unskip\
\newblock
\APACrefYearMonthDay{2008}{}{}.
\newblock
{\BBOQ}\APACrefatitle {Spatio-Temporal Correlations and Visual Signalling in a
  Complete Neuronal Population} {Spatio-temporal correlations and visual
  signalling in a complete neuronal population}.{\BBCQ}
\newblock
\APACjournalVolNumPages{Nature}{454}{7207}{995-999}.
\PrintBackRefs{\CurrentBib}

\bibitem [\protect \citeauthoryear {%
Potjans%
\ \BBA {} Diesmann%
}{%
Potjans%
\ \BBA {} Diesmann%
}{%
{\protect \APACyear {2014}}%
}]{%
potjansCelltypeSpecificCortical2014}
\APACinsertmetastar {%
potjansCelltypeSpecificCortical2014}%
\begin{APACrefauthors}%
Potjans, T\BPBI C.%
\BCBT {}\ \BBA {} Diesmann, M.%
\end{APACrefauthors}%
\unskip\
\newblock
\APACrefYearMonthDay{2014}{}{}.
\newblock
{\BBOQ}\APACrefatitle {The Cell-Type Specific Cortical Microcircuit: Relating
  Structure and Activity in a Full-Scale Spiking Network Model} {The cell-type
  specific cortical microcircuit: Relating structure and activity in a
  full-scale spiking network model}.{\BBCQ}
\newblock
\APACjournalVolNumPages{Cerebral Cortex (New York, N.Y.:
  1991)}{24}{3}{785-806}.
\PrintBackRefs{\CurrentBib}

\bibitem [\protect \citeauthoryear {%
Ramirez%
\ \BBA {} Paninski%
}{%
Ramirez%
\ \BBA {} Paninski%
}{%
{\protect \APACyear {2014}}%
}]{%
ramirezFastInferenceGeneralized2014}
\APACinsertmetastar {%
ramirezFastInferenceGeneralized2014}%
\begin{APACrefauthors}%
Ramirez, A\BPBI D.%
\BCBT {}\ \BBA {} Paninski, L.%
\end{APACrefauthors}%
\unskip\
\newblock
\APACrefYearMonthDay{2014}{}{}.
\newblock
{\BBOQ}\APACrefatitle {Fast Inference in Generalized Linear Models via Expected
  Log-Likelihoods} {Fast inference in generalized linear models via expected
  log-likelihoods}.{\BBCQ}
\newblock
\APACjournalVolNumPages{Journal of Computational
  Neuroscience}{36}{2}{215--234}.
\PrintBackRefs{\CurrentBib}

\bibitem [\protect \citeauthoryear {%
Rule%
, Schnoerr%
, Hennig%
\BCBL {}\ \BBA {} Sanguinetti%
}{%
Rule%
\ \protect \BOthers {.}}{%
{\protect \APACyear {2019}}%
}]{%
ruleNeuralFieldModels2019}
\APACinsertmetastar {%
ruleNeuralFieldModels2019}%
\begin{APACrefauthors}%
Rule, M\BPBI E.%
, Schnoerr, D.%
, Hennig, M\BPBI H.%
\BCBL {}\ \BBA {} Sanguinetti, G.%
\end{APACrefauthors}%
\unskip\
\newblock
\APACrefYearMonthDay{2019}{{\APACmonth{11}}}{}.
\newblock
{\BBOQ}\APACrefatitle {Neural Field Models for Latent State Inference:
  {{Application}} to Large-Scale Neuronal Recordings} {Neural field models for
  latent state inference: {{Application}} to large-scale neuronal
  recordings}.{\BBCQ}
\newblock
\APACjournalVolNumPages{PLOS Computational Biology}{15}{11}{e1007442}.
\PrintBackRefs{\CurrentBib}

\bibitem [\protect \citeauthoryear {%
Salvatier%
, Wiecki%
\BCBL {}\ \BBA {} Fonnesbeck%
}{%
Salvatier%
\ \protect \BOthers {.}}{%
{\protect \APACyear {2016}}%
}]{%
salvatierProbabilisticProgrammingPython2016}
\APACinsertmetastar {%
salvatierProbabilisticProgrammingPython2016}%
\begin{APACrefauthors}%
Salvatier, J.%
, Wiecki, T\BPBI V.%
\BCBL {}\ \BBA {} Fonnesbeck, C.%
\end{APACrefauthors}%
\unskip\
\newblock
\APACrefYearMonthDay{2016}{}{}.
\newblock
{\BBOQ}\APACrefatitle {Probabilistic Programming in {{Python}} Using {{PyMC3}}}
  {Probabilistic programming in {{Python}} using {{PyMC3}}}.{\BBCQ}
\newblock
\APACjournalVolNumPages{PeerJ Computer Science}{2}{}{e55}.
\PrintBackRefs{\CurrentBib}

\bibitem [\protect \citeauthoryear {%
Schwalger%
\ \BBA {} Chizhov%
}{%
Schwalger%
\ \BBA {} Chizhov%
}{%
{\protect \APACyear {2019}}%
}]{%
schwalgerMindLastSpike2019}
\APACinsertmetastar {%
schwalgerMindLastSpike2019}%
\begin{APACrefauthors}%
Schwalger, T.%
\BCBT {}\ \BBA {} Chizhov, A\BPBI V.%
\end{APACrefauthors}%
\unskip\
\newblock
\APACrefYearMonthDay{2019}{{\APACmonth{10}}}{}.
\newblock
{\BBOQ}\APACrefatitle {Mind the Last Spike \textemdash{} Firing Rate Models for
  Mesoscopic Populations of Spiking Neurons} {Mind the last spike \textemdash{}
  firing rate models for mesoscopic populations of spiking neurons}.{\BBCQ}
\newblock
\APACjournalVolNumPages{Current Opinion in Neurobiology}{58}{}{155-166}.
\PrintBackRefs{\CurrentBib}

\bibitem [\protect \citeauthoryear {%
Schwalger%
, Deger%
\BCBL {}\ \BBA {} Gerstner%
}{%
Schwalger%
\ \protect \BOthers {.}}{%
{\protect \APACyear {2017}}%
}]{%
schwalgerTheoryCorticalColumns2017}
\APACinsertmetastar {%
schwalgerTheoryCorticalColumns2017}%
\begin{APACrefauthors}%
Schwalger, T.%
, Deger, M.%
\BCBL {}\ \BBA {} Gerstner, W.%
\end{APACrefauthors}%
\unskip\
\newblock
\APACrefYearMonthDay{2017}{}{}.
\newblock
{\BBOQ}\APACrefatitle {Towards a Theory of Cortical Columns: {{From}} Spiking
  Neurons to Interacting Neural Populations of Finite Size} {Towards a theory
  of cortical columns: {{From}} spiking neurons to interacting neural
  populations of finite size}.{\BBCQ}
\newblock
\APACjournalVolNumPages{PLOS Computational Biology}{13}{4}{e1005507}.
\PrintBackRefs{\CurrentBib}

\bibitem [\protect \citeauthoryear {%
Skilling%
}{%
Skilling%
}{%
{\protect \APACyear {2006}}%
}]{%
skillingNestedSamplingGeneral2006}
\APACinsertmetastar {%
skillingNestedSamplingGeneral2006}%
\begin{APACrefauthors}%
Skilling, J.%
\end{APACrefauthors}%
\unskip\
\newblock
\APACrefYearMonthDay{2006}{}{}.
\newblock
{\BBOQ}\APACrefatitle {Nested Sampling for General {{Bayesian}} Computation}
  {Nested sampling for general {{Bayesian}} computation}.{\BBCQ}
\newblock
\APACjournalVolNumPages{Bayesian analysis}{1}{4}{833--859}.
\PrintBackRefs{\CurrentBib}

\bibitem [\protect \citeauthoryear {%
Sussillo%
\ \BBA {} Barak%
}{%
Sussillo%
\ \BBA {} Barak%
}{%
{\protect \APACyear {2012}}%
}]{%
sussilloOpeningBlackBox2012}
\APACinsertmetastar {%
sussilloOpeningBlackBox2012}%
\begin{APACrefauthors}%
Sussillo, D.%
\BCBT {}\ \BBA {} Barak, O.%
\end{APACrefauthors}%
\unskip\
\newblock
\APACrefYearMonthDay{2012}{}{}.
\newblock
{\BBOQ}\APACrefatitle {Opening the {{black box}}: {{Low}}-{{dimensional
  dynamics}} in {{high}}-{{dimensional recurrent neural networks}}} {Opening
  the {{black box}}: {{Low}}-{{dimensional dynamics}} in {{high}}-{{dimensional
  recurrent neural networks}}}.{\BBCQ}
\newblock
\APACjournalVolNumPages{Neural Computation}{25}{3}{626-649}.
\PrintBackRefs{\CurrentBib}

\bibitem [\protect \citeauthoryear {%
Talts%
, Betancourt%
, Simpson%
, Vehtari%
\BCBL {}\ \BBA {} Gelman%
}{%
Talts%
\ \protect \BOthers {.}}{%
{\protect \APACyear {2018}}%
}]{%
taltsValidatingBayesianInference2018}
\APACinsertmetastar {%
taltsValidatingBayesianInference2018}%
\begin{APACrefauthors}%
Talts, S.%
, Betancourt, M.%
, Simpson, D.%
, Vehtari, A.%
\BCBL {}\ \BBA {} Gelman, A.%
\end{APACrefauthors}%
\unskip\
\newblock
\APACrefYearMonthDay{2018}{{\APACmonth{04}}}{}.
\newblock
{\BBOQ}\APACrefatitle {Validating {{Bayesian Inference Algorithms}} with
  {{Simulation}}-{{Based Calibration}}} {Validating {{Bayesian Inference
  Algorithms}} with {{Simulation}}-{{Based Calibration}}}.{\BBCQ}
\newblock
\APACjournalVolNumPages{arXiv:1804.06788 [stat]}{}{}{}.
\PrintBackRefs{\CurrentBib}

\bibitem [\protect \citeauthoryear {%
Team%
\ \protect \BOthers {.}}{%
Team%
\ \protect \BOthers {.}}{%
{\protect \APACyear {2016}}%
}]{%
thetheanodevelopmentteamTheanoPythonFramework2016}
\APACinsertmetastar {%
thetheanodevelopmentteamTheanoPythonFramework2016}%
\begin{APACrefauthors}%
Team, T\BPBI T\BPBI D.%
, {Al-Rfou}, R.%
, Alain, G.%
, Almahairi, A.%
, Angermueller, C.%
, Bahdanau, D.%
\BDBL {}Zhang, Y.%
\end{APACrefauthors}%
\unskip\
\newblock
\APACrefYearMonthDay{2016}{}{}.
\newblock
{\BBOQ}\APACrefatitle {Theano: {{A Python}} Framework for Fast Computation of
  Mathematical Expressions} {Theano: {{A Python}} framework for fast
  computation of mathematical expressions}.{\BBCQ}
\newblock
\APACjournalVolNumPages{arXiv:1605.02688 [cs]}{}{}{}.
\PrintBackRefs{\CurrentBib}

\bibitem [\protect \citeauthoryear {%
{van Haasteren}%
}{%
{van Haasteren}%
}{%
{\protect \APACyear {2014}}%
}]{%
vanhaasterenMarginalLikelihoodCalculation2014}
\APACinsertmetastar {%
vanhaasterenMarginalLikelihoodCalculation2014}%
\begin{APACrefauthors}%
{van Haasteren}, R.%
\end{APACrefauthors}%
\unskip\
\newblock
\APACrefYearMonthDay{2014}{}{}.
\newblock
{\BBOQ}\APACrefatitle {Marginal {{likelihood calculation}} with {{MCMC
  methods}}} {Marginal {{likelihood calculation}} with {{MCMC
  methods}}}.{\BBCQ}
\newblock
\BIn{} R.~{van Haasteren}\ (\BED), \APACrefbtitle {Gravitational {{Wave
  Detection}} and {{Data Analysis}} for {{Pulsar Timing Arrays}}}
  {Gravitational {{Wave Detection}} and {{Data Analysis}} for {{Pulsar Timing
  Arrays}}}\ (\BPG~99-120).
\newblock
\APACaddressPublisher{{Berlin, Heidelberg}}{{Springer Berlin Heidelberg}}.
\PrintBackRefs{\CurrentBib}

\bibitem [\protect \citeauthoryear {%
Vogels%
, Rajan%
\BCBL {}\ \BBA {} Abbott%
}{%
Vogels%
\ \protect \BOthers {.}}{%
{\protect \APACyear {2005}}%
}]{%
vogelsNeuralNetworkDynamics2005}
\APACinsertmetastar {%
vogelsNeuralNetworkDynamics2005}%
\begin{APACrefauthors}%
Vogels, T\BPBI P.%
, Rajan, K.%
\BCBL {}\ \BBA {} Abbott, L\BPBI F.%
\end{APACrefauthors}%
\unskip\
\newblock
\APACrefYearMonthDay{2005}{}{}.
\newblock
{\BBOQ}\APACrefatitle {Neural {{Network Dynamics}}} {Neural {{Network
  Dynamics}}}.{\BBCQ}
\newblock
\APACjournalVolNumPages{Annual Review of Neuroscience}{28}{1}{357-376}.
\PrintBackRefs{\CurrentBib}

\bibitem [\protect \citeauthoryear {%
Waibel%
, Hanazawa%
, Hinton%
, Shikano%
\BCBL {}\ \BBA {} Lang%
}{%
Waibel%
\ \protect \BOthers {.}}{%
{\protect \APACyear {1989}}%
}]{%
waibelPhonemeRecognitionUsing1989}
\APACinsertmetastar {%
waibelPhonemeRecognitionUsing1989}%
\begin{APACrefauthors}%
Waibel, A.%
, Hanazawa, T.%
, Hinton, G.%
, Shikano, K.%
\BCBL {}\ \BBA {} Lang, K\BPBI J.%
\end{APACrefauthors}%
\unskip\
\newblock
\APACrefYearMonthDay{1989}{}{}.
\newblock
{\BBOQ}\APACrefatitle {Phoneme Recognition Using Time-Delay Neural Networks}
  {Phoneme recognition using time-delay neural networks}.{\BBCQ}
\newblock
\APACjournalVolNumPages{IEEE Transactions on Acoustics, Speech, and Signal
  Processing}{37}{3}{328-339}.
\PrintBackRefs{\CurrentBib}

\bibitem [\protect \citeauthoryear {%
Wallace%
, Benayoun%
, van Drongelen%
\BCBL {}\ \BBA {} Cowan%
}{%
Wallace%
\ \protect \BOthers {.}}{%
{\protect \APACyear {2011}}%
}]{%
wallaceEmergentOscillationsNetworks2011}
\APACinsertmetastar {%
wallaceEmergentOscillationsNetworks2011}%
\begin{APACrefauthors}%
Wallace, E.%
, Benayoun, M.%
, van Drongelen, W.%
\BCBL {}\ \BBA {} Cowan, J\BPBI D.%
\end{APACrefauthors}%
\unskip\
\newblock
\APACrefYearMonthDay{2011}{}{}.
\newblock
{\BBOQ}\APACrefatitle {Emergent {{oscillations}} in {{networks}} of
  {{stochastic spiking neurons}}} {Emergent {{oscillations}} in {{networks}} of
  {{stochastic spiking neurons}}}.{\BBCQ}
\newblock
\APACjournalVolNumPages{PLOS ONE}{6}{5}{e14804}.
\PrintBackRefs{\CurrentBib}

\bibitem [\protect \citeauthoryear {%
Wilson%
\ \BBA {} Cowan%
}{%
Wilson%
\ \BBA {} Cowan%
}{%
{\protect \APACyear {1972}}%
}]{%
wilsonExcitatoryInhibitoryInteractions1972}
\APACinsertmetastar {%
wilsonExcitatoryInhibitoryInteractions1972}%
\begin{APACrefauthors}%
Wilson, H\BPBI R.%
\BCBT {}\ \BBA {} Cowan, J\BPBI D.%
\end{APACrefauthors}%
\unskip\
\newblock
\APACrefYearMonthDay{1972}{}{}.
\newblock
{\BBOQ}\APACrefatitle {Excitatory and {{inhibitory interactions}} in
  {{localized populations}} of {{model neurons}}} {Excitatory and {{inhibitory
  interactions}} in {{localized populations}} of {{model neurons}}}.{\BBCQ}
\newblock
\APACjournalVolNumPages{Biophysical Journal}{12}{1}{1-24}.
\PrintBackRefs{\CurrentBib}

\bibitem [\protect \citeauthoryear {%
Wood%
}{%
Wood%
}{%
{\protect \APACyear {2010}}%
}]{%
woodStatisticalInferenceNoisy2010}
\APACinsertmetastar {%
woodStatisticalInferenceNoisy2010}%
\begin{APACrefauthors}%
Wood, S\BPBI N.%
\end{APACrefauthors}%
\unskip\
\newblock
\APACrefYearMonthDay{2010}{}{}.
\newblock
{\BBOQ}\APACrefatitle {Statistical Inference for Noisy Nonlinear Ecological
  Dynamic Systems} {Statistical inference for noisy nonlinear ecological
  dynamic systems}.{\BBCQ}
\newblock
\APACjournalVolNumPages{Nature}{466}{7310}{1102-1104}.
\PrintBackRefs{\CurrentBib}

\bibitem [\protect \citeauthoryear {%
Zhao%
\ \BBA {} Park%
}{%
Zhao%
\ \BBA {} Park%
}{%
{\protect \APACyear {2016}}%
}]{%
zhaoInterpretableNonlinearDynamic2016}
\APACinsertmetastar {%
zhaoInterpretableNonlinearDynamic2016}%
\begin{APACrefauthors}%
Zhao, Y.%
\BCBT {}\ \BBA {} Park, I\BPBI M.%
\end{APACrefauthors}%
\unskip\
\newblock
\APACrefYearMonthDay{2016}{}{}.
\newblock
{\BBOQ}\APACrefatitle {Interpretable {{nonlinear dynamic modeling}} of {{neural
  trajectories}}} {Interpretable {{nonlinear dynamic modeling}} of {{neural
  trajectories}}}.{\BBCQ}
\newblock
\BIn{} D\BPBI D.~Lee, M.~Sugiyama, U\BPBI V.~Luxburg, I.~Guyon\BCBL {}\ \BBA {}
  R.~Garnett\ (\BEDS), \APACrefbtitle {Advances in {{Neural Information
  Processing Systems}} 29} {Advances in {{Neural Information Processing
  Systems}} 29}\ (\BPGS\ 3333--3341).
\newblock
\APACaddressPublisher{}{{Curran Associates, Inc.}}
\PrintBackRefs{\CurrentBib}

\end{thebibliography}
\end{document}